\documentclass[useAMS,usenatbib,usegraphicx]{mn2e}

\newif\ifAMStwofonts

\usepackage{lscape}
\usepackage{subfigure}

\newcommand{\kms}{km\,s$^{-1}$}

\newcommand{\around}{$\sim$}

\newcommand{\Msun}{M$_{\odot}$}

\newcommand{\MHI}{M$\rm _{HI}$}

\newcommand{\Halpha}{H$\alpha$}

\newcommand{\Hbeta}{H$\beta$}

\newcommand{\Hdelta}{H$\delta$}

\newcommand{\Hgamma}{H$\gamma$}

\newcommand{\Lya}{Ly$\alpha$}

\newcommand{\microJy}{$\rm \mu$Jy}

\newcommand{\HI}{H{\sc i}}

\newcommand{\HII}{H{\sc ii}}

\newcommand{\degrees}{$^{\circ}$}

\newcommand{\AIPS}{{\sc aips}}

\newcommand{\Rtwo}{R$_{200}$}

\newcommand\apj{ApJ}
\newcommand\apjl{ApJ}
\newcommand\aap{A\&A}
\newcommand\aaps{A\&AS}

\title[\HI\ gas around Abell 370]{The \HI\ gas content of galaxies around Abell 370, a galaxy cluster at z~=~0.37}

\author[Lah et al.]
{Philip Lah$^1$\thanks{E-mail: plah@mso.anu.edu.au}, 
Michael B. Pracy$^1$,
Jayaram N. Chengalur$^2$, 
Frank H. Briggs$^{1}$, 
  \newauthor 
Matthew Colless$^3$,   
Roberto De Propris$^4$, 
Shaun Ferris$^1$,
Brian P. Schmidt$^1$
  \newauthor 
and Bradley E. Tucker$^1$ \\
\\
$^1$ Research School of Astronomy \& Astrophysics, The Australian National University, Weston Creek, ACT 2611, Australia \\ 
$^2$ National Centre for Radio Astrophysics, Post Bag 3, Ganeshkhind, Pune 411 007, India \\
$^3$ Anglo-Australian Observatory, PO Box 296, Epping, NSW 2111, Australia \\
$^4$ Cerro Tololo Inter-American Observatory, Casilla 603, La Serena, Chile \\
}  

\date{Accepted ...
      Received ...;
      in original form ...}

\pagerange{\pageref{firstpage}--\pageref{lastpage}}
\pubyear{2009}

\begin{document}

\maketitle

\label{firstpage}


\begin{abstract}

We used observations from the Giant Metrewave Radio Telescope to measure the atomic hydrogen gas content of 324 galaxies around the galaxy cluster Abell~370 at a redshift of z~=~0.37 (a look-back time of \around 4~billion years).  The \HI\
21-cm emission from these galaxies was measured by coadding their signals using precise optical redshifts obtained with the Anglo-Australian Telescope.  {The average \HI\ mass measured for all 324 galaxies is $(6.6 \pm 3.5) \times 10^9$~\Msun, while the average \HI\ mass measured for the 105 optically blue galaxies is $(19.0 \pm 6.5) \times 10^9$~\Msun. The significant quantities of gas found around Abell~370, suggest that there has been substantial evolution in the gas content of galaxy clusters since redshift z~=~0.37. The total amount of atomic hydrogen gas found around Abell~370 is up to \around 8~times more than that seen around the Coma cluster, a nearby galaxy cluster of similar size.}  Despite this higher gas content, Abell~370 shows the same trend as nearby clusters, that galaxies close to the cluster core have lower \HI\ gas content than galaxies further away where the galaxy density is lower.  The optically blue galaxies contain the majority of the \HI\ gas surrounding the cluster.  However, there is evidence that the optically red galaxies contain appreciable quantities of \HI\ gas within their central regions.  The Abell~370 galaxies have \HI\ mass to optical light ratios similar to local galaxy samples and have the same correlation between their star formation rate and \HI\ mass as found in nearby galaxies.  The average star formation rate derived from [OII] emission and from de-redshifted 1.4~GHz radio continuum for the Abell~370 galaxies also follows the correlation found in the local universe. The large amounts of \HI\ gas found around the cluster can easily be consumed entirely by the observed star formation rate in the galaxies over the \around 4~billion years (from z~=~0.37) to the present day.  Abell~370 appears set to evolve into a gas poor system similar to galaxy clusters observed in the local universe.

\end{abstract}




\section{Introduction}
\label{Introduction}

Galaxy properties, such as morphology and star formation rate, are found to depend on environmental density.  These effects probably originate from a density-dependent quenching of star formation which can come about via the consumption or removal of the available gas supply in the galaxies.  As such, the amount of \HI\ gas available to galaxies to fuel further star formation is likely to be a fundamental parameter in understanding the effect of environment.  Galaxies in nearby clusters and their surroundings show strong evidence for environmental effects on their gas content.  However, at higher redshifts the gas content of clusters has not been studied in detail.  Consequently, our understanding of the effect of environment on galaxy evolution has been incomplete.

As the number density of galaxies increases, the rate of star formation in the galaxies decreases.  This star formation--density correlation can be seen locally in the variation of the fraction of blue galaxies with galaxy density \citep{pimbblet02,depropris04} and in the fraction of emission line galaxies with galaxy density \citep{hashimoto98,lewis02,gomez03,balogh04,kauffmann04}.  At higher redshift the blue fraction of galaxies in the dense environment of clusters increases with redshift \citep{butcher84} as does the fraction of galaxies with emission lines \citep{couch87,balogh98,balogh99,poggianti99,dressler04}.  The amount of star formation in galaxies in general increases with redshift, with there being an order of magnitude increase in the cosmic star formation rate density between the present time and z~\around~1 \citep{lilly96,madau96,hopkins04}.  The trend that with higher galaxy density there are fewer galaxies with ongoing star formation continues to at least z~\around~0.8 \citep{poggianti08}.

In the local universe the proportion of galaxies which are ellipticals or S0 galaxies increases with galaxy number density and there is a corresponding decrease in spirals galaxies \citep{dressler80}.  This density--morphology relationship also evolves with redshift.  While the fraction of elliptical galaxies in dense cluster environments stays reasonably constant, the fraction of S0 galaxies decreases from present values by a factor of 2--3 by z~\around~0.5 and there is a proportional increase in the spiral galaxy fraction at the same time \citep{dressler97,fasano00}.

The physical mechanisms considered when attempting to understand these environment trends all involve processes that effect the gas content of the galaxies, the fuel supply for star formation.  These proposed physical mechanisms include: 

\vspace{0.3em}
\hspace{-1.5em}
(i) galaxy mergers and strong gravitational galaxy--galaxy interactions \citep{toomre72}.  

\hspace{-1.5em}
(ii) galaxy harassment, which is the cumulative effect of tidal forces from many weak galaxy encounters \citep{richstone76,farouki81,moore96,moore98}.  

\hspace{-1.5em}
(iii) interactions between a galaxy and the inter-galactic medium (IGM), which includes ram pressure stripping, viscous stripping and thermal evaporation \citep{gunn72,quilis00}.  

\hspace{-1.5em}
(iv) strangulation, which is the removal of any envelope of hot gas surrounding galaxies that was destined to cool and accrete on to the galaxy to fuel further star formation \citep{larson80,diaferio01}.  
\vspace{0.3em}

These mechanisms effect the gas either by stimulating star formation (which rapidly consumes the gas, locking it up in stars) or by removing the gas from the galaxy (and ionising it) or by preventing further accretion.  Unlike the other mechanisms listed which quench star formation relatively quickly (\around $10^7$~years), strangulation causes a slow decline in star formation over longer timescales ($>$1~Gyr) \citep{poggianti04}.

Neutral atomic hydrogen gas (\HI) is a large component of the gas content of galaxies and can be directly quantified from \HI\ 21-cm emission.  In the central regions of nearby clusters, late-type galaxies are found to be \HI\ deficient compared to similar galaxies in the field \citep{haynes84}.  This effect continues well outside the cluster cores with a gradual change in the \HI\ content of galaxies with galaxy density \citep{solanes01}.  Galaxies in nearby clusters show evidence of disruption of their \HI\ gas with the presence of unusual asymmetric \HI\ gas distributions, spatial offsets between the \HI\ and optical disks, and tails of \HI\ gas streaming away from the galaxies \citep{cayatte90,bravo-alfaro00,chung07}.  {Despite these gas depletion trends with galaxy number density in the local universe, the high \HI\ mass galaxies are found to trace the same galaxy densities as the optical galaxies while the low \HI\ mass galaxies trace low density environments \citep{basilakos07}.}  This pattern is likely to continue at higher redshift making the search for high \HI\ mass galaxies easier around optical overdensities of galaxies.

Observing \HI\ 21-cm emission from galaxies at cosmological distances (z~$>~0.1$) is difficult due to the weak flux of the line.  Only a single galaxy with \HI\ 21-cm emission was detected in the rich cluster Abell~2218 at z~=~0.18 using 216~hours with the Westerbork Synthesis Radio Telescope (WSRT) \citep{zwaan01}.  Again only a single galaxy with \HI\ 21-cm emission was detected in the galaxy cluster Abell~2192 at z~=~0.19 using \around 80~hours with the Very Large Array (VLA) \citep{verheijen04}.  After an upgrade of the WSRT, a new pilot study was able to detect \HI\ 21-cm emission from 19 galaxies in Abell~963 at z~=~0.21 using 240~hours and 23 galaxies in Abell~2192 at z~=~0.19 using 180~hours \citep{verheijen07}.  Additionally, the upgraded Arecibo radio telescope has detected \HI\ 21-cm emission at redshifts between z~=~0.17 to 0.25 from \around 20 isolated galaxies selected in the optical from the Sloan Digital Sky Survey \citep{catinella08}.  While the numbers of \HI\ 21-cm emission galaxies directly detected at cosmological distances has been increasing with time, the numbers are still small and limited to the most gas rich systems.  

To quantify the \HI\ gas content of large numbers of galaxies at cosmologically interesting redshifts,  we have been coadding the \HI\ 21-cm emission from multiple galaxies using their observed optical positions and redshifts.  The coadding technique for measuring neutral atomic hydrogen gas has been used previously in galaxy cluster Abell~2218 at $\rm z = 0.18$ \citep{zwaan00} and in Abell~3128 at $\rm z = 0.06$ \citep{chengalur01}.  We expanded this technique to field galaxies with active star formation at z~=~0.24 \citep{lah07}. 

The goal of this work is to quantify the evolution of the \HI\ gas in galaxies in a variety of environments.  At moderate redshifts the angular extent of galaxy clusters is sufficiently small enough that galaxy environments from the dense cluster core to almost field galaxy densities can be observed in a single radio telescope pointing.  In this paper we are applying the \HI\ coadding technique to galaxies surrounding Abell~370, a galaxy cluster at z~=~0.37 (a look-back time of \around 4.0~billion years).   Abell~370 is a large galaxy cluster, similar in size and mass to the nearby Coma cluster.  The \HI\ gas content of galaxies at different distances from the Abell~370 cluster core are examined, in particular that of galaxies inside and outside the extent of the hot, X-ray emitting, intracluster gas as well as that of galaxies inside and outside the cluster \Rtwo\ radius (the radius at which the galaxy density is 200 times the general field).

In this paper, the cluster centre of Abell~370 has been set as the mid point between the two cD galaxies which is at right ascension (R.A.) $\rm 02^{h}39^{m}52.90^{s}$ declination (Dec.) $\rm -01^{\circ}34^{\prime}37.5^{\prime\prime}$~J2000.  This value is close to the centre determined from X-ray measurements \citep{ota04} and a good match to the velocity and surface density distribution of the galaxy data used in this paper.  The redshift of the cluster centre has been set as z~=~0.373 based on the galaxy redshift distribution we observed.  



We adopt the consensus cosmological parameters of $\rm \Omega_{\lambda} = 0.7$, $\rm \Omega_{M} = 0.3$ and $\rm H_0 = 70 \ km \, s^{-1} \, Mpc^{-1}$ throughout the paper. 

The structure of this paper is as follows. Section~\ref{The_Optical_Data} details the optical imaging and optical spectroscopy of the Abell~370 galaxies.  Section~\ref{The_Radio_Data} details the radio observations and data reduction.  Section~\ref{HI_all_galaxies} presents the measurement of the \HI\ 21-cm emission signal from all the Abell~370 galaxies with usable redshifts.  Section~\ref{The_HI_subsamples} presents the \HI\ signal from different subsamples of these galaxies.   Section~\ref{Comparison_of_the_HI_measurements_with_the_literature} details the comparison of the \HI\ results for Abell~370 with various literature measurements.  Section~\ref{Star_Formation_Rate_Results} elaborates on the star formation properties of the Abell~370 galaxies. Finally, Section~\ref{Conclusion} presents a summary and discussion of the results. 


\begin{table*} 

\centering

\begin{centering}

\begin{tabular}[b]{|c|c|c|c|}  

\hline 

\ &
Type of &
Date of &
Duration of \\

Telescope  & 
Observation &
Observations &
Observations \\

\hline %

Giant Metrewave Radio Telescope (GMRT) &
radio spectroscopy &
\ \ \ \ \ 10--17 August 2003 &
63 hours \\

Australian National University (ANU) 40 inch &
optical imaging &
\ \ \ \ 1--2 November 2005 &
\ 2 nights \\

Australian National University (ANU) 40 inch &
optical imaging &
19--24 September 2006 & 
\ 5 nights \\

Anglo-Australian Telescope (AAT) &
optical spectroscopy &
\ \ \ 11--14 October 2006 &
\ 4 nights \\

\hline 

\end{tabular}

\caption{List of telescope observations used in this paper.} 

\label{observations_log}  

\end{centering}
\end{table*}



\section{The Optical Data} 
\label{The_Optical_Data}

\subsection{The optical imaging and galaxy target selection} 
\label{The_optical_imaging} 

 
\begin{figure}  

  \begin{center}  
  \leavevmode  
		
    \includegraphics[width=8cm]{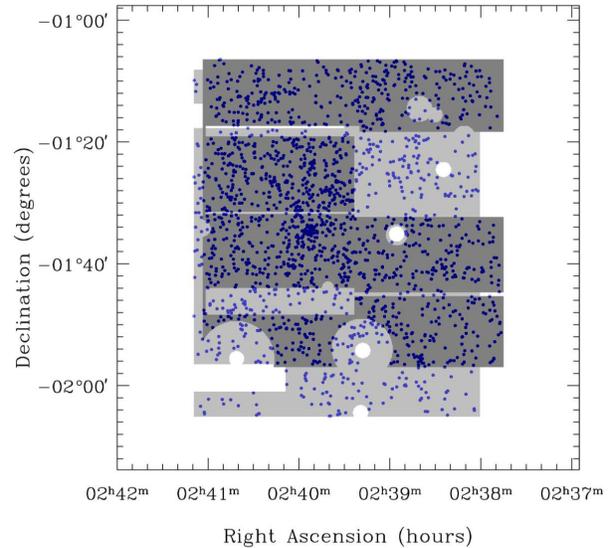}
  
   \end{center}

   \caption{This figure shows the extent of the optical imaging taken with the ANU 40~inch telescope.  The dark grey areas were observed in 2005 November; the light grey areas are the additional areas covered in 2006 September.  The circles are the regions around bright stars where galaxy identification could not be made.  The circles are larger in the 2005 November observations as the better seeing caused the brightest stars to saturate the CCD.  The small points are the 1877 objects for which redshifts were obtained during the AAOmega run in 2006 October.}

   \label{exclusion_WFI_galaxies_all}

\end{figure}


Wide-field and moderately deep imaging of the galaxy cluster Abell~370 and its surroundings were obtained using the Wide Field Imager (WFI) on the Australian National University (ANU) 40~inch telescope at Siding Spring Observatory.  There were two separate observing runs: the first in 2005 November 1--2, and the second in 2006 September 19--24.  Both sets of observations were taken through the standard broad passband filters $V$, $R$ and $I$. 

The Wide Field Imager has a 52~by 52~arcmin field-of-view which is sampled by a mosaic of 8 2k$\times$4k CCDs.  One of these detectors was not functioning leaving a 12~arcmin by 24~arcmin gap in the imaged field.  In 2005 November only a single WFI pointing was observed, leaving in a gap in the imaging due to the missing CCD.  The conditions during this run were close to photometric and the seeing in the final combined imaging was \around 2.1~arcsec.  The average $3 \sigma$ surface brightness limit of this imaging was 25.5~mag~arcsec$^{-2}$ in $V$, 25.3 in $R$ and 24.2 in $I$.  The 2006 September WFI observations consisted of two pointings that filled in the gap from the missing WFI detector and extended the imaged region.  The observing conditions were poor during this run with the seeing in the final combined images being \around 2.8~arcsec.  The average $3 \sigma$ surface brightness limit of this later imaging was 24.1~mag~arcsec$^{-2}$ in $V$, 23.8 in $R$ and 22.7 in $I$.  The extent of the WFI imaging from both runs is shown in Fig.~\ref{exclusion_WFI_galaxies_all}.  The total imaged area was \around 51 by \around 59 arcmin.

Targets were selected from the WFI imaging for optical spectroscopic follow up.  Initial target selection was done on the $I$~band images with confirmation of the targets in the $R$ and $V$~bands.  There were many spurious object detections around bright stars, diffraction spikes and chip defects.  To remove these a series of rectangular and circular exclusion regions were defined around the problem areas as seen in Fig.~\ref{exclusion_WFI_galaxies_all}. 

Objects brighter than a $V$~band total magnitude of 19.2 were removed from list of targets for spectroscopic follow up.  This limit is 0.6 magnitudes fainter than the cD galaxies in Abell~370 and no other cluster members are expected to be this luminous (Note: the two cD were included for spectroscopic follow up as additional targets).  Objects brighter than $V$~band total magnitude of 21.7 were used for target selection.  A small number of objects were removed from the target list that were substantially redder than galaxies belonging to the red sequence of Abell~370 and hence likely to be stars or distant background galaxies.

Aperture magnitudes were measured using 3~arcsec diameter apertures.  Total magnitudes were measured using the best magnitude parameter of SExtractor.  All magnitudes were calibrated on the Vega system.  Comparisons between the total and aperture magnitudes were used to remove stars from the galaxy sample.  The optical magnitudes of the galaxies were corrected for the foreground extinction from the Galaxy using the Schlegel--Finkbeiner--Davis Galactic reddening map \citep{schlegel98}.  No correction was done for the internal extinction of the galaxies and it is assumed that any correction would have minimal effect on our results.  Separate lists of targets were made from each set of the WFI imaging.  Only after applying all the above selection criteria were the target catalogues merged.  {The photometry is consistent for objects that appear in both catalogues.}


\subsection{The optical spectroscopic data}
\label{The_optical_spectroscopic_data}

Over 4 nights from 2006 October 11--14, the targets were observed with AAOmega, the fibre-fed, dual-beam spectrograph on the Anglo-Australian Telescope.  Eleven fibre configurations were observed for 2 hours each.  Optical spectra were obtained for a total of 2347 unique targets.  The combination of the blue and red arms of the spectrograph provided continuous wavelength coverage from \around 3700~\AA\ to \around 8800~\AA.  At the redshift of the cluster (z~=~0.37) this covers the rest frame wavelength region from \around 2700~\AA\ to \around 6400~\AA\ which includes the emission lines of [OII]$\lambda$3727, \Hbeta, [OIII]$\lambda$4959 and [OIII]$\lambda$5007 (but not \Halpha) as well as the absorption features Ca H\&K, Na, Mg, \Hdelta, G--band and \Hgamma. 

 
\begin{figure}  

  \begin{center}  
  \leavevmode  
		
    \includegraphics[width=8cm]{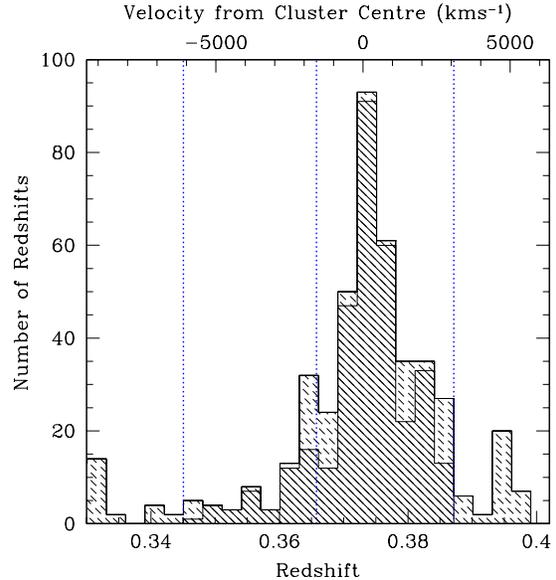}
  
   \end{center}

   \caption{This figure shows the distribution of the 450 redshifts around the galaxy cluster Abell~370.  The vertical dotted lines are the GMRT frequency limits converted to their \HI\ redshift; the central dotted line is the boundary between the upper and lower sidebands of the GMRT radio data.  The histogram shaded with the unbroken line is the distribution of the 324 redshifts used in \HI\ coadding.  The histogram shaded with the broken line is the distribution of the unusable redshifts (see Section \ref{HI_all_galaxies} for details).  The top x-axis shows the velocity from the cluster centre, giving an indication of the peculiar motion of the galaxies within the cluster.  The cluster centre is at $\rm z_{cl}$~=~0.373 and the listed velocity includes the cosmological correction (i.e.~it is divided by $\rm 1+z_{cl}$). } 

   \label{hist_z_A370}

\end{figure}


The data were reduced using the Anglo-Australian Observatory's {\sc 2dfdr} data reduction pipeline\footnote{Anglo-Australian Observatory website \\http://www.aao.gov.au/AAO/2df/aaomega/aaomega.html}  
and the redshift determination was done with the interactive software package {\sc runz}.  In total there were 1877 objects with secure redshifts (includes both galaxies and some foreground stars),  a redshift completeness of 80~per~cent.  The objects range in redshift from z~\around~0 to 1.2 and the error in the redshifts is \around 70~\kms.  There are 450 galaxies with redshifts between z~=~0.33 and z~=~0.40  (see Fig.~\ref{hist_z_A370} for the redshift distribution);  324 of these galaxies were usable for \HI\ coadding (see Section \ref{HI_all_galaxies} for details on this selection).


\subsection{The optical properties of the Abell 370 galaxies}
\label{The_optical_properties_of_the_Abell_370_galaxies}

 
\begin{figure}  

  \begin{center}  
  \leavevmode  
		
    \includegraphics[width=8cm]{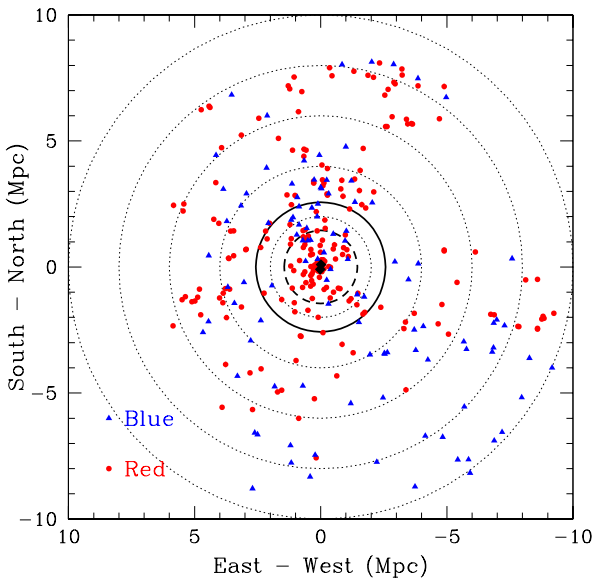}
 
   \end{center}

   \caption{This figure shows the 324 galaxies used in the \HI~coadding plotted as projected distance in Mpc from the galaxy cluster centre.  The triangular points are the blue galaxies and the circular points are the red galaxies.  The two diamond points near the centre are the two cD galaxies.  The dashed circle is the 3$\sigma$ extent of the X-ray gas \citep{ota04} and is at a radius of 1.45~Mpc.  The solid circle is the R$_{200}$ radius of 2.57~Mpc.  The faint, dotted circles are at radii of 2, 4, 6, 8 \& 10 Mpc from the cluster centre. 
}

   \label{A370_radius_Mpc}

\end{figure}


Fig.~\ref{A370_radius_Mpc} shows the 324 galaxies used in the \HI\ coadding plotted as projected distance in Mpc from the galaxy cluster centre.  Plotted on this figure is the extent of the hot intracluster gas, the X-ray significance radius.  This is the radius from the cluster centre where the X-ray emission surface brightness has fallen to three times the background sky level (the 3$\sigma$ extent of the X-ray gas).  For the cluster Abell~370 this radius is at 1.45 Mpc (4.7 arcmin) \citep{ota04}.  Also plotted on this figure is the \Rtwo\ radius, the radius at which the cluster is 200 times denser than the general field \citep{carlberg97}.  For Abell~370 this is \Rtwo~=~2.57~Mpc which was derived from the cluster velocity dispersion of 1263~\kms.  The cluster velocity dispersion of Abell~370 was measured from the redshifts obtained in this work \citep{pracy08}.

For galaxies near the redshift of the cluster at z~=~0.37 inter-band K-corrections were performed using the methodology of \citet{kim96} and \citet{schmidt98}.  The observed $R$~band magnitudes were K-corrected to rest frame $B$~band and the observed $I$~band magnitudes to rest frame $V$~band.  Input spectra were created using linear combinations of pairs of galaxy template spectra that reproduce the observed galaxy $V-I$ colour within the photometric uncertainties.  The final K-correction value is derived from the mean of the corrections derived from these input spectra.

 
\begin{figure*}  

  \begin{center}  
  \leavevmode  

    \includegraphics[width=8.0cm]{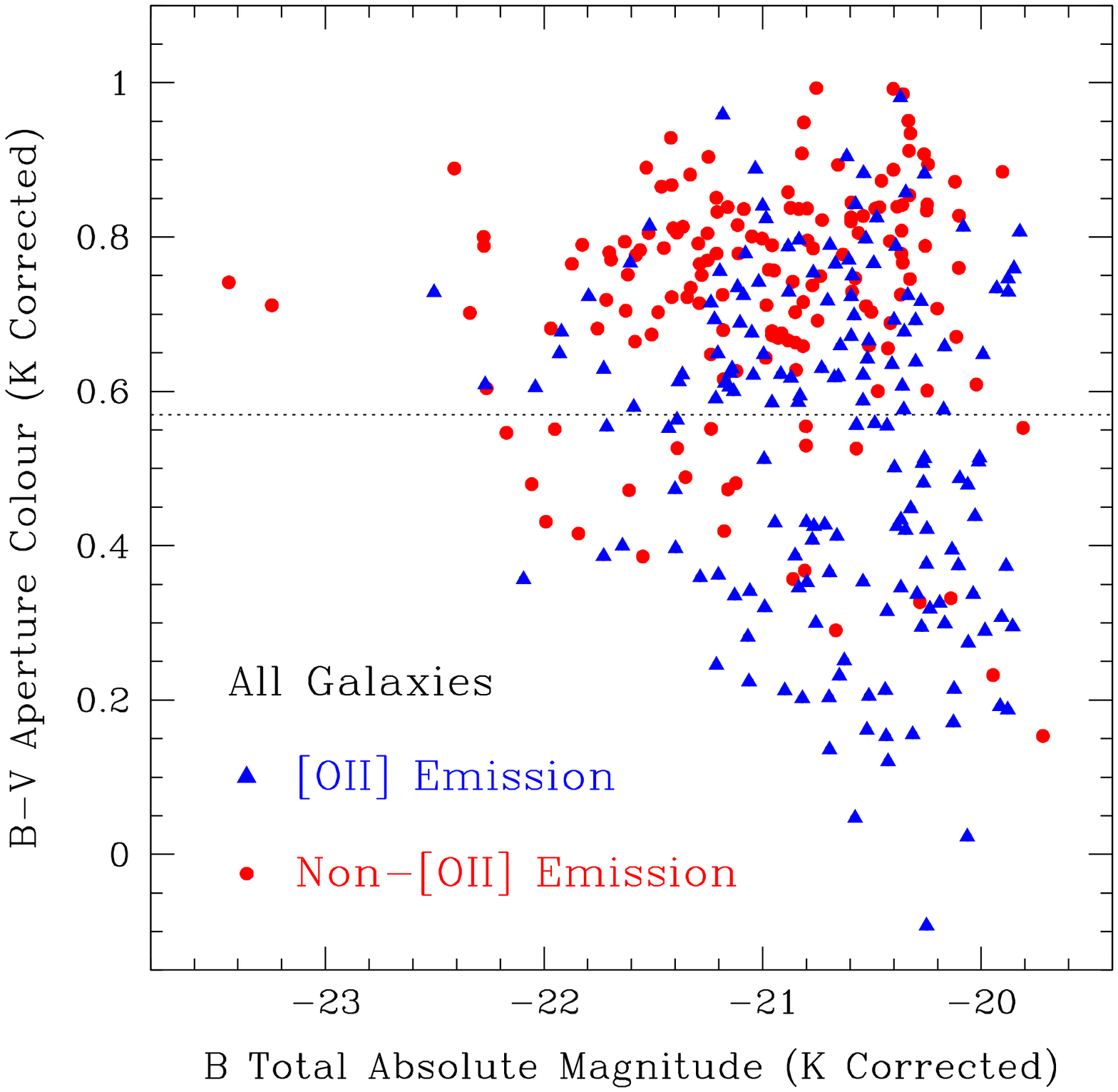}
    \includegraphics[width=8.0cm]{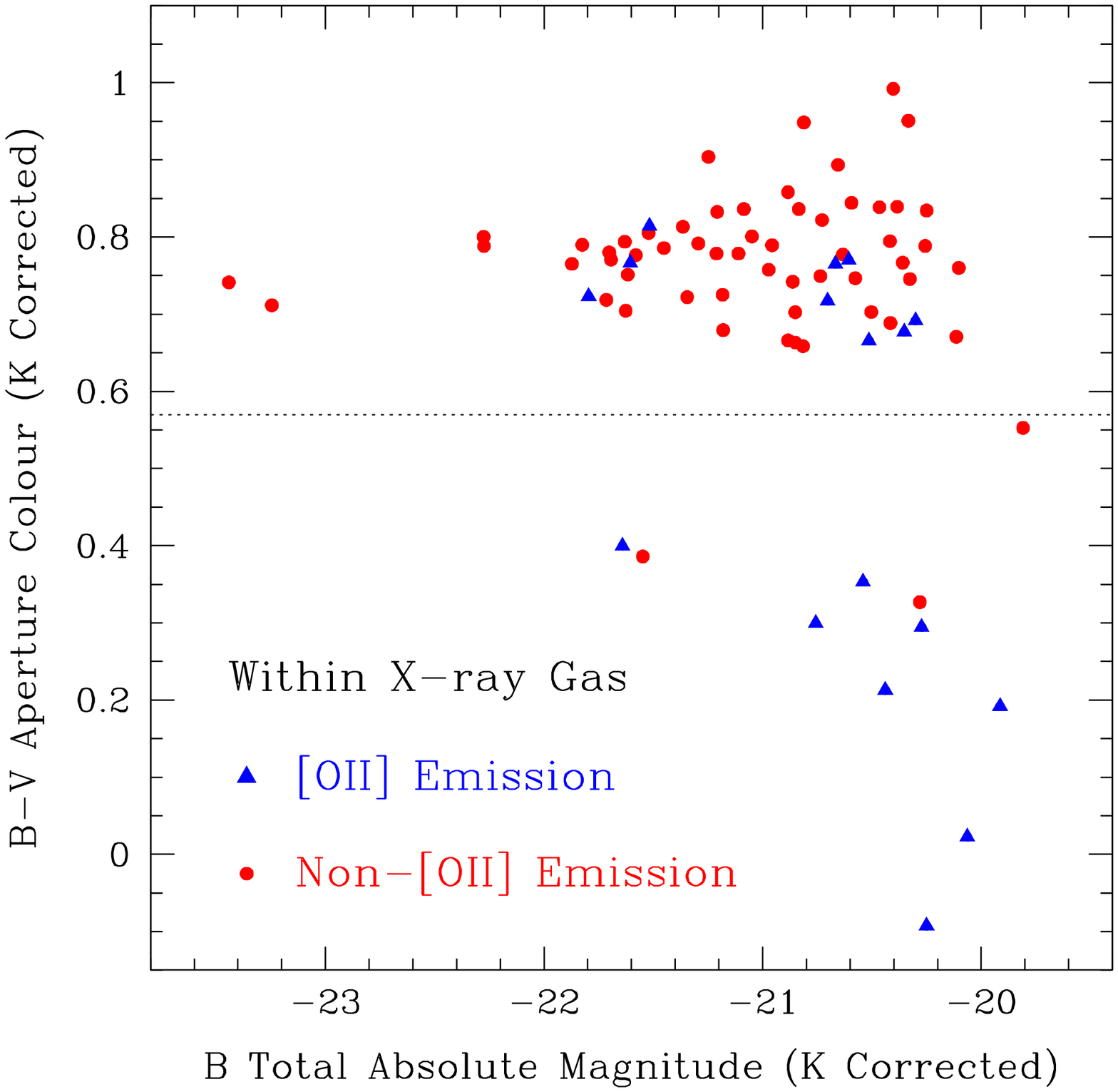}

   \end{center}

   \caption{In the left panel is the $B-V$ aperture colour vs.\ $B$~band total absolute magnitude for the 324 galaxies used in the \HI\ coadding.  The right panel shows the same for the 75 galaxies close to the galaxy cluster centre (within a projected distance from the cluster centre of 1.45~Mpc, the 3$\sigma$ extent of the X-ray gas).  The triangular points are galaxies with measured [OII] equivalent widths~$\rm > 5$~\AA\ and the circular points are those with [OII] equivalent widths~$\rm \le 5$~\AA.  The dotted line is the dividing line used to separate the blue and red galaxies at $B-V$~=~0.57.}

   \label{colour_mag_diagram}

\end{figure*}


The $B-V$ colour vs.\ $B$~band magnitude diagram for all the 324 galaxies used in the \HI\ coadding can be seen in the left panel of Fig.~\ref{colour_mag_diagram} and for those galaxies within 1.45~Mpc of the galaxy cluster centre (the X-ray significance radius) in the right panel.  A least absolute deviation regression fit was made to the cluster ridge-line in the colour--magnitude diagram containing those galaxies within 1.45~Mpc of the cluster centre.  The slope of this linear fit was close to zero, so a fixed colour term for the ridge-line of $B-V$~=~0.77 was used.  This almost zero slope is probably due to the limited magnitude range used and the large errors in the optical magnitudes (\around 0.1 mag).  The galaxies were divided into blue and red galaxies using the Butcher--Oemler condition, i.e.~the blue galaxies are those at least 0.2 magnitudes bluer than the cluster ridge-line \citep{butcher84}.  This sets the division at $B-V$~=~0.57.  In Fig.~\ref{A370_radius_Mpc}, which shows the projected distance distribution of the galaxies,  the blue and red galaxies are plotted using different symbols.  Near the cluster core the red galaxies dominate.  Away from the cluster centre the number of blue galaxies increases though a large number of these outer galaxies are red.  The blue fraction of galaxies in the cluster using the Butcher--Oemler criteria is $0.132 \pm 0.003$ within the measured R$_{30}$ radius of 0.669~Mpc (2.17~arcmin).  (R$_{30}$ is defined as the radius containing 30~per~cent of the projected galaxies that lie within 3-Mpc of the cluster centre.)   The blue fraction found in the majority of nearby clusters is \around 0.03 \citep{butcher84}, substantially lower than that found in Abell~370.

 
\begin{figure}  

  \begin{center}  
  \leavevmode  
		
    \includegraphics[width=8cm]{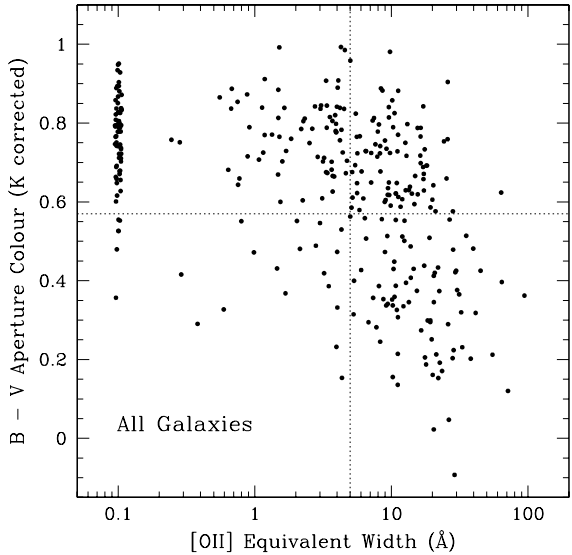}
 
   \end{center}

   \caption{This figure shows the $B-V$ aperture colour vs.\ [OII] equivalent width for all the 324 galaxies used in the \HI\ coadding.  The 68 galaxies that have measured [OII] equivalent widths equal to or less than zero have been given a value of \around 0.1~\AA\ so that they can be plotted on the log scale used here.  The dashed vertical line is the [OII] cut used to separate emission from non-emission galaxies (at 5~\AA) and the horizontal line is the colour cut used to separate blue and red galaxies (at 0.57~mag).}

   \label{OII_BV_colour}

\end{figure}


The equivalent width of the [OII]$\lambda$3727 emission line was measured from the AAOmega optical spectra using the standard flux-summing technique.  The galaxies were divided into an [OII] emission and non-[OII] emission samples at an equivalent width of 5~\AA.  This is roughly the $2 \sigma$\ limit for the equivalent width measurements.  It was not possible to identify the active galactic nuclei (AGN) in this sample as the optical spectra at z~=~0.37 did not include the \Halpha\ and [NII] lines, which are required for the standard AGN diagnostic test \citep{baldwin81}.  In the colour--magnitude diagrams of Fig.~\ref{colour_mag_diagram} the [OII] emission and non-[OII] emission galaxies are plotted using different symbols.  Around the cluster centre (the right panel of Fig.~\ref{colour_mag_diagram}) there are few [OII] emission galaxies.  For all 324 galaxies (the left panel of Fig.~\ref{colour_mag_diagram}) the majority of the blue galaxies have [OII] emission but the red galaxies are almost evenly split between emission and non-emission galaxies.  The $B-V$ colour vs.\ [OII] equivalent width for all the 324 galaxies used in the \HI\ coadding can be seen in the Fig.~\ref{OII_BV_colour}.  A trend is seen with bluer galaxies having a larger [OII] equivalent width.  However there is a large amount of scatter in this relationship which is mostly due to real astrophysics variation, i.e.\ not due to random statistical errors.

The star formation rate for the galaxies at redshifts near Abell~370 was calculated from the [OII]$\lambda 3727$ emission line.  The [OII] line flux for the galaxies was estimated using the measured equivalent width from the optical spectra and the broad-band photometry from the optical imaging.  A correction for internal dust extinction in the galaxies was made to the [OII] line flux by assuming the canonical 1~mag extinction at H$\alpha$ \citep{kennicutt83,niklas97}.  This was converted to an extinction at the wavelength of [OII]$\lambda 3727$ using the extinction relation of \citet{calzetti97}.  This correction increases the [OII] flux by a factor of 5.15.  The [OII] line luminosity was converted to a star formation rate using Equation~(4) of \citet{kewley04}.  The errors in the derived star formation rates are based on a combination of the statistical errors in the spectral line fits, the intrinsic error in the conversion from [OII] luminosity to star formation rate, and the scatter in the internal dust extinction correction which is of order 50~per~cent \citep{kennicutt83}.  The dust correction dominates the error and limits the precision for the measured star formation rate of individual galaxies to a percentage error of at best 55~per~cent.  However, the average star formation rate of the galaxies can be derived with significantly higher precision, assuming the error is purely random, i.e.~there are no systematic offsets.  

For extended detail on the optical data reduction and analysis see \citet{pracy08}.


\subsection{The comparison of the properties of galaxy subsamples}
\label{The_overlap_of_the_galaxy_subsamples}

 
\begin{figure*}   

  \begin{center}  
  \leavevmode  

  \subfigure{		
    \includegraphics[width=6.2cm]{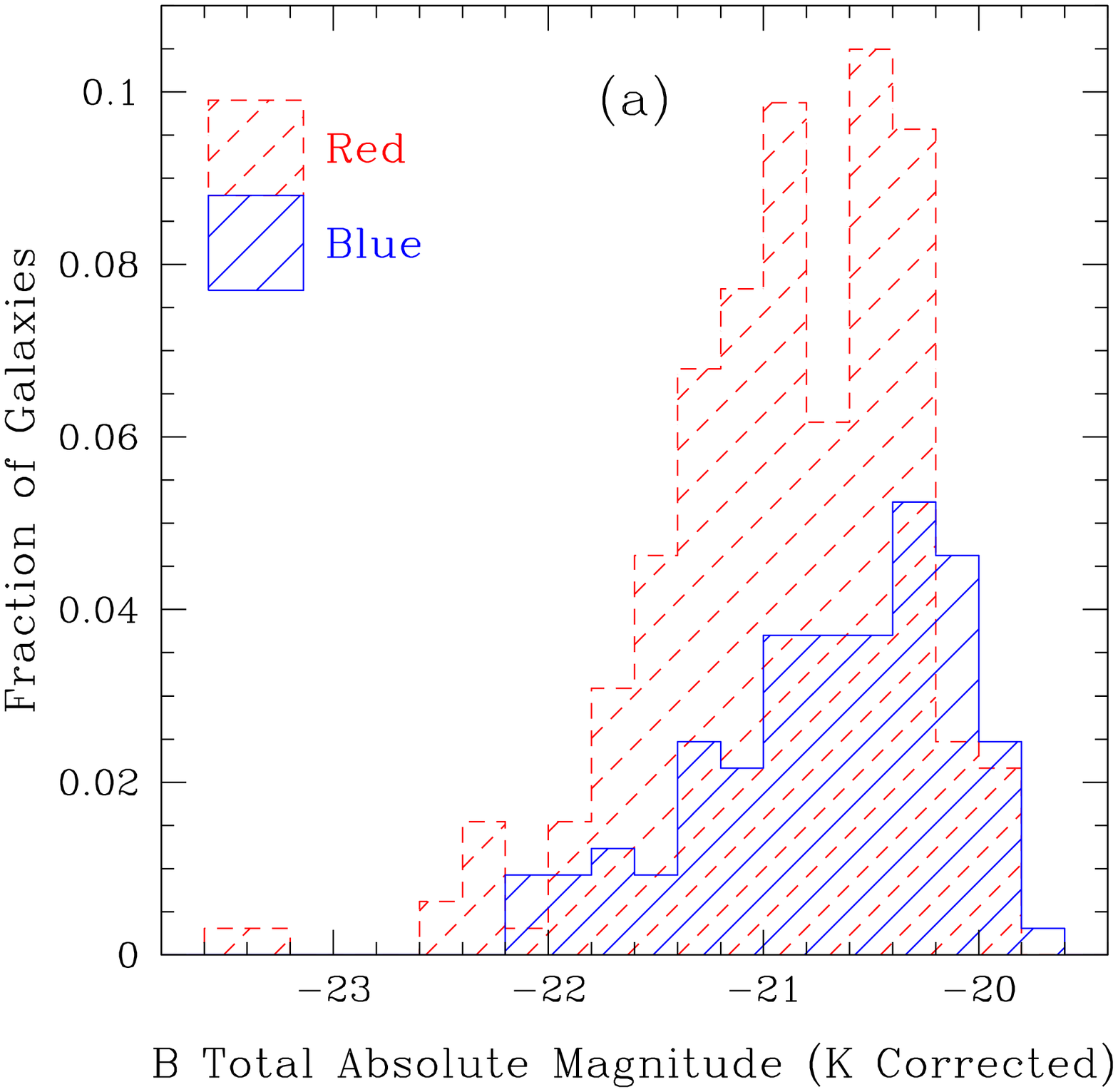} 
    \includegraphics[width=6.2cm]{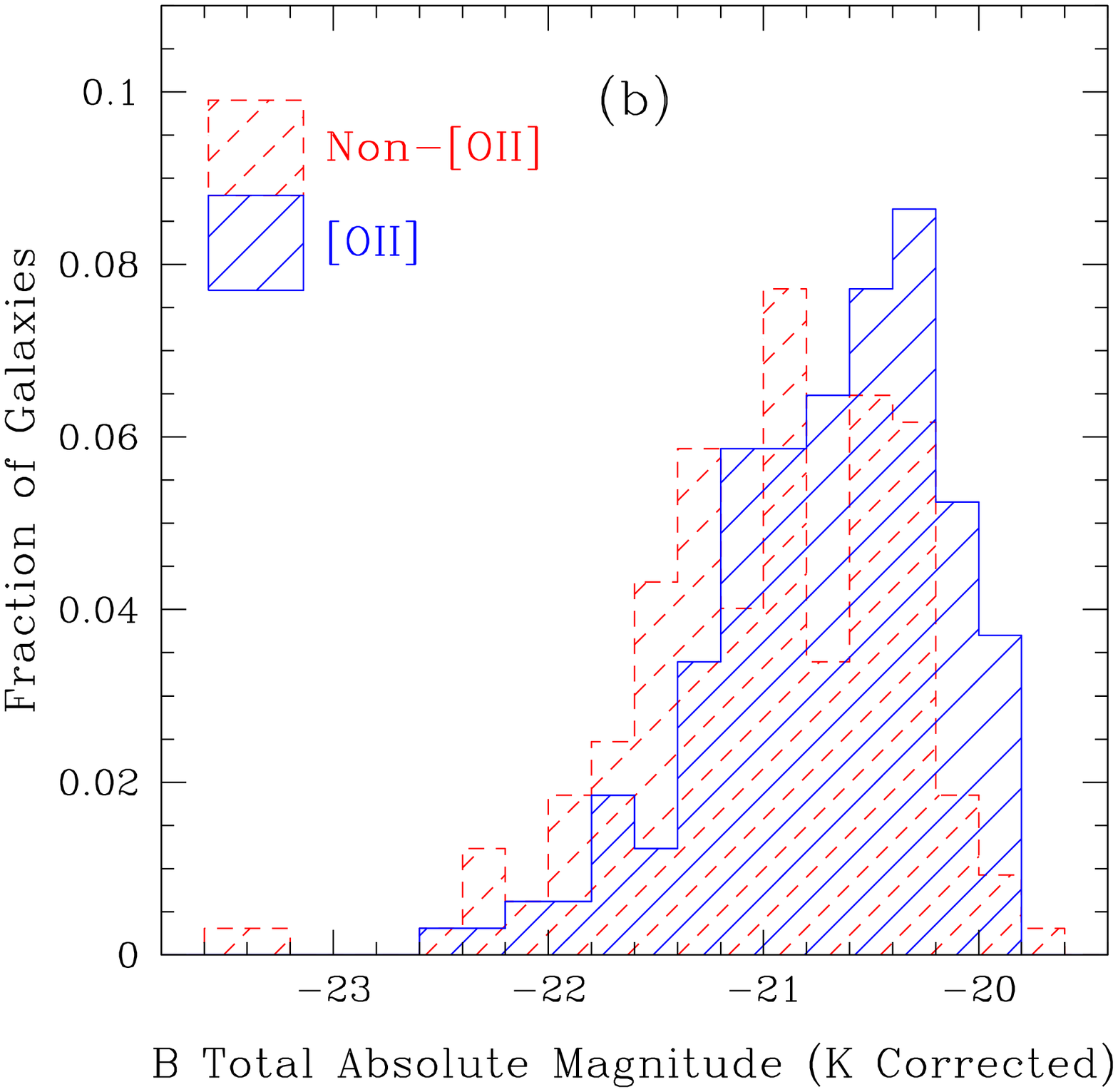}
    \includegraphics[width=6.2cm]{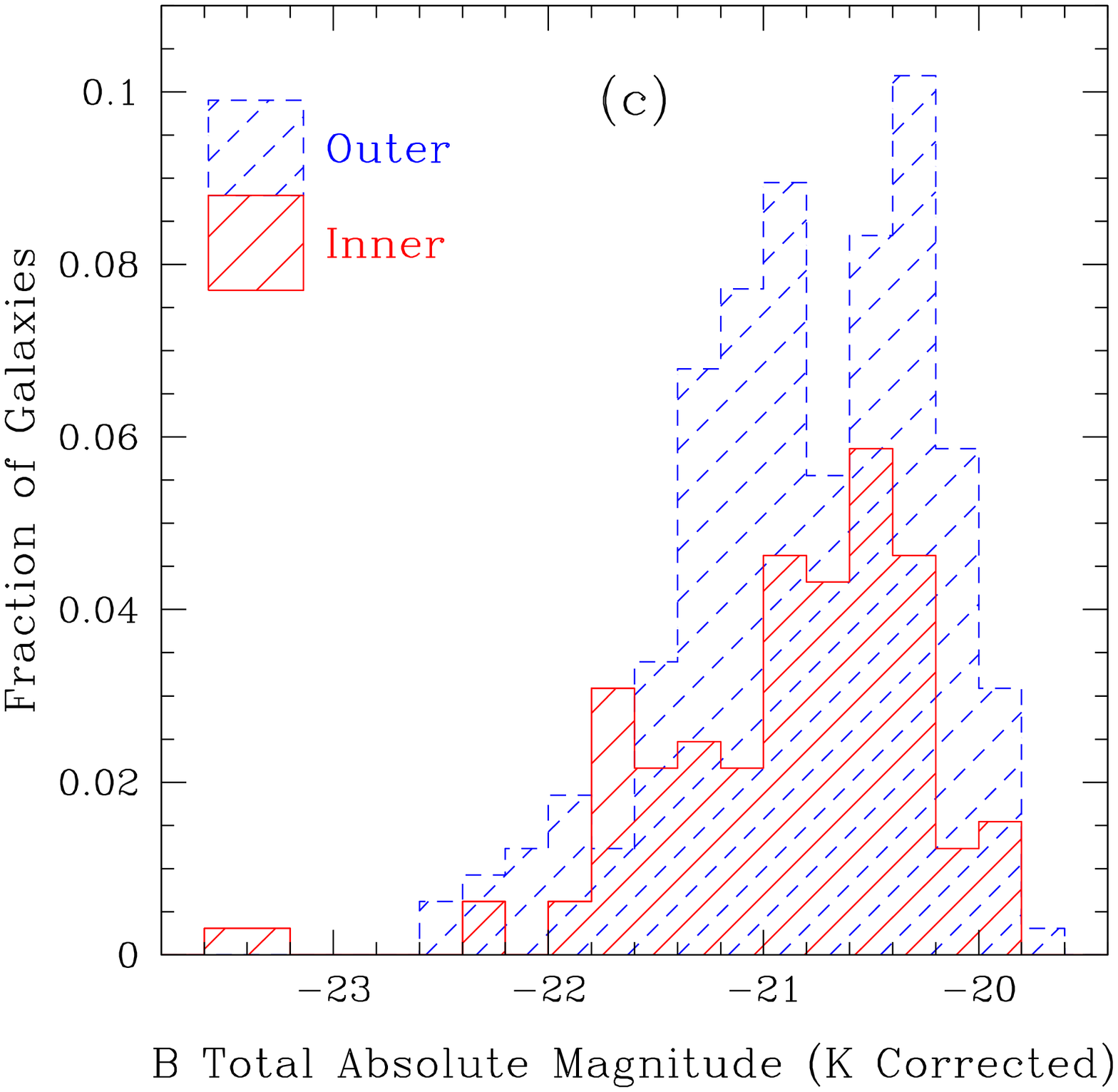}
   } 

  \subfigure{	
    \includegraphics[width=6.2cm]{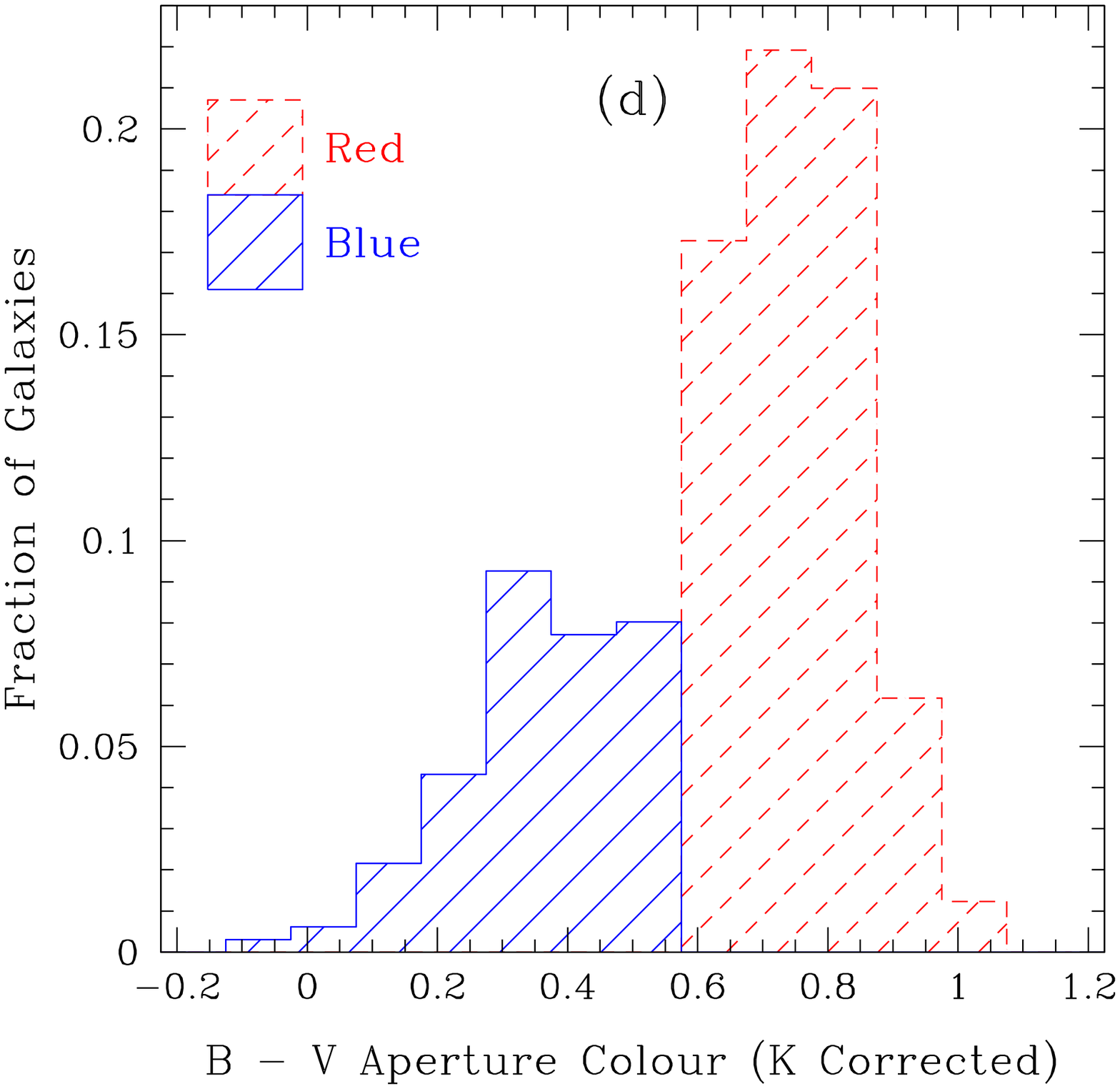}
    \includegraphics[width=6.2cm]{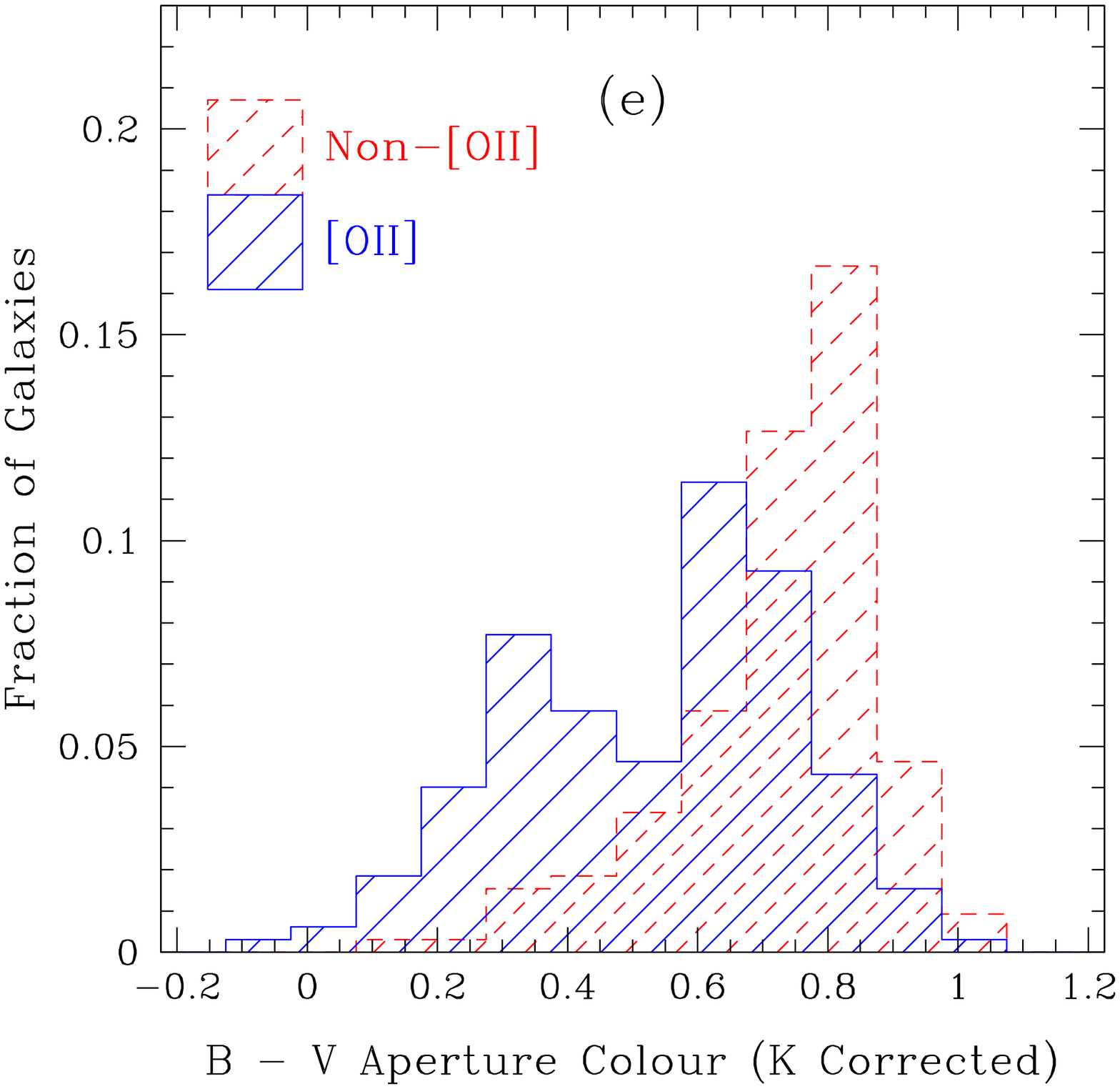} 
    \includegraphics[width=6.2cm]{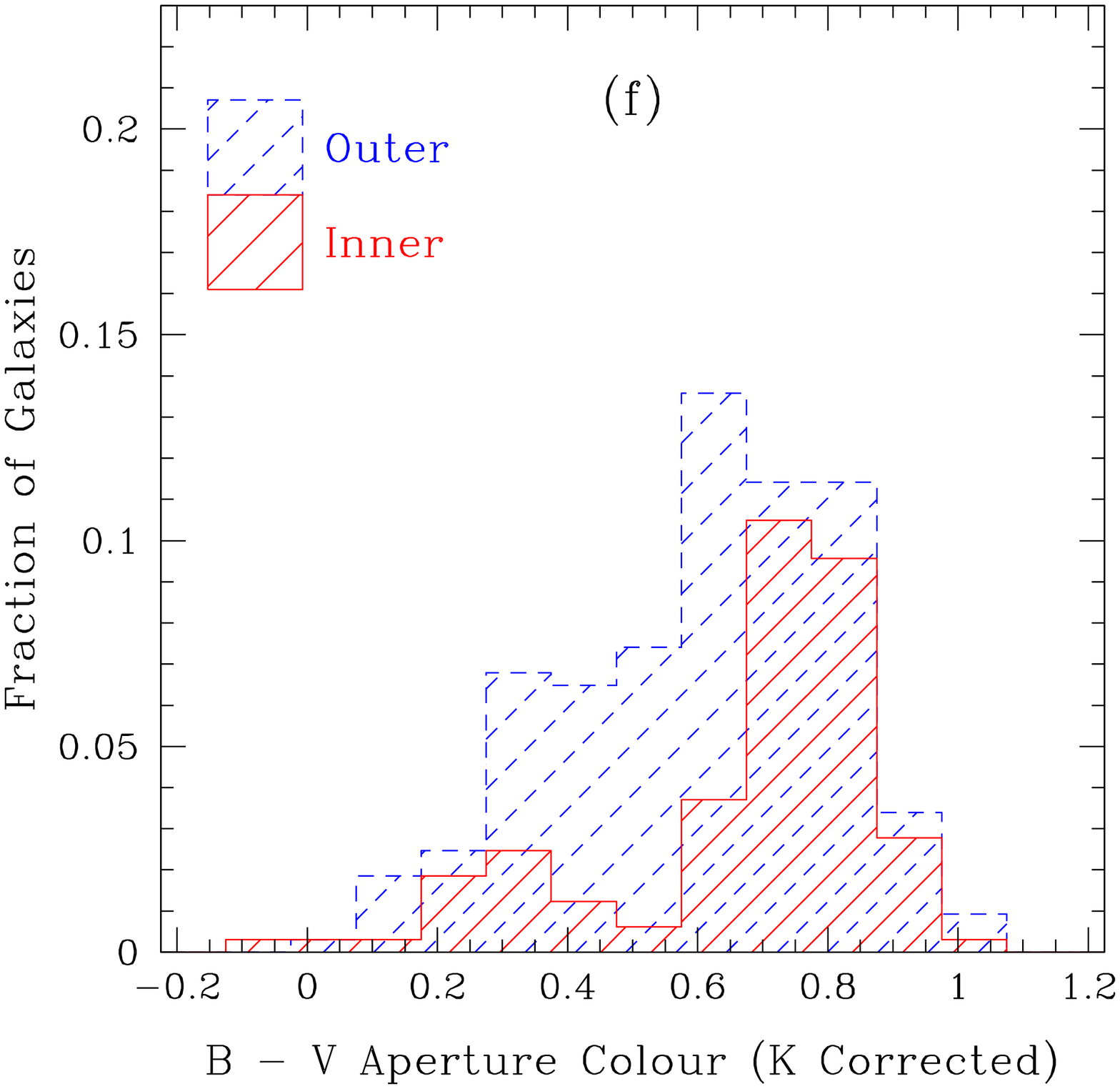} 
   } 

  \subfigure{	
    \includegraphics[width=6.2cm]{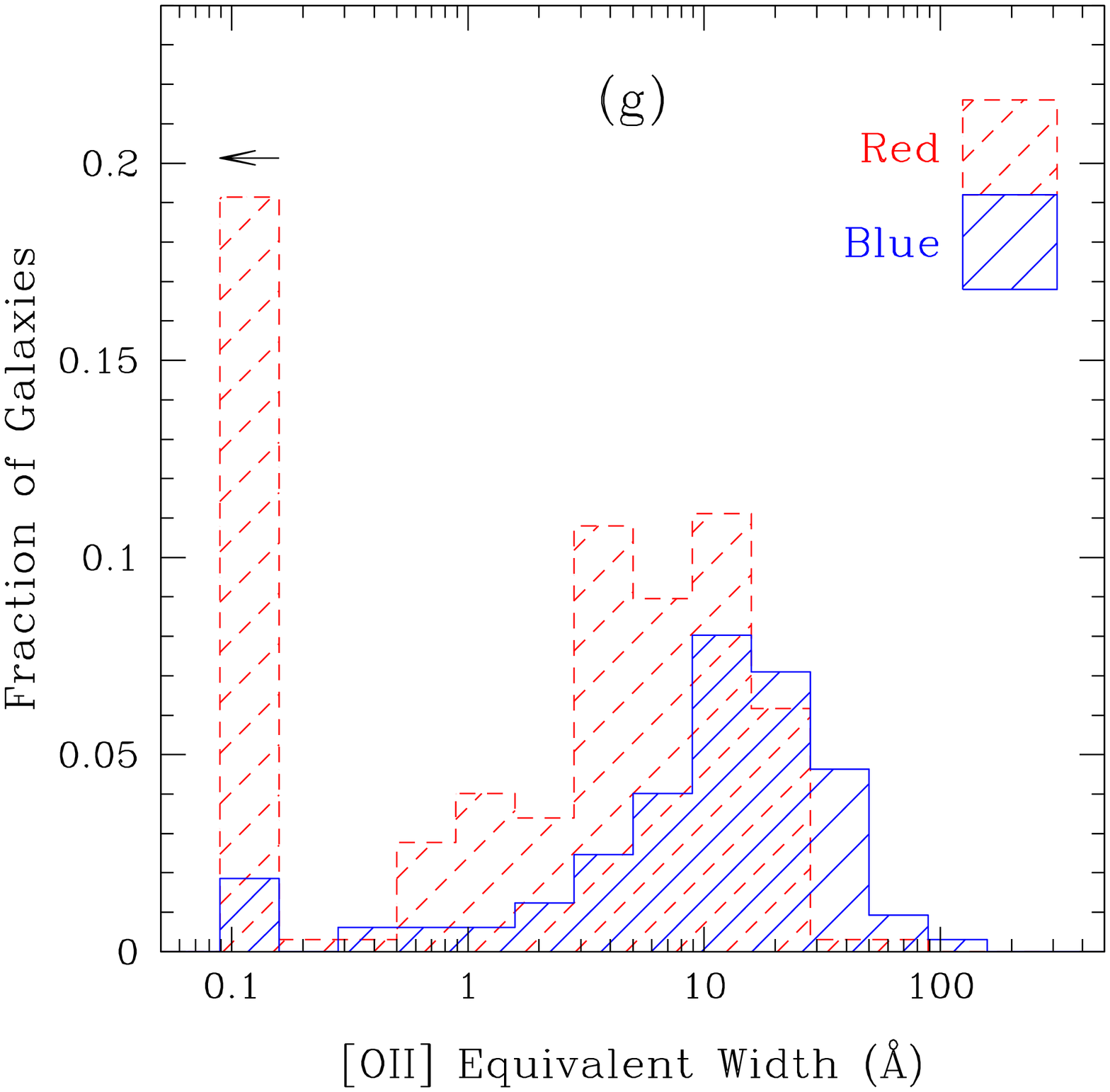}
    \includegraphics[width=6.2cm]{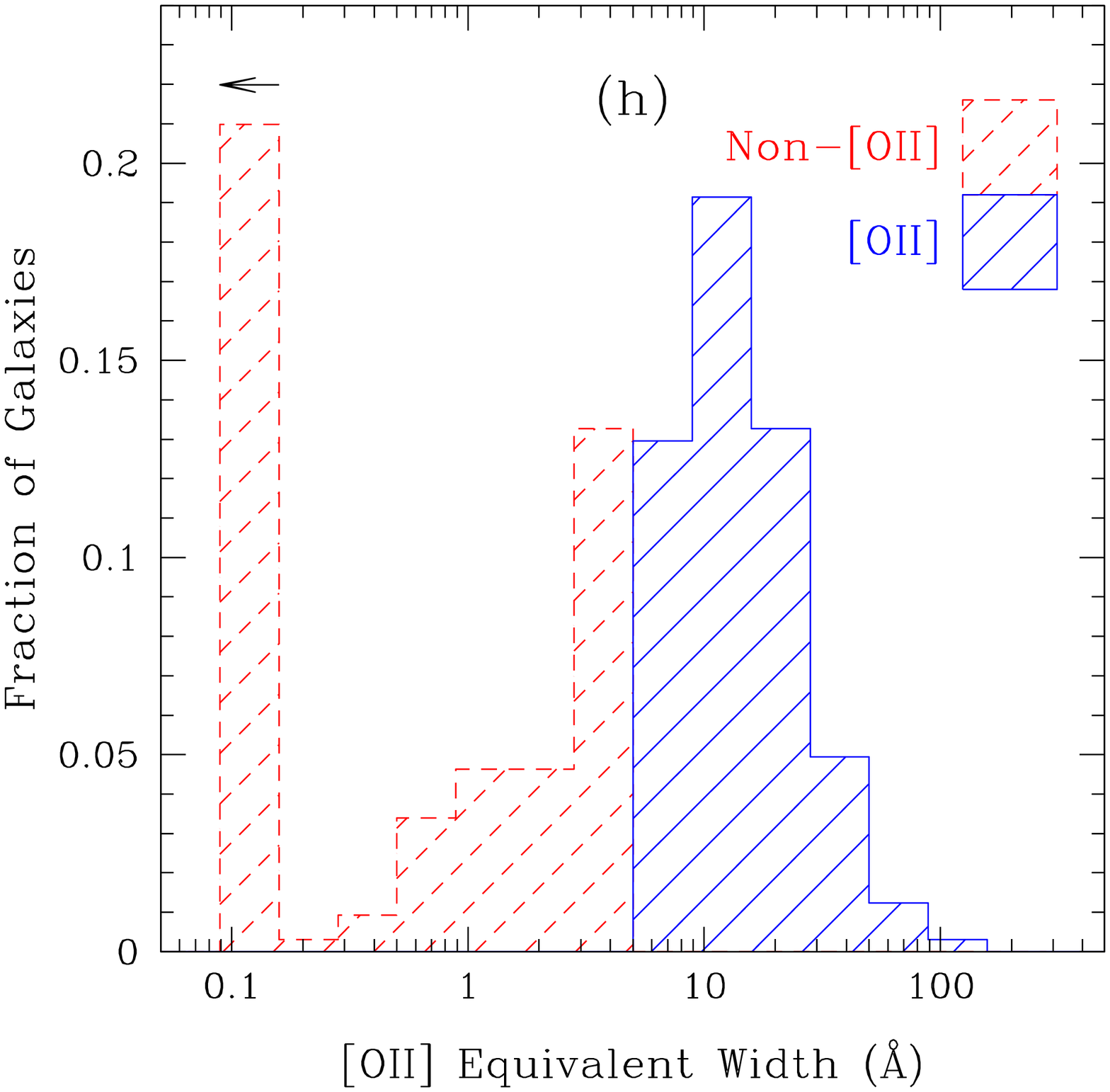} 
    \includegraphics[width=6.2cm]{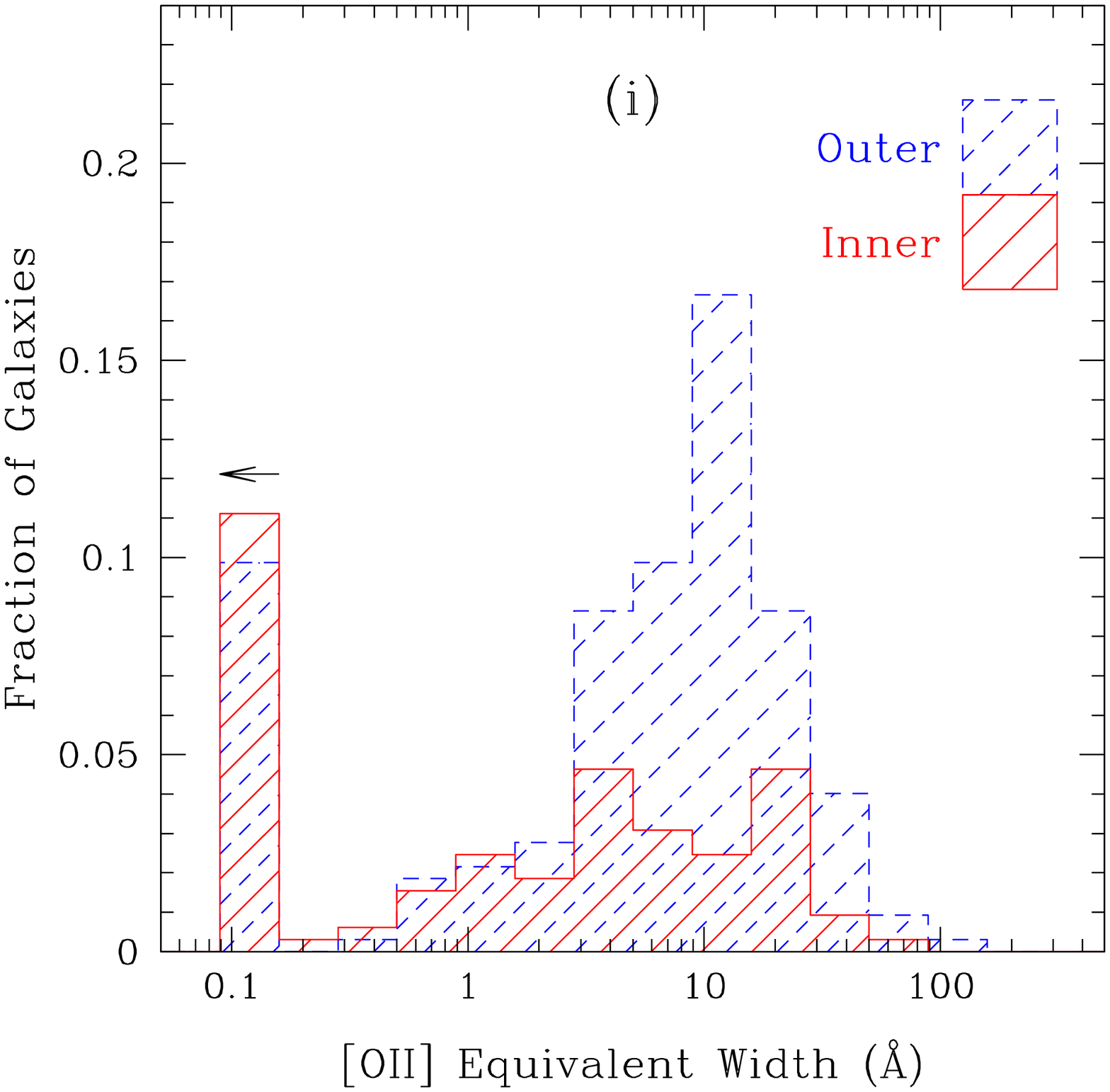} 
   } 

   \end{center}

   \caption{This figure shows the distribution in various parameters of the galaxies within the examined subsamples.  The distributions are given as the fraction of the total number of galaxies (324).   The parameters considered are the $B$~band total absolute magnitude (top panels), the $B-V$ colour (middle panels) and [OII] equivalent width (bottom panels).  The subsamples are the red and blue galaxies (left panels), [OII] and non-[OII] emission galaxies (centre panels), and the inner and outer cluster galaxies (right panels).  In the bottom panels, galaxies with [OII] equivalent widths that are zero or negative have been given a value of 0.1~\AA\ to allow them to be plotted on the log scale used.  In the [OII] equivalent width histograms, these galaxies are in the last bin which is marked with an arrow on top.}

   \label{hist_overlap_ratio}

\end{figure*}


In this work the Abell~370 galaxies have been broken up into a number of subsamples based on their optical colour, spectroscopic properties and location in the cluster (their galaxy environment).  The galaxies are divided into optically blue and red galaxy subsamples at $B-V$~=~0.57, as discussed previously.  Out of the 324 galaxies used in the \HI\ coadding, there are 219 red galaxies and 105 blue galaxies.  The galaxies are divided into [OII] emission and [OII] non-emission galaxy subsamples at [OII] equivalent width of 5~\AA, as discussed previously. Out of the 324 galaxies used in the \HI\ coadding, there are 156 non-[OII] emission galaxies and 168 [OII] emission galaxies.  An inner subsample of galaxies were selected that lay within a projected distance of 2.57~Mpc (the \Rtwo\ radius) of the cluster centre.  To remove a handful of galaxies that were clearly foreground to the cluster, galaxies in this subsample were also required to lie within 4 times the velocity dispersion of 1263~\kms\ of the cluster centre ($\rm z_{cl}$~=~0.373), a redshift range of z~=~0.356 to 0.390.  Using these criteria there are 110 galaxies in the inner subsample, leaving 214 galaxies in an outer subsample, away from the cluster centre.

These subsamples will be important in Section \ref{The_HI_subsamples}, where the \HI\ content of each galaxy subsample is considered.  Each of the Abell~370 subsamples share a number of galaxies in common.   Understanding the overlap between the subsamples is critical to interpreting the different \HI\ measurements made.  The distribution of various properties for each of the subsamples is shown in Fig.~\ref{hist_overlap_ratio}.  

The top panels of Fig.~\ref{hist_overlap_ratio} shows distribution of the $B$~band total absolute magnitude of the galaxies broken up into colour, emission type and cluster location in the left, middle and right panels respectively.  The blue galaxies, on average, tend to be slightly fainter than the red galaxies.  The [OII] and non-[OII] emission galaxies span similar ranges in absolute magnitude as do the outer and inner subsamples.  In this comparison, we are ignoring the two red cD galaxies that are both brighter than -23~$B$~band total absolute magnitude. 

The middle panels of Fig.~\ref{hist_overlap_ratio} show the distribution of the $B-V$ colour for the galaxies broken into the different subsamples.  In panel~(d), which breaks the galaxies into blue and red colour, the $B-V$ colour shows only that there are considerably more red galaxies than blue galaxies.  Panel (e), which displays the emission subsamples, shows that the [OII] emission galaxies span the complete colour range from blue to red.  The non-[OII] emission galaxies are much more tightly grouped, with most being red in colour with only a small tail of blue galaxies.   Panel~(f) shows the colour distribution of the inner and outer cluster galaxies.  Not surprisingly the inner sample is dominated by red galaxies while the outer galaxies have a more even distribution of colours, though there are still more red galaxies than blue galaxies (136 red to 84 blue) in the outer sample.

The bottom panels of Fig.~\ref{hist_overlap_ratio} shows the distribution of the [OII] equivalent width of the galaxies broken into the different subsamples.  Panel~(g) shows the [OII] equivalent width distribution for the blue and red galaxies.  While the galaxies with the very largest equivalent widths are dominated by the blue galaxies, there is a fair amount of overlap between the red and blue galaxies in the intermediate equivalent width values.  Below the 5~\AA\ (the cutoff used in the [OII] samples) the galaxies are dominated by the red galaxies with the blue galaxies tailing off.  Of the 168 galaxies above the 5~\AA\ cutoff there are 87 red galaxies and 81 blue galaxies; below the cutoff there are 132 red galaxies and only 24 blue galaxies.  In panel~(h) the [OII] equivalent width is broken up into the [OII] emission and non-[OII] emission samples at the 5~\AA\ value.  The most common equivalent width value for galaxies with [OII] emission lies around 15~\AA.  Panel~(i) shows the distribution of the [OII] equivalent width for the outer and inner cluster galaxies.  There is  a greater fraction of [OII] emission galaxies in the less dense region away from the cluster centre.  This agrees with the star formation--density relationship found by \cite{balogh98} in clusters from $\rm 0.18 < z < 0.55$ using [OII] emission.


\section{The Radio Data} 
\label{The_Radio_Data}

 
\begin{figure}  

  \begin{center}  
  \leavevmode  
		
    \includegraphics[width=8cm]{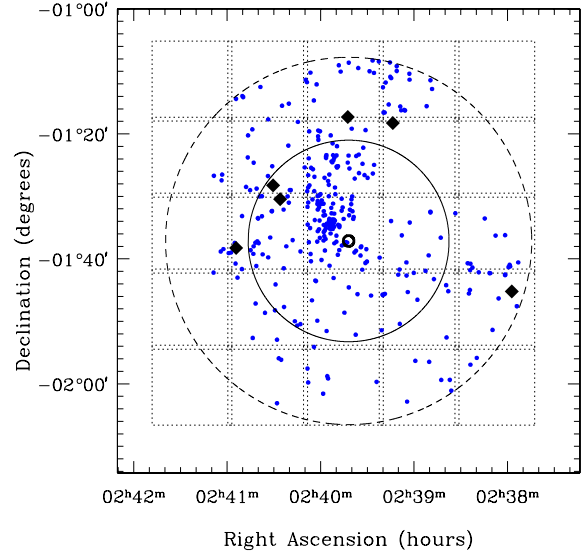}
  
   \end{center}

   \caption{This figure shows the pointing and primary beam size of the GMRT observations with the 324 optical galaxies used in the \HI\ coadding (the small points).  The large diamond points are the radio continuum sources on which self calibration was performed.  The small, bold circle is the centre of the GMRT pointing. The unbroken circle is the primary beam FWHM diameter of 32.2~arcmin and the dashed circle is the 10~per~cent primary beam level with diameter of 58.8~arcmin.  The dotted, overlapping, square grid shows the 25 facets used to tile the sky in the radio imaging. The centre of this grid pattern was created slightly offset from the GMRT pointing to ensure that the all of the 10~per~cent level of the primary beam lay within the tiled region.}


   \label{A370_positions}

\end{figure}


Radio observations of the galaxy cluster Abell~370 were carried out in 2003 August 10--17 using the Giant Metrewave Radio Telescope (GMRT) in India.  A total of 63~hours of telescope time was used, with 34~hours of on-source integration after the removal of the slewing time, flux and phase calibrator scans.   After flagging, \around 63~per~cent of the visibilities for the lower sideband and \around 50~per~cent of the visibilities for the upper sideband of this on-source data remained (50~per~cent of the total GMRT visibilities provides an equivalent sensitivity as an integration with \around 21 of the 30 GMRT antennas working perfectly for the entire time).

The total observing bandwidth of 32~MHz was split into two 16~MHz-wide sidebands covering the frequency range from 1024~MHz to 1056~MHz which is a redshift range $\rm 0.345 < z < 0.387$ for \HI\ 21-cm emission.  The pointing centre of the GMRT observations was R.A.~$\rm 2^{h}39^{m}42.0^{s}$ Dec.~$\rm -01^{\circ}37^{\prime}08^{\prime\prime}$~J2000 (see Fig.~\ref{A370_positions}).  This is 3.7~arcmin from the cluster centre.   The reason for the offset was to bring the strong radio continuum source 4C~-02.13 closer to centre of the GMRT primary beam and out of the sidelobes of the primary beam.  The data has two polarisations and 128 spectral channels per sideband, giving a channel spacing of 0.125~MHz (36.0~\kms).  Primary flux calibration was done using periodic observations of 3C48, which has a flux density at 1040~MHz of 20.18~Jy.  Phase calibration was done using scans on the VLA calibrator source 0323+055 for which our observations give a flux density of $3.723 \pm 0.061$~Jy.

The data reduction was primarily done using \AIPS.  Each sideband of data was processed separately.  Flux and phase calibration was determined using the bandpass calibration task BPASS, and the solution was interpolated to the data from the calibrator scans.  No normalisation was done before determining the solutions.  The regular phase calibrator scans throughout the observations allow us to correct for any time variability of the bandpass shape.  We have used an imaging pixel size of 0.75~arcsec and a imaging robustness value of 0.  {The synthesised beam size (resolution) of the data is \around 3.3~arcsec.}  For primary beam correction, we assumed a Gaussian beam with FWHM of 32.2~arcmin at 1040~MHz\footnote{National Centre for Radio Astrophysics website \\ http://www.ncra.tifr.res.in/\around ngk/primarybeam/beam}.

When making continuum images only channels 11-110 (out of the 128 in each sideband) were used. In order to avoid bandwidth smearing, the visibilities were averaged into a new data set consisting of 10 channels, each of which was the average of 10 of the original channels.   This new 10 channel {\it u~--~v}~data file was made into a single channel continuum image using the \AIPS\ task IMAGR, which combined the channels using a frequency-dependent primary beam correction based on the effective antenna diameter of 37.5~m.  This correction assumes that the primary beam is a uniformly illuminated disk of the specified diameter.  The value of this diameter was estimated using the known GMRT primary beam size. 

Self-calibration of the data was done using the 6 brightest radio continuum sources in the field.  These sources have raw flux density values (not corrected for the primary beam shape) ranging from 18~mJy to 80~mJy.  The self-calibration of GMRT data sometimes does not converge quickly, probably as a consequence of the GMRT hybrid configuration. Radio continuum sources that lack coherence in the synthesis image due to phase errors tend to remain defocused during self calibration.  To fix this problem, slightly extended sources were replaced in the first self-calibration loop with point sources with the same centroid and flux density as the original source. Further self-calibration loops were done using the clean components in the traditional manner.  The observational data for each day were initially self calibrated alone using 4 loops of self-calibration (2 loops of phase calibration and 2 amplitude and phase calibration loops).  At this point the six brightest continuum source were subtracted from the {\it u~--~v}~data, and the data were flagged to exclude any visibility residuals that exceeded a threshold set by the system noise statistics.  The continuum sources were then added back into that day's data.

After this detailed editing process, all the {\it u~--~v}~data were combined and self-calibration was repeated on the entire combined data set to ensure consistency between the observations on the different days.  A large continuum image of the entire field was then made.  The \AIPS\ imaging routine assumes that the sky is flat (it ignores the vertical `w' term).  This is an appropriate assumption only for small fields.   In order to reduce the distortion over our large field it was necessary to break the image up into 25~facets (see Fig.~\ref{A370_positions}).  Each of these square facets was 12.8~arcmin on a side (1024~pixels) and each overlapped by 38.4~arcsec.   The total combined field size was 61.4~arcmin on each side.  These overlapping facets cover the region within the 10~per~cent primary beam level (58.8~arcmin in diameter).  Clean boxes were put around all the discernible continuum sources during the imaging process.

In this initial continuum image, faint sidelobes from the brighter radio sources remained visible which took the form of narrow strips running north-south from the continuum sources.   These are residuals from the dirty synthesised beam for the equatorial field that have not been fully removed in the cleaning process.  In order to improve the quality of the image a process of peeling was used to remove the bright sources and these artifacts from the data.  This involved removing all continuum sources from the {\it u~--~v}~data except for one of the bright sources.  A full round of self calibration was then performed on this new set of {\it u~--~v}~data on the single bright source alone.  The new calibration derived from the self calibration (its \AIPS\ SN table) was then applied to the original {\it u~--~v}~data (the data with all the continuum sources still in it).  The new model of the bright source was then subtracted from this {\it u~--~v}~data.  Finally the effect of the new calibration was removed from the {\it u~--~v}~data to restore it to its previous calibration by applying the inverse SN table created using the \AIPS\ task CLINV.  This process was done for each of the six brightest sources in turn, working from east to west across the field, removing the sources from the {\it u~--~v}~data.

The final continuum image was made from this new {\it u~--~v}~data set.  This image has an RMS of 20~\microJy.  The astrometry of the radio continuum sources in the data were checked against their positions in the VLA FIRST survey \citep{becker95}.  They show good astrometric agreement of $-0.27 \pm 0.11$~arcsec in Right Ascension and $0.24 \pm 0.31$~arcsec in Declination.  Several objects showed good alignment between their optical and radio continuum components suggesting that the optical and radio astrometry are in good agreement.  

The remaining radio continuum sources were subtracted from the {\it u~--~v}~data using the \AIPS\ task UVSUB.  The final spectral data cube was made from this {\it u~--~v}~data set.  While UVSUB removed most of the continuum emission, some small residuals remained.  To remove these final traces, a linear fit to the continuum across frequency was subtracted in the final data cube using the \AIPS\ task IMLIN. This introduces a small bias to the data cube, since any line emission from the galaxies would be included in the fit.  The correction for this bias is described in Section~\ref{HI_all_galaxies}.  In the final data cubes, the median RMS was 152~\microJy\ per channel in the lower sideband and 174~\microJy\ per channel in the upper sideband.  


\section{Measuring the \HI\ 21-cm emission signal from the Abell~370 galaxies}
\label{HI_all_galaxies}

Directly detecting the \HI\ 21-cm emission from even the most gas rich galaxy in the observed radio data for Abell~370 is unlikely.  For a galaxy to be directly detected at $5 \sigma$ significance it would need to have a \HI\ mass of at least \MHI~=~\around $5 \times 10^{10}$~\Msun, have a velocity width of 300~\kms and the signal to be all contained within a diameter of \around 3.3~arcsec (17~kpc at z~=~0.37).  In the volume probed by our observations it is unlikely to find a galaxy with such a large \HI\ mass.  Additionally the \HI\ gas in such a galaxy would likely extend out to a much greater diameter, making it even harder to detect in our data.  A search of the radio data cube for \HI\ direct detections was made using the {\sc Duchamp} software developed by Matthew Whiting of the Australia Telescope National Facility.  Nothing of significance was found as expected.  

Instead of direct detection, the \HI\ 21-cm emission signal from multiple galaxies has been coadded to increase the signal to noise of the measurement.  This stacked signal can then be used to quantify the total \HI\ gas content of the galaxies.  The galaxies are located in the radio data using their measured optical positions and redshifts.  

The expected frequency of the \HI\ 21-cm emission from each galaxy is determined from their optical redshift.  Of the 1877 optical redshifts obtained 324 are usable for \HI\ coadding.  The optical redshift distribution for galaxies around the cluster at z~=~0.37 can be seen in Fig.~\ref{hist_z_A370}.  Only redshifts that lay within the \HI\ frequency range covered by the radio data of z~=~0.3451 to 0.3871 could be used in the coadding.  The galaxies used for coadding were also limited to those inside the 10~per~cent GMRT primary beam level.  Redshifts that lie within 7 channels (0.875~MHz) of the boundary between the two radio sidebands (this is z~\around~0.3658) were also excluded.  Finally redshifts that lay in the 9 channels (1.125~MHz) surrounding some strong radio interference in the lower sideband were excluded from the coadding.  This strong radio interference is located at \around 1030~MHz (\HI\ redshift z~=~0.379).

The \HI\ 21-cm emission signal of a galaxy is spread over a velocity width (frequency width) determined by the motion of the \HI\ gas within the galaxy. If the gas in the galaxy is rotating in a disk, then the inclination of the disk to the observer will have a substantial effect on the observed velocity width.  A rotating disk galaxy will have its largest velocity width when its disk is viewed edge on by an observer and its smallest velocity width when its disk is viewed face on.  Assuming minimal \HI\ self-absorption, galaxies with the same \HI\ mass but different disk inclinations should have the same integrated \HI\ flux.  However the peak \HI\ flux of these galaxies will vary greatly depending on their disk inclinations.  The peak flux will be a maximum when its disk is face on, i.e.~when the velocity width is at a minimum.

Assuming a random distribution of disk orientations, 50~per~cent of disk galaxies will have inclinations to the observer of greater than 60\degrees\ (edge on) and only 13~per~cent galaxies will have inclinations less than 30\degrees\ (face on).  By selecting the galaxies in the optical, a higher proportion of the disk galaxies in our sample will be edge on compared to \HI\ selected samples.  This is due to a combination of this preferred orientation and because edge on galaxies have higher optical surface brightnesses making them easier to detect than face on systems.  As a result of this selection effect, the galaxies in our sample with the most \HI\ gas (the spiral disk galaxies) are likely to have large \HI\ velocity widths.  Unfortunately it is not possible to measure the inclination of the disk galaxies in our sample due to the combination of the poor seeing in our optical imaging and the small angular size of the galaxies at redshift of z~=~0.37.  

As we do not detect the individual \HI\ 21-cm emission of the galaxies we cannot measure their individual galaxy velocity widths. Instead we have to make some reasonable assumptions and define a velocity width that should encompass all the coadded \HI\ 21-cm emission signal from the galaxies.  From the HIPASS survey \citep{meyer04}, a galaxy with \MHI~=~\around $10^9$~\Msun\ has a maximum velocity width of \around 250~\kms\ and a galaxy with \MHI~=~\around $10^{10}$~\Msun\ has a maximum velocity width of \around 400~\kms (w$_{50}$).  Galaxies with low disk inclinations can have markedly narrower velocity widths than these maxima.   Using the optical Tully-Fisher relationship from \citet{mcgaugh00} and assuming no inclination correction, the median estimated velocity width (W$_{20}$) based on the $B$~band absolute magnitudes of the Abell~370 galaxies is \around 320 \kms.  

The statistical uncertainty in the optical redshifts obtained for Abell~370 is \around 70~\kms.  {When coadding the galaxies this redshift uncertainty will broaden the resulting \HI\ emission signal.  This redshift broadening is taken into account by increasing the \HI\ velocity width used by $\pm \, 2 \sigma$ (280~\kms). To ensure that all the \HI\ signal from all the combined galaxies is measured a velocity width of 600~\kms\ was used.  This velocity width takes into account the width of the larger galaxies in the sample as well as the effect of the redshift broadening.} The uncertainty due to noise in the measured coadded \HI\ 21-cm emission signal is approximately proportional to the square root of the velocity width used.  A narrower velocity width will have a smaller estimated error but will likely miss some of the \HI\ signal.

When making the final data cube, a linear fit to the spectrum through each sky pixel was subtracted to remove the residuals left from the continuum sources (see the discussion on IMLIN in Section \ref{The_Radio_Data}).  Any \HI\ signal would be included in the calculation of this linear fit across frequency and this would create an over-estimation in the continuum fit, creating a bias in the final \HI\ spectrum.  To remove this bias, the \HI\ spectrum for each galaxy has a new linear fit made across all frequency channels except those corresponding to a 600~\kms\ velocity width around the galaxy at its redshift (these should be the channels that would contain any \HI\ signal).  This fit is then subtracted from the data, removing any bias created by IMLIN.  This correction increases the \HI\ flux density measured in the coadded spectra of different subsample of galaxies from anywhere from 20~to 40~per~cent.

 
\begin{figure}  

  \begin{center}  
  \leavevmode  

    \includegraphics[width=8cm]{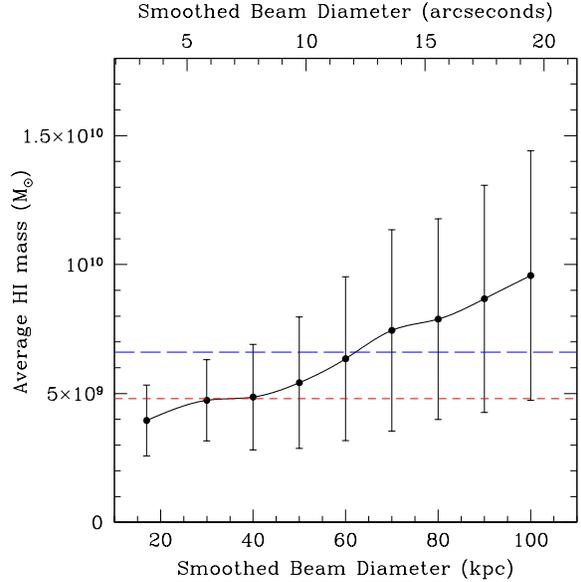} 

   \end{center}

   \caption{
This figure shows the average \HI\ mass for all 324 Abell~370 galaxies measured in the different smoothed synthesised beam size data.  The short dashed line shows the \HI\ mass for the mid-smoothing measurement and the long dashed line shows the \HI\ mass for the large smoothing measurement (see text for details). }

   \label{HI_mass_SM}

\end{figure}


Variance weighting is used when coadding the separate \HI\ spectra.  The variance is calculated from the known RMS per frequency channel in the radio data cube, factoring in the primary beam correction for galaxies away from the beam centre.  Variance weighting provides the optimal signal to noise for the coadded \HI\ signal.  However it does introduce a potential bias as those galaxies located near the cluster centre are given higher weight as they are close to the GMRT pointing centre.  The measured \HI\ 21-cm emission flux density can be converted to the mass of atomic hydrogen that produced the signal by the following relation:

\begin{equation}
  \rm M_{HI} = \frac{236}{( \ 1 + z \ )} 
  \left ( \frac{S_{v}}{ \ mJy \ }\right ) 
  \left ( \frac{d_{L}}{ \ Mpc \ } \right )^2  
  \left ( \frac{\Delta V}{ \ km \, s^{-1} }  \right ),
\end{equation}

\noindent where $\rm S_{v}$ is the \HI\ emission flux density averaged across the velocity width $\rm \Delta V$ and $\rm d_L$ is the luminosity distance to the source.  This equation assumes that the cloud of atomic hydrogen gas has a spin temperature well above the cosmic background temperature, that collisional excitation is the dominant process, and that the cloud is optically thin \citep{wieringa92}.  No correction for \HI\ self absorption has been made to any of the measurements.   \HI\ self-absorption may cause an underestimation of the \HI\ flux by as much as 15~per~cent.  However this value is extremely uncertain \citep{zwaan97}. In this analysis the galaxies were all assumed to be at the distance of the cluster centre at $\rm z_{cl}$~=~0.373, a luminosity distance of $\rm d_L$~=~2000~Mpc.  

It is not possible to measure the projected extent of the \HI\ gas on the sky for the galaxies, as we do not detect their individual \HI\ 21-cm emission.  The calibrator convention for synthesis images means that the peak specific intensity of an unresolved source is equal the total flux density of that source.  This means, that if a galaxy is unresolved by the GMRT synthesised beam, we can take the value of the specific intensity at the optical position of the galaxy as a measure of the total \HI\ flux density of the galaxy.  The GMRT synthesised beam has a FWHM \around 3.3 arcsec at 1040~MHz which corresponds to \around 17~kpc at z~=~0.37.   Unfortunately all but the very smallest of the galaxies in our sample are likely to be resolved by this synthesised beam size.  The solution to this problem is to smooth the radio data to larger synthesised beam sizes until the galaxies are unresolved and then measure their \HI\ signal.  

Gaussian smoothing in the image plane is equivalent to tapering (multiplying by a Gaussian) in the {\it u~--~v} plane. This effectively reduces the weight of the longer GMRT baselines in the image plane.  Thus smoothing to larger synthesised beam sizes increases the RMS noise in the radio data.  However, the measured \HI\ flux density increases for galaxies that are now unresolved in the new smoothed data.  From the initial radio data with synthesised beam size \around 3.3 arcsec (\around 17~kpc at z~=~0.373), smoothed data sets were created with circular synthesised beam sizes in 10~kpc steps, i.e.~equal to 30, 40, 50, 60, 70, 80, 90 and 100~kpc (5.8, 7.8, 9.7, 11.7, 13.6, 15.5, 17.5 and 19.4~arcsec).  The smoothing was done with the \AIPS\ task SMOTH.  The RMS per channel in the radio data increases from \around 160~\microJy\ at a synthesised beam size of 17~kpc (\around 3.3~arcsec), to \around 280~\microJy\ at 50~kpc (9.7~arcsec), and to \around 550~\microJy\ at 100~kpc (19.4~arcsec).  

The coadded \HI\ 21-cm flux density for all 324 galaxies in the sample was measured in each set of smoothed data and converted to the equivalent average \HI\ mass as seen in Fig.~\ref{HI_mass_SM}.  In this figure the measured \HI\ mass can be seen to rise with increasing synthesised beam size.  This indicates that the \HI\ gas extends beyond the inner regions of the galaxies that is probed by the smaller synthesised beam sizes.  As the synthesised beam size increases, the error also increases making it difficult to precisely define the extent of the \HI~signal. 
 
 
\begin{figure}  

  \begin{center}  
  \leavevmode  
		
    \includegraphics[width=8cm]{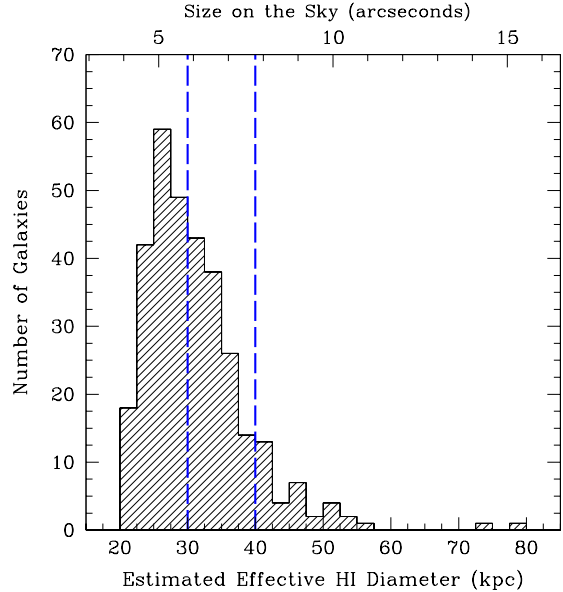}
 
   \end{center}

   \caption{This figure shows the estimated \HI\ effective diameter (the diameter within which half the \HI\ mass is contained) for the 324 galaxies that are used in the \HI\ coadding.  The diameter was derived from the relationship between \HI\ size and optical magnitude found for spiral and irregular galaxies in the field \citep{broeils97}.  The dashed lines mark the values used to split the galaxies into the small, medium and large categories.  The top x-axis displays the size the galaxies would appear on the sky at the redshift of the cluster ($\rm z_{cl}$~=~0.373).}

   \label{hist_galaxy_size}


\end{figure}


An estimate of the projected size of the \HI\ gas in a galaxy can be used to determine the minimum synthesised beam size that would leave the galaxy unresolved. This enables a definite measurement of the total \HI\ gas as well as reducing the error introduced by smoothing the data.  An estimate of the \HI\ size of a galaxy can be made using the correlation found between the optical $B$~band magnitude and \HI\ size of spiral and irregular galaxies in the field at low redshift \citep{broeils97}.  This relationship is:

\begin{equation}
  \rm \log(D_{eff}) = - \ (0.1588 \pm 0.011)\ B_{abs}\ - \ (1.827 \pm 0.22).
  \label{Deff}
\end{equation}
\noindent where $\rm D_{eff}$ is the diameter within which half the \HI\ mass of the galaxy is contained and is measured in kpc.  Fig.~\ref{hist_galaxy_size} shows the distribution of the estimated effective \HI\ diameter for the 324 galaxies around Abell~370.  The galaxies are broken up into three groups based on their estimated $\rm D_{eff}$.  Small galaxies are defined as those with $\rm D_{eff} \le 30$~kpc, medium galaxies as those with 30~kpc~$\rm < D_{eff} \le 40$~kpc and large galaxies as those with $\rm D_{eff} > 40$~kpc.   The 30~kpc value corresponds to an absolute $B$~band magnitude of -20.8 and the 40~kpc value to -21.6.  There are 168 small galaxies, 121 medium sized galaxies and 35 large galaxies in the sample.

Galaxies that have $\rm 2 \times D_{eff}$~=~synthesised beam size have peak specific intensity values that are \around 90~per~cent of their total flux density (assuming a Gaussian shape to the \HI\ gas spatial distribution). In previous coadding \HI\ work looking at star-forming field galaxies at z~=~0.24 \citep{lah07} this was the criterion used with the estimated $\rm D_{eff}$ to determine the maximum synthesised beam size in which a galaxy would be unresolved.  However, the relationship found by \citet{broeils97} was for spiral and irregular galaxies in the field, while the galaxies in Abell~370 are a mixture of early and late-type galaxies in a variety of environments.  This criterion will likely over estimate the true extent of the \HI\ gas for many of the galaxies.

Three different measurements of the average \HI\ gas mass are made to reflect this uncertainty in the \HI\ extent of the galaxies.  The first coadded \HI\ mass measurement is made by measuring the specific intensity at the optical positions of the galaxy in the original unsmoothed GMRT radio data (synthesised beam size of 17~kpc, \around 3.3~arcsec).  For all 324 galaxies the measured average \HI\ mass in this unsmoothed measurement is $(4.0 \pm 1.4) \times 10^9$~\Msun.  The second coadded \HI\ mass measurement is made by first breaking the galaxies into their small, medium and large size groups based on their estimated $\rm D_{eff}$.  The \HI\ flux density for each group of galaxies is then measured in the radio data that has $\rm D_{eff} \le$~the~smoothed synthesised beam size (30, 40 and 50~kpc beam sizes).  The three measured values are then combined to give a single average \HI\ flux density.  For all 324 galaxies the measured average \HI\ mass in this mid-smoothed measurement is $(4.8 \pm 1.8) \times 10^9$~\Msun.  A third measurement is made similarly, except each group of galaxies is measured in the radio data that has $\rm 2 \times D_{eff} \le $~the~smoothed synthesised beam size (60, 80 and 100~kpc beams sizes).  This large smoothing measurement should give an accurate reflection of the total \HI\ gas mass for the galaxies that have similar \HI\ extents to the field spiral and irregular galaxies observed by \citet{broeils97}.  For all 324 galaxies the measured average \HI\ mass in this large smoothed measurement is $(6.6 \pm 3.5) \times 10^9$~\Msun.  

 
\begin{figure*}  

  \begin{center}  
  \leavevmode  
		
    \includegraphics[height=7cm]{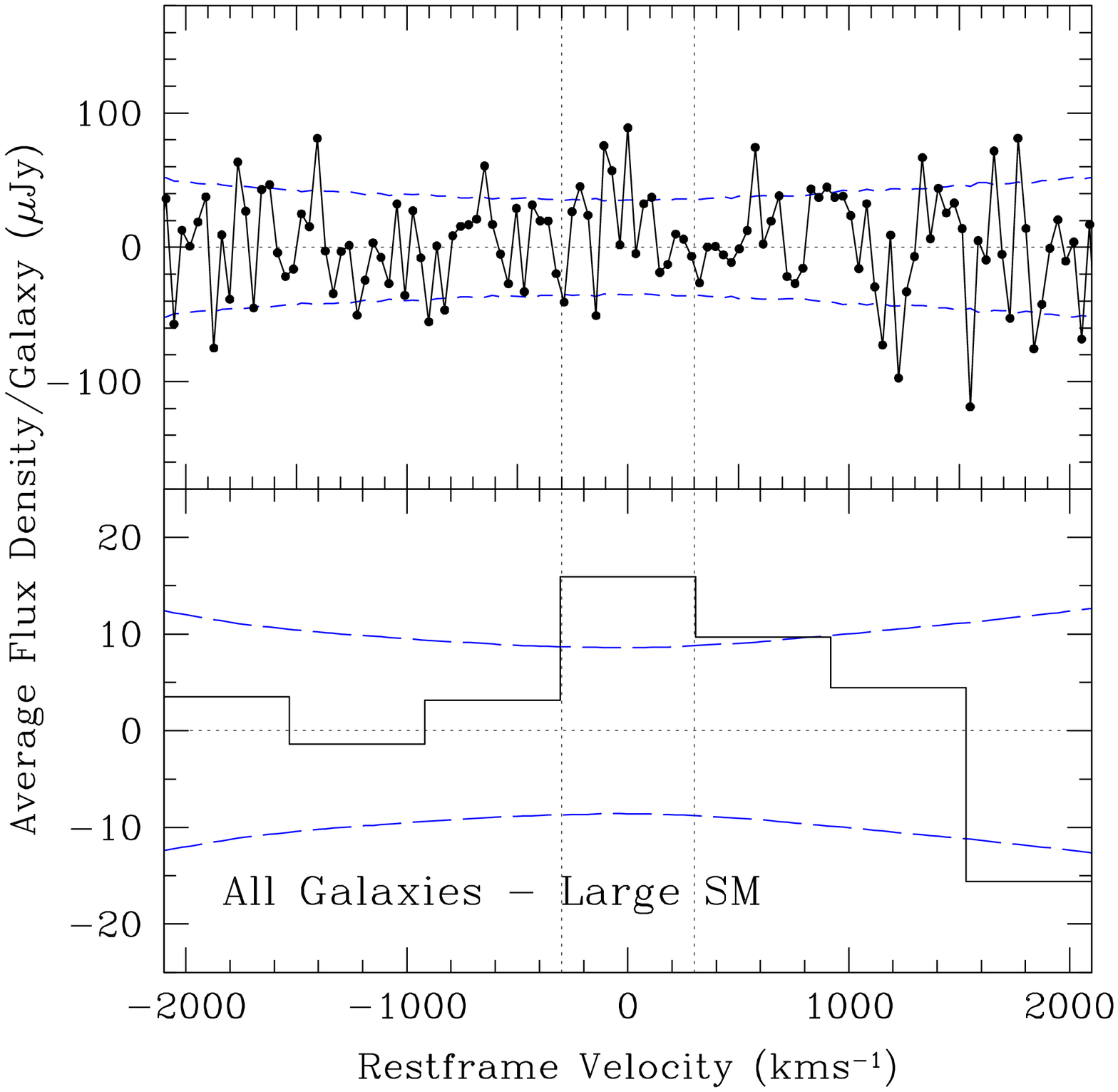}
    \includegraphics[height=7cm]{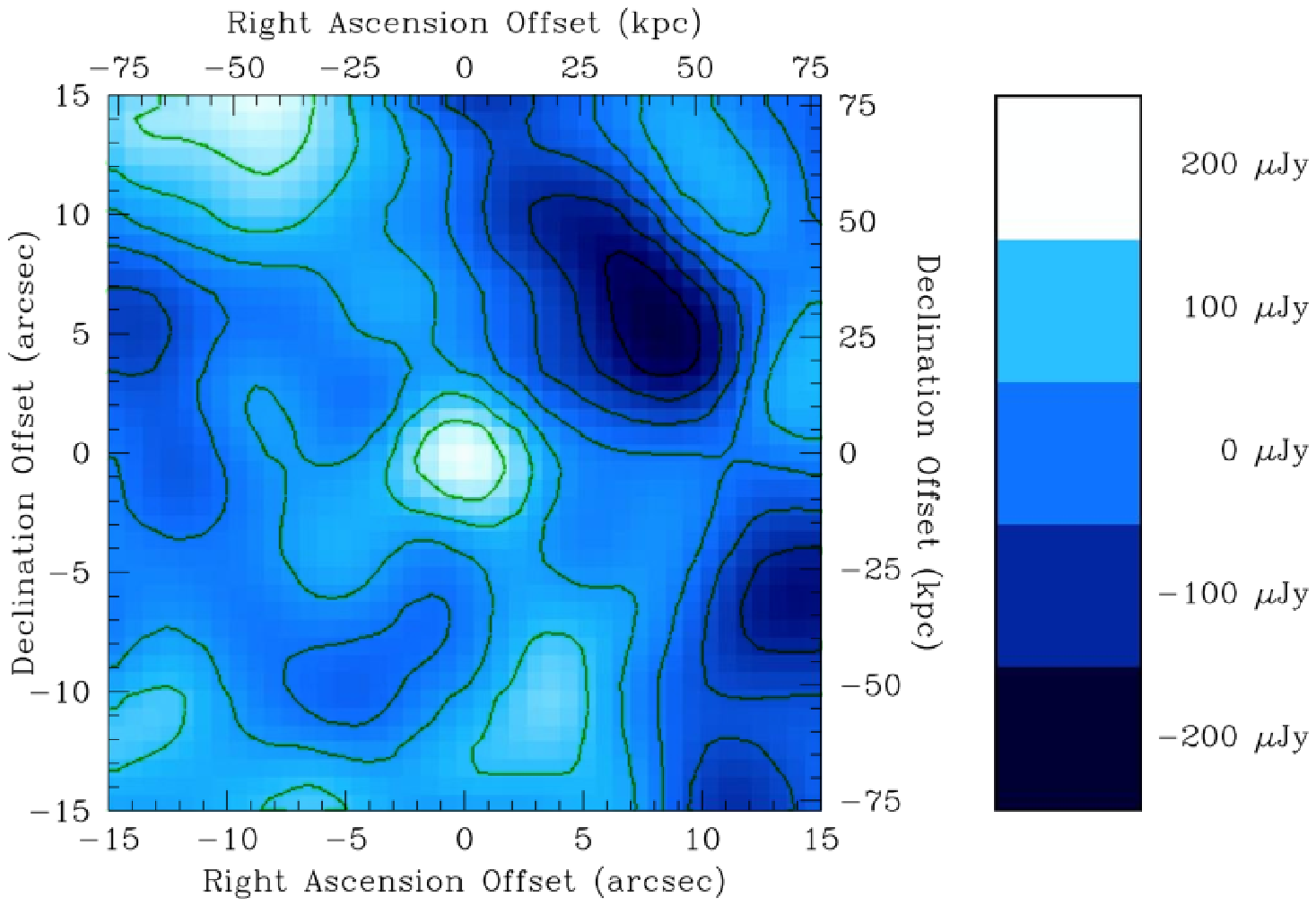}

   \end{center}

   \caption{The left panel shows the average \HI\ galaxy spectrum created from coadding the signal of all 324 galaxies using the large spatial smoothing.  The top spectrum has no smoothing or binning and has a velocity step size of 36.0~\kms .  The bottom spectrum has been binned to 600~\kms .  This is the velocity width that the combined \HI\ signal of the galaxies is expected to span.  For both spectra the $1\sigma$ error is shown as dashed lines above and below zero.
The right panel shows the average \HI\ image made by coadding the data cube around each of the 324 galaxy and binning across 17 spectral channels (600~\kms).  The radio data used for this image is that smoothed to a synthesised beam size of 5.8~arcsec (30~kpc at z~=~0.373).  The image size is shown in both arcsec and in kpc on opposite axes.  The contour levels are -150, -100, -50, 0, 50, 100 and 150~\microJy.
}


   \label{HI_spectrum_all}

\end{figure*}


The left panel in Fig.~\ref{HI_spectrum_all} shows the weighted average \HI\ spectrum from coadding the signal from all 324 galaxies using the large smoothing criteria.  To estimate the error in the \HI\ measurements, a series of artificial galaxies with random positions and random \HI\ redshifts were used to create coadded random spectra.  From many such artificial spectra a good estimate of the noise level in the measured real \HI\ spectrum could be determined.  As seen in Fig.~\ref{HI_spectrum_all}, the noise level increases with increasing velocity offset from the centre of the coadded spectrum.  This is because some galaxies lie at redshifts (frequencies) near the edges of the radio data cube.  When adding the spectra of these galaxies to the total, there is no data for velocities that correspond to frequencies that lie off the edge of the data cube.  These velocities with no data are given zero weight in the coadded sum resulting in a higher noise level at these velocities in the final coadded spectrum.  

The right panel in Fig.~\ref{HI_spectrum_all} shows the coadded \HI\ image for all galaxies after averaging the frequency channels across the velocity width of 600~\kms.  The radio data used for this image was that smoothed to a synthesised beam size of 30~kpc (5.8~arcsec), i.e.~the galaxies have not been broken up into the groups of small, medium and large.  As such the signal to noise is not the maximum measured. 


\section{The \HI\ 21-cm emission signal from subsamples of galaxies}
\label{The_HI_subsamples}


\begin{table*} 

\centering

\begin{centering}

\begin{tabular}[b]{|c|c|c|c|c|c|}  

\hline 

\ &
\ &
Number &
HI Mass &
HI Mass &
HI Mass \\ 

Galaxy &
Selection &
of &
Unsmoothed &
Mid-Smoothing &
Large Smoothing \\ 

Sample &
Criteria &
Galaxies &
($10^9$ \Msun) &
($10^9$ \Msun) &
($10^9$ \Msun) \\

\hline

All &
-- &
324 &
\ $4.0 \pm 1.4$ \ (2.9$\sigma$)  &
\ $4.8 \pm 1.8$ \ (2.7$\sigma$)  &
\ $6.6 \pm 3.5$ \ (1.9$\sigma$)  \\

Red &
$ B-V > 0.57$ &
219 &
\ $2.8 \pm 1.6$ \ (1.8$\sigma$)  &
\ $2.6 \pm 2.1$ \ (1.2$\sigma$)  &
\ $1.4 \pm 4.2$ \ (0.3$\sigma$)  \\

Blue &
$ B-V \le 0.57$  &
105 &
\ $7.1 \pm 2.7$ \ (2.6$\sigma$)  &
 $10.0 \pm 3.3$ \ (3.0$\sigma$)  &
 $19.0 \pm 6.5$ \ (2.9$\sigma$)  \\

Blue, Outside X-ray Gas &
$ B-V \le 0.57$ \& \rm d$\rm _{cl} > 1.45$ Mpc &
\ 94 &
\ $8.0 \pm 3.2$ \ (2.5$\sigma$)  &
 $11.1 \pm 3.9$ \ (2.8$\sigma$)  &
 $23.0 \pm 7.7$ \ (3.0$\sigma$)  \\

Non-[OII] Emission &
$\rm [OII] \le 5 \ \AA $ &
156 &
\ $4.3 \pm 1.8$ \ (2.4$\sigma$)  &
\ $5.2 \pm 2.4$ \ (2.2$\sigma$)  &
\ $2.3 \pm 4.8$ \ (0.5$\sigma$)  \\

[OII] Emission &
$\rm [OII] > 5 \ \AA $ &
168 &
\ $3.6 \pm 2.1$ \ (1.7$\sigma$)  &
\ $4.3 \pm 2.6$ \ (1.7$\sigma$)  &
 $11.4 \pm 5.2$ \ (2.2$\sigma$)  \\

Inner &
d$\rm _{cl} \le R_{200}$ \& z $\le$ 0.357 &
110 &
\ $3.6 \pm 1.7$ \ (2.2$\sigma$)  &
\ $3.9 \pm 2.2$ \ (1.8$\sigma$)  &
\ $3.3 \pm 4.4$ \ (0.8$\sigma$)  \\

Outer &
d$\rm _{cl} > R_{200}$ or z $>$ 0.357 &
214 &
\ $4.6 \pm 2.4$ \ (1.9$\sigma$)  &
\ $6.4 \pm 3.1$ \ (2.1$\sigma$)  &
 $12.1 \pm 6.1$ \ (2.0$\sigma$)  \\

Within 8 Mpc &
d$\rm _{cl} \le 8$ Mpc &
220 &
\ $3.6 \pm 1.4$ \ (2.6$\sigma$)  &
\ $4.3 \pm 1.9$ \ (2.3$\sigma$)  &
\ $5.1 \pm 3.7$ \ (1.4$\sigma$)  \\

Blue, Within 8 Mpc &
$ B-V \le 0.57$ \& d$\rm _{cl} \le 8$ Mpc &
\ 58 &
\ $8.5 \pm 3.0$ \ (2.8$\sigma$)  &
 $11.6 \pm 3.7$ \ (3.1$\sigma$)  &
 $17.5 \pm 7.3$ \ (2.4$\sigma$)  \\

\hline 

\end{tabular}

\caption{This table lists the average galaxy \HI\ mass for subsamples of the Abell~370 galaxies measured using the three different smoothed radio data combinations.  The `\HI\ Mass Unsmoothed' is the measurement made using only the original GMRT resolution radio data.  The `\HI\ Mass Mid-Smoothing' measurements are made by coadding different smoothed radio data such that the galaxies have estimated $\rm D_{eff} \le$~smoothed synthesised beam size and the `\HI\ Mass Large Smoothing' using $\rm 2 \times D_{eff} \le $~smoothed synthesised beam size.  In brackets next to each value is the signal to noise of the measurement.  d$\rm _{cl}$ is the projected distance in Mpc from the cluster centre and R$_{200}$ is 2.57~Mpc for Abell~370.  The two measurements `Within 8~Mpc' subsamples are for comparison with the Coma cluster in Section~\ref{Comparison_of_the_HI_measurements_with_the_literature}.  The selection for these subsamples is more complicated than that listed above (see Fig.~\ref{radius_vs_redshift}).  See the text for further details on the subsample selection and smoothing size.} 

\label{HI_mass_measurments}  

\end{centering}
\end{table*}


 
\begin{figure}  

  \begin{center}  
  \leavevmode  
	
   

   \includegraphics[width=8cm, bb= 18 184 592 718, angle=270]{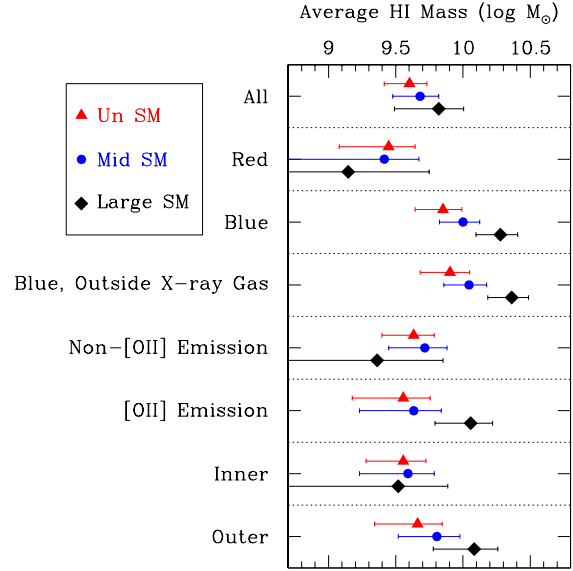}
 
   \end{center}

   \caption{This figure displays the average galaxy \HI\ mass for the different subsamples of galaxies measured using the different smoothed data combinations.   The `Un SM' are the Unsmoothed values,  `Mid SM' the Mid-Smoothing values and the `Large SM' are the Large Smoothing values.}

   \label{HI_mass}

\end{figure}


The 324 Abell~370 galaxies used in the \HI\ coadding are a mixture of early and late-type galaxies in a variety of environments.  It is interesting to examine the average \HI\ content of subsamples of galaxies selected by their optical colour, spectroscopic properties or location in the cluster (their galaxy environment).  The definition of the major galaxy subsamples considered can be found in Section~\ref{The_overlap_of_the_galaxy_subsamples}, which also details the overlap in their optical properties.  The \HI\ mass measurements for various subsamples of the Abell~370 galaxies can be seen in Table~\ref{HI_mass_measurments}.  The table lists three average \HI~masses for each subsample; these measurements are the same as used previously for all the galaxies, i.e.~unsmoothed, mid-smoothed and large smoothed measurements.  These average \HI\ mass measurements are displayed in Fig.~\ref{HI_mass}.  From this figure it can be seen that there are differences not only in the quantity of the \HI\ gas between the subsamples but also differences in where the \HI\ gas is located within the galaxies.  

 
\begin{figure*}  

  \begin{center}  
  \leavevmode  

  \subfigure{		
    \includegraphics[width=6cm]{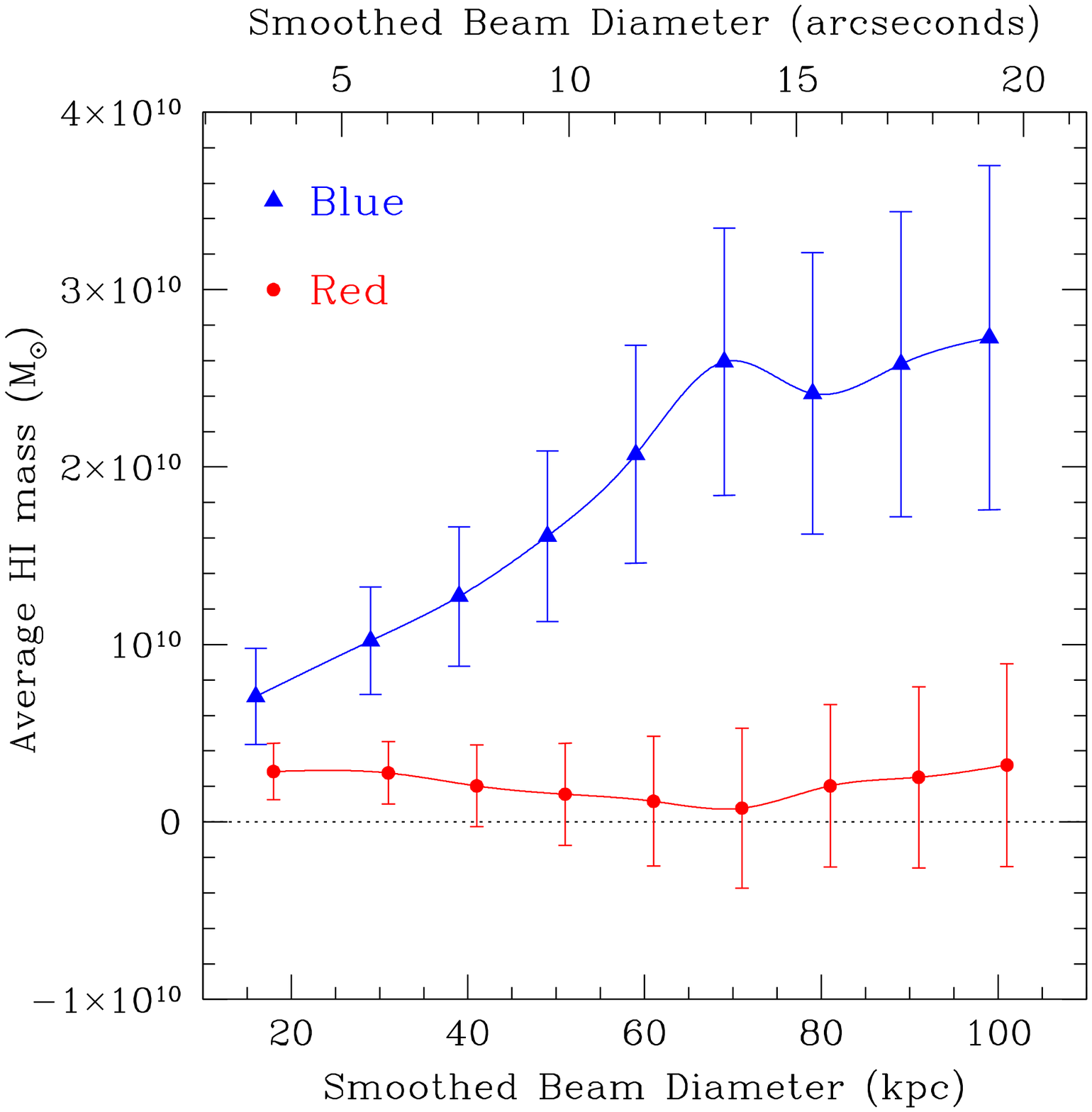} 
    \includegraphics[width=6cm]{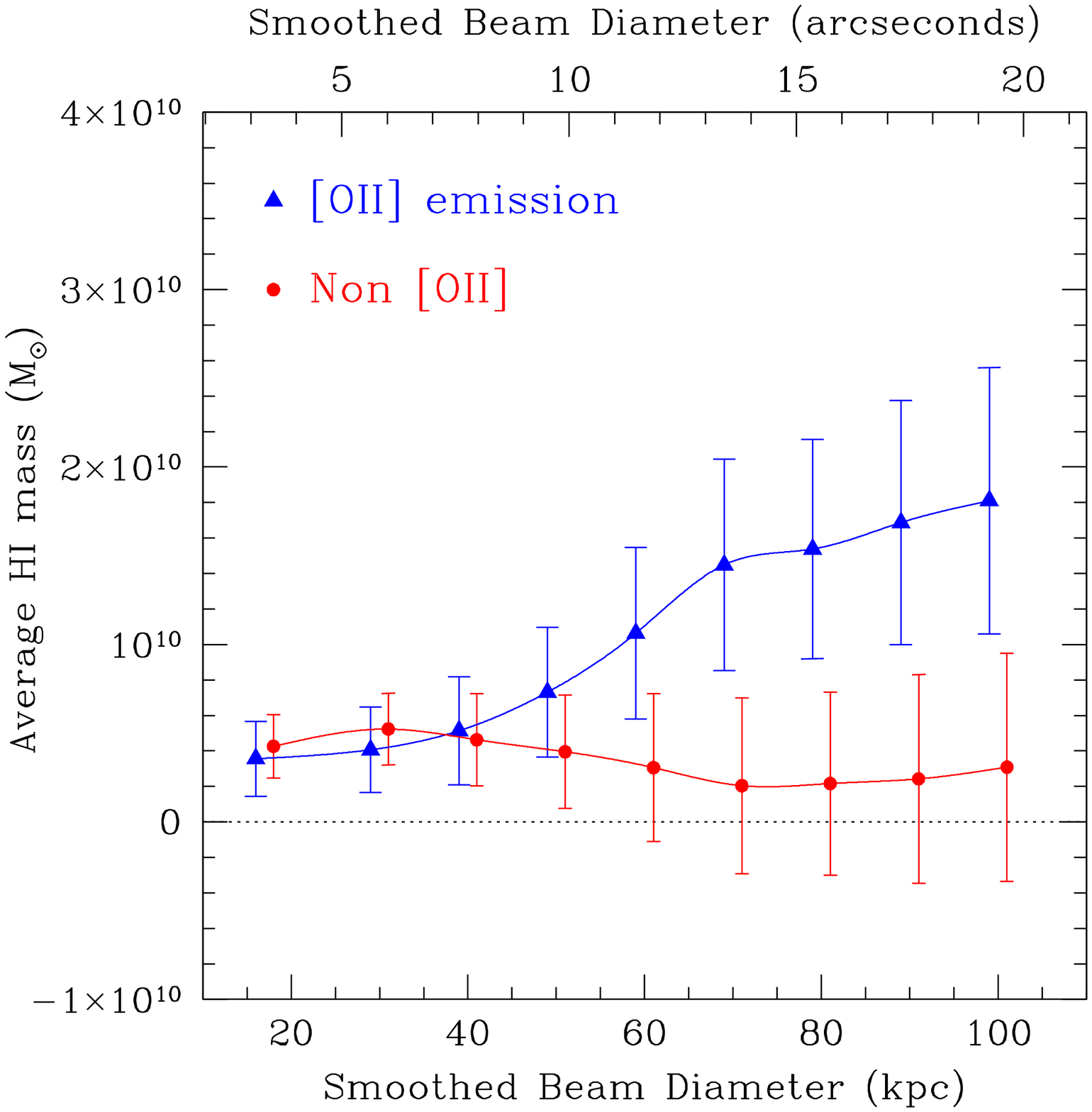}
    \includegraphics[width=6cm]{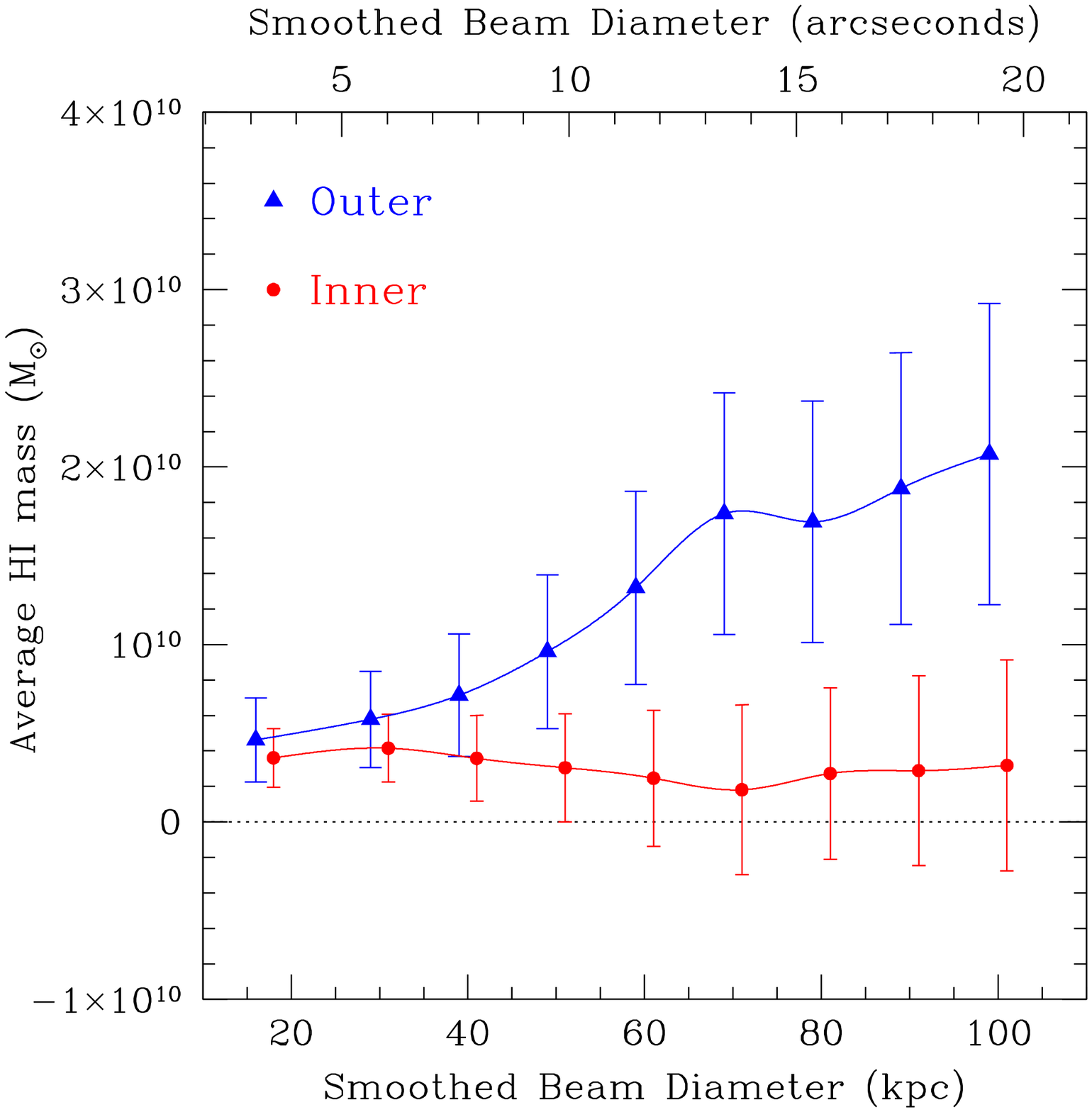}
   } 

   \end{center}

   \caption{This figure shows the average \HI\ mass as measured in the different smoothed synthesised beam size data for various galaxy subsamples.  The left panel shows the blue and red subsamples, the middle panel the [OII] emission and non-[OII] emission subsamples and the right panel the inner and outer subsamples.  The points in each panel for the two subsamples have been slightly offset in the x-direction to prevent obscuration by overlapping values.}

   \label{HI_mass_SM_comparison}

\end{figure*}


 
\begin{figure*}  

  \begin{center}  
  \leavevmode  

  \subfigure{		
    \includegraphics[width=6cm]{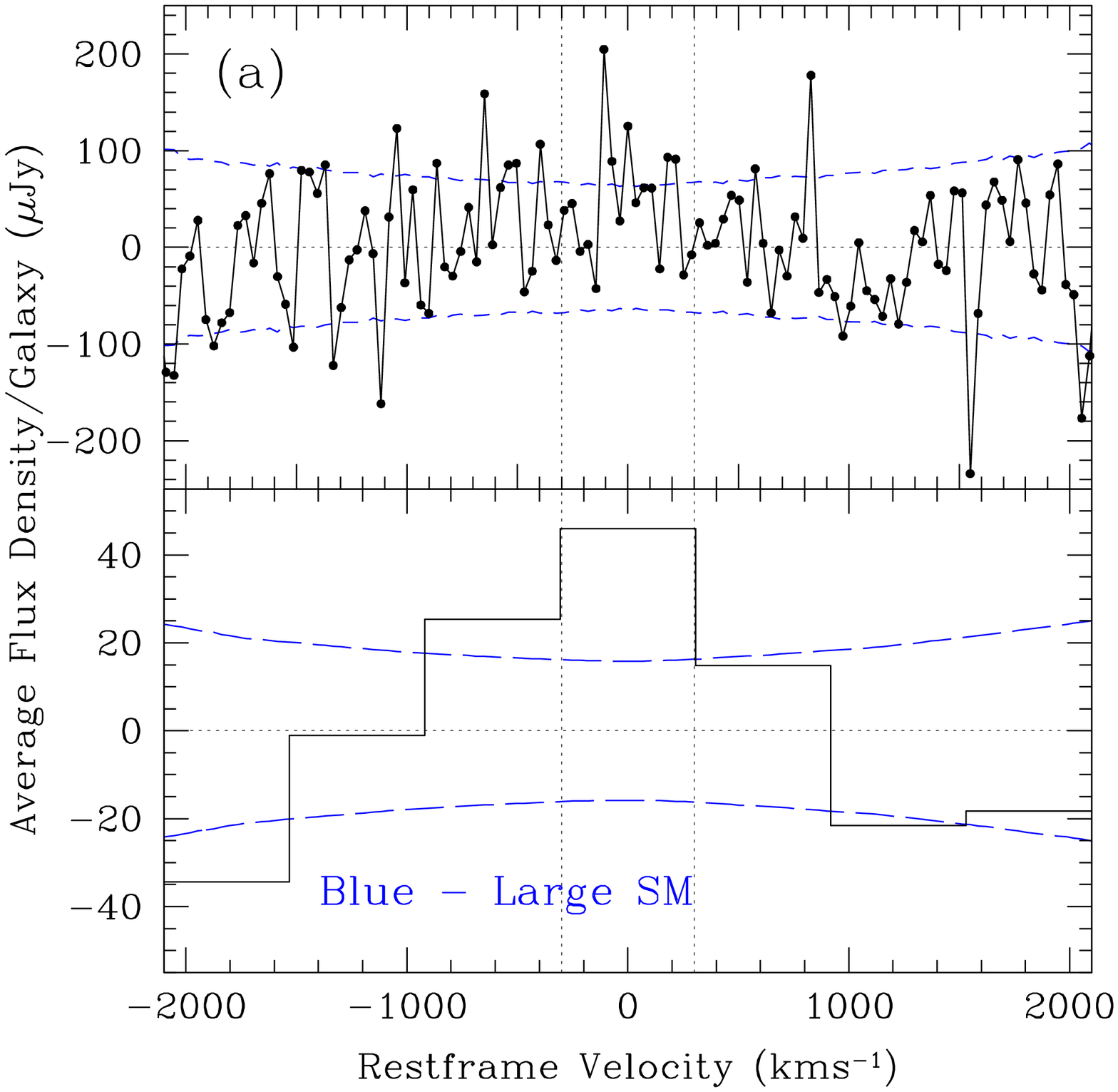} 
    \includegraphics[width=6cm]{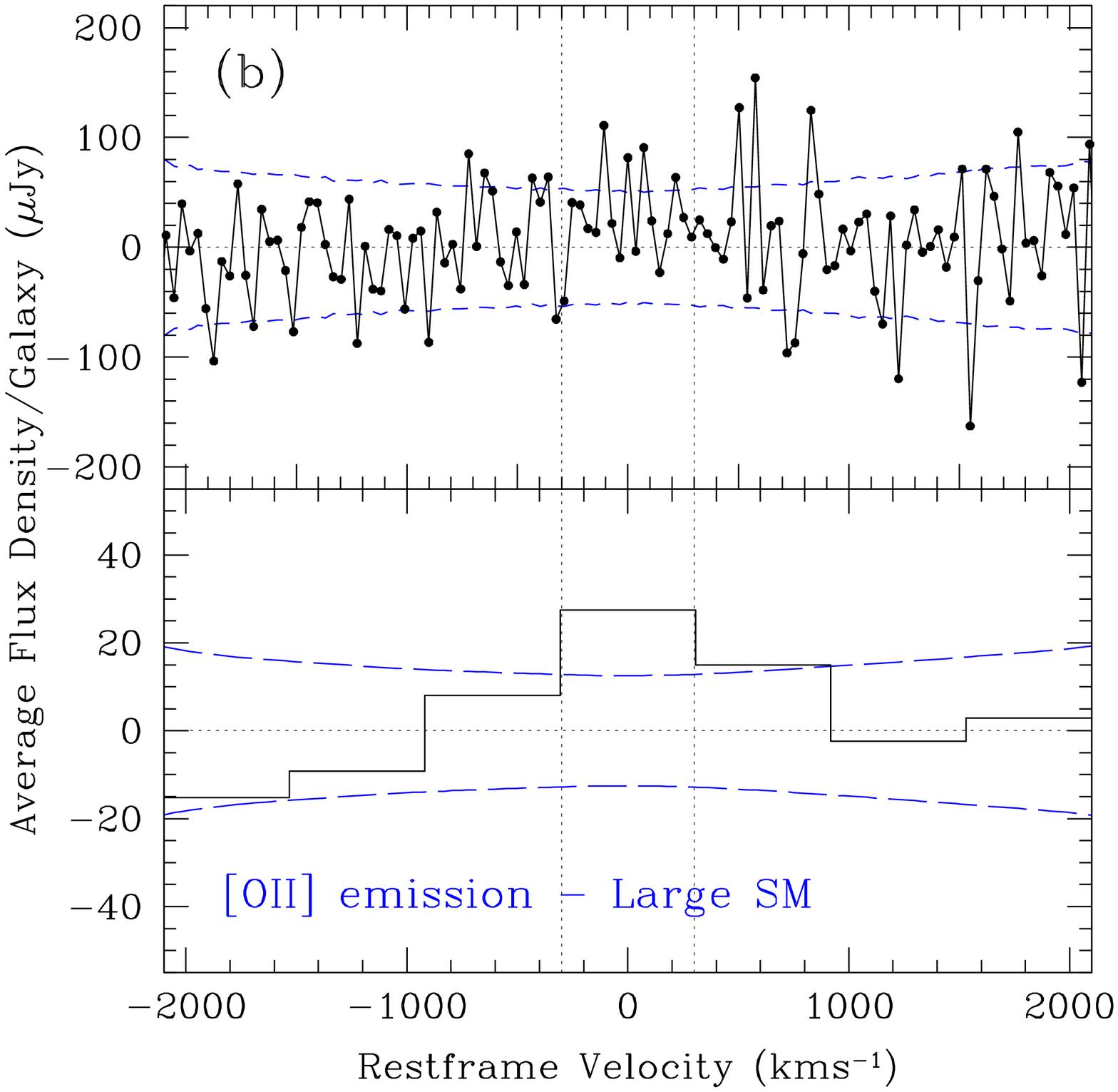} 
    \includegraphics[width=6cm]{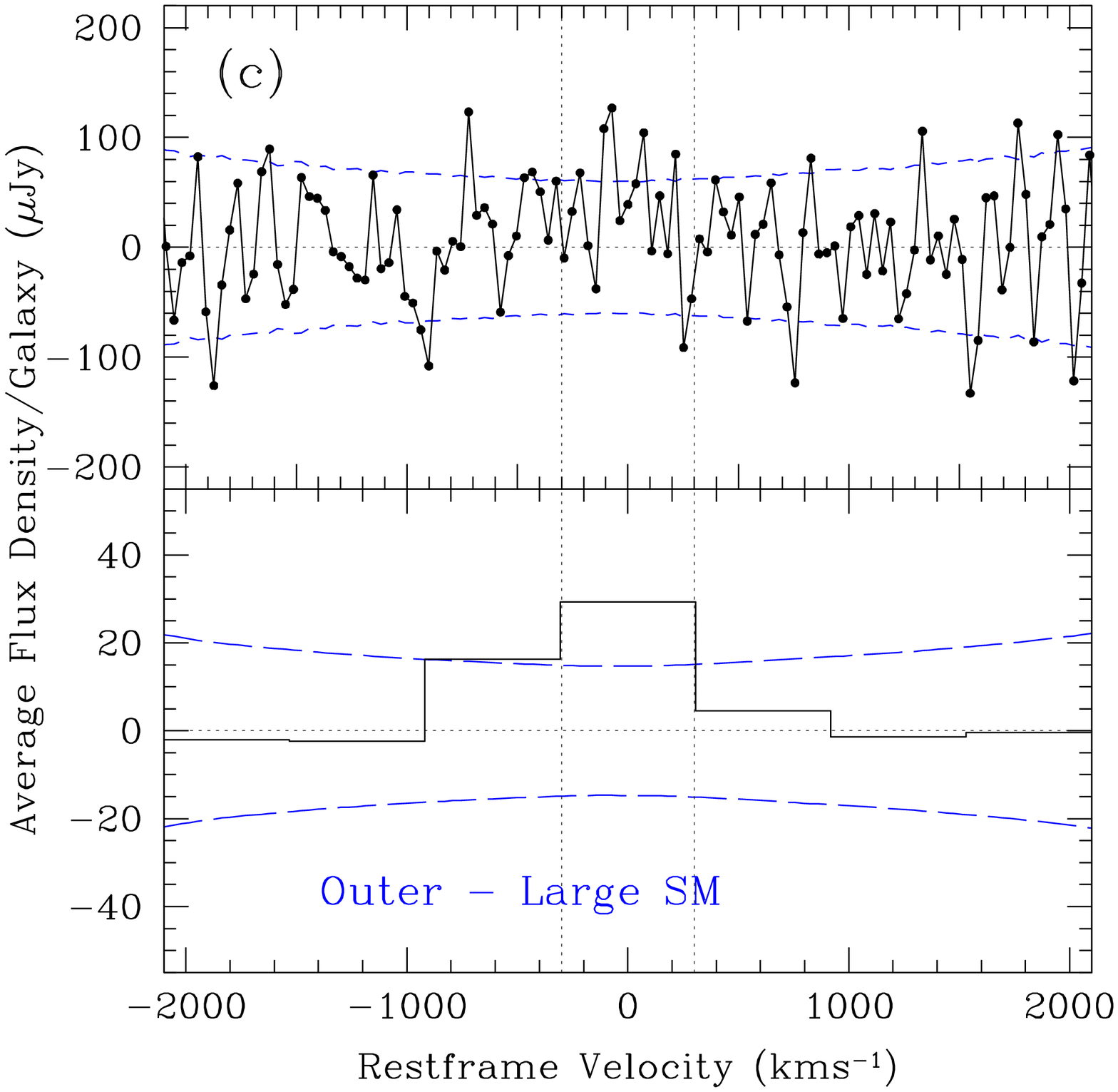}
   } 

  \subfigure{
    \includegraphics[width=6cm]{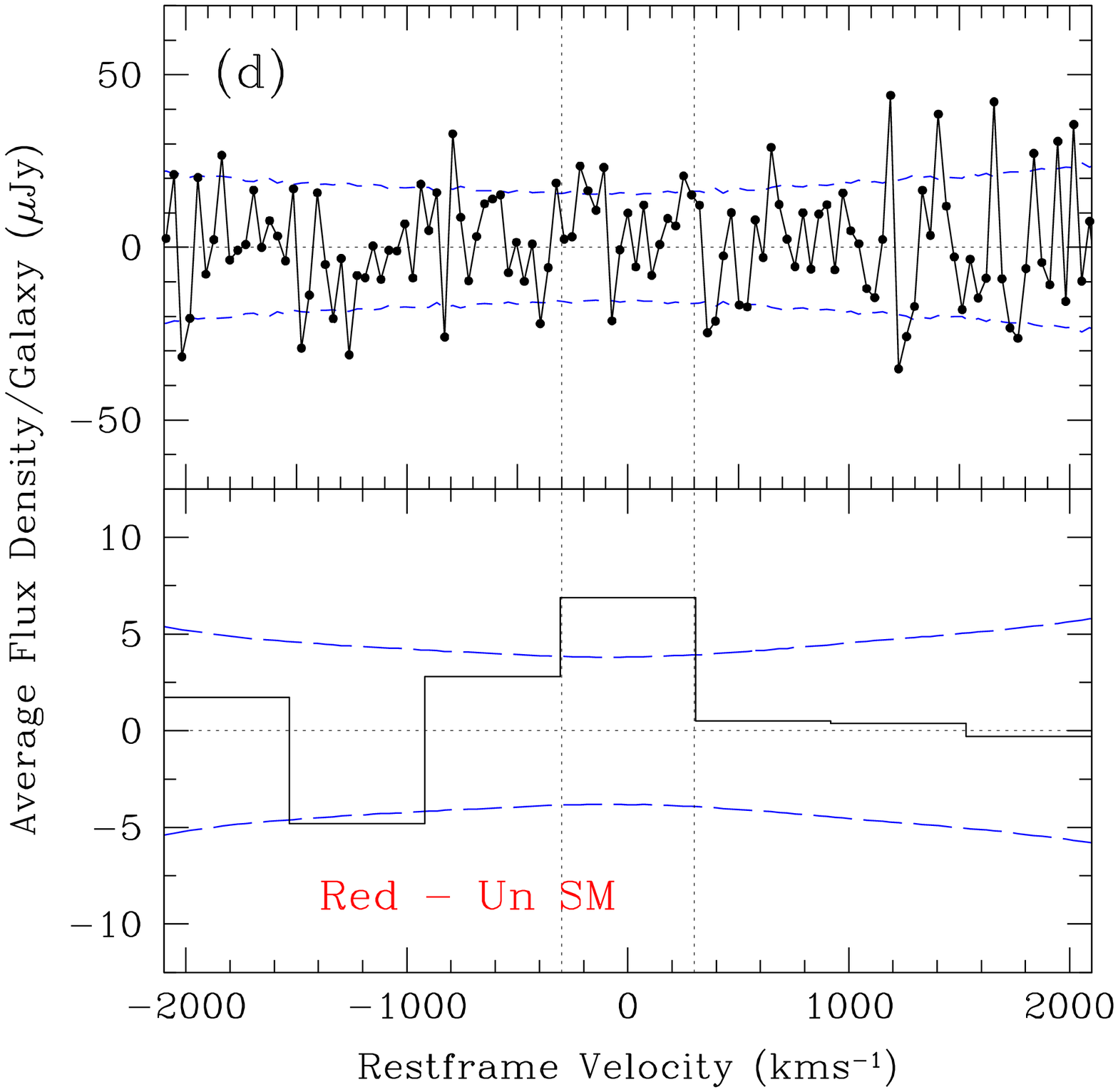}
    \includegraphics[width=6cm]{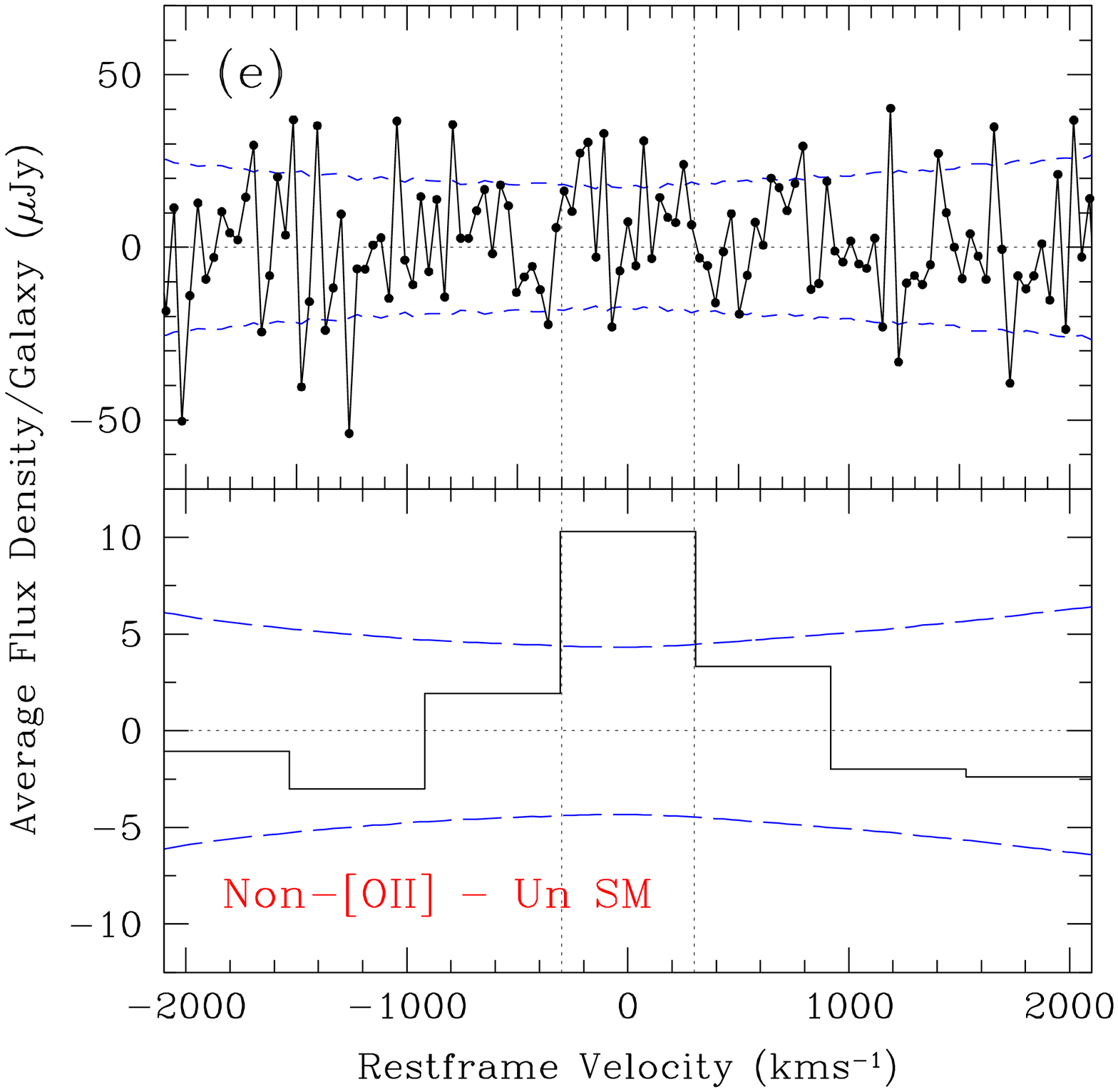}
    \includegraphics[width=6cm]{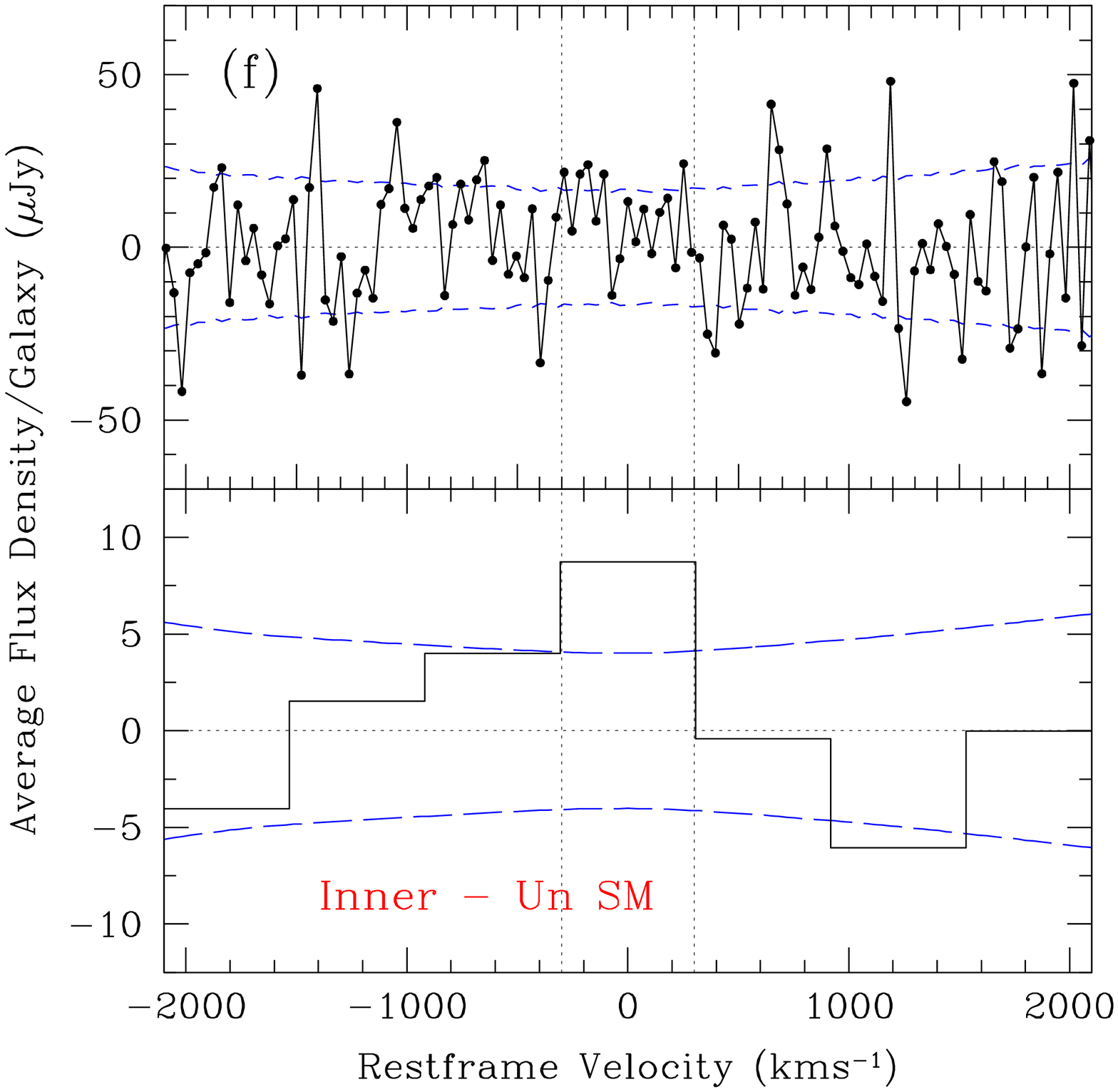} 
   } 

   \end{center}

   \caption{
The average \HI\ galaxy spectrum created from coadding the signal of galaxies in different subsamples.  
The top panels use the large smoothing criterion; the bottom panels the unsmoothed criterion. 
Panel~(a) is for the 105 blue galaxies, panel~(b) is for the 168 [OII] emission galaxies and the panel~(c) is for those 214 outer galaxies (those away from the cluster centre).  
Panel~(d) is for the 219 red galaxies, panel~(e) for the 156 non-[OII] emission galaxies and panel~(f) is for the 110 inner galaxies (those close to the cluster centre).  
In each sub-window the top spectrum has no smoothing or binning and has a velocity step size of 36.0~\kms .  The bottom spectrum in each sub-window has been binned to 600~\kms .  This is the velocity width that the combined \HI\ signal of the galaxies is expected to span.  For both spectra the $1\sigma$ error is shown as dashed lines above and below zero.  Note that the y-axis scale for the top and bottom panels are substantially different.
}

   \label{HI_spectrum_multiple}

\end{figure*}


\subsection{The blue and red subsamples}
\label{The_blue_and_red_subsamples}

The separation of the Abell~370 galaxies into optically blue and red subsamples was done using the Butcher--Oemler criterion, i.e.~the blue galaxies in Abell~370 have $B-V$~colour~$\le 0.57$ (see Section~\ref{The_optical_properties_of_the_Abell_370_galaxies}).  

In the coadded signal from the 219 red galaxies the unsmoothed average \HI\ mass measurement is $(2.8 \pm 1.6) \times 10^9$~\Msun.  Measurements made using the higher smoothing criteria show no increase in the average \HI\ mass and the higher noise in these measurements overwhelms any signal as can be seen in Fig.~\ref{HI_mass}.  These results suggest that any substantial quantities of \HI\ gas in the red galaxies must lie within the central regions of the galaxies, i.e.~closer than 8.5~kpc to the centre of the galaxies based on the unsmoothed measurement.  It is likely that a fair number of these red galaxies have no \HI\ gas.  An examination of the data shows that there is no immediately obvious handful of galaxies that have the majority of this \HI\ signal.   

In contrast to the red galaxies, coadding the blue galaxies gives rise to a strong detection of \HI\ 21-cm emission with the observed signal increasing appreciable in the higher smoothing measurements (see Fig.~\ref{HI_mass}).  This is similar to the trend seen in \HI\ rich galaxies in the local universe, which contain large amounts of \HI\ gas that extend beyond their visible stellar discs \citep{broeils97}.  The average \HI\ mass measured for the 105 blue galaxies using the large smoothing criteria is $(19.0 \pm 6.5) \times 10^9$~\Msun.  The large smoothing criteria, which is based on the \HI\ galaxy sizes of spiral and irregular galaxies in the field, appears to be a good fit to \HI\ gas content of the blue galaxies. 

The very different way the \HI\ gas is distributed in the blue and red galaxies can be seen in greater detail in the left panel of Fig.~\ref{HI_mass_SM_comparison}.  This figure traces the change in \HI\ mass as a function of smoothed synthesised beam size for the blue and red galaxies, i.e.~not using the galaxy size groupings.  The blue galaxies show steady increase in \HI\ gas with smoothing size until it appears to end \around 70~kpc.  The red galaxies show some \HI\ signal at the lowest smoothing sizes but the higher beam sizes appear to just add noise obscuring any signal.  The coadded \HI\ spectrum for the blue galaxies using the large smoothing criteria can be seen in panel~(a) of Fig.~\ref{HI_spectrum_multiple} and the coadded \HI\ spectrum for the red galaxies using the unsmoothed criteria in panel~(d) of this figure.

 
\begin{figure*}  

  \begin{center}  
  \leavevmode  
		
    \includegraphics[height=7cm]{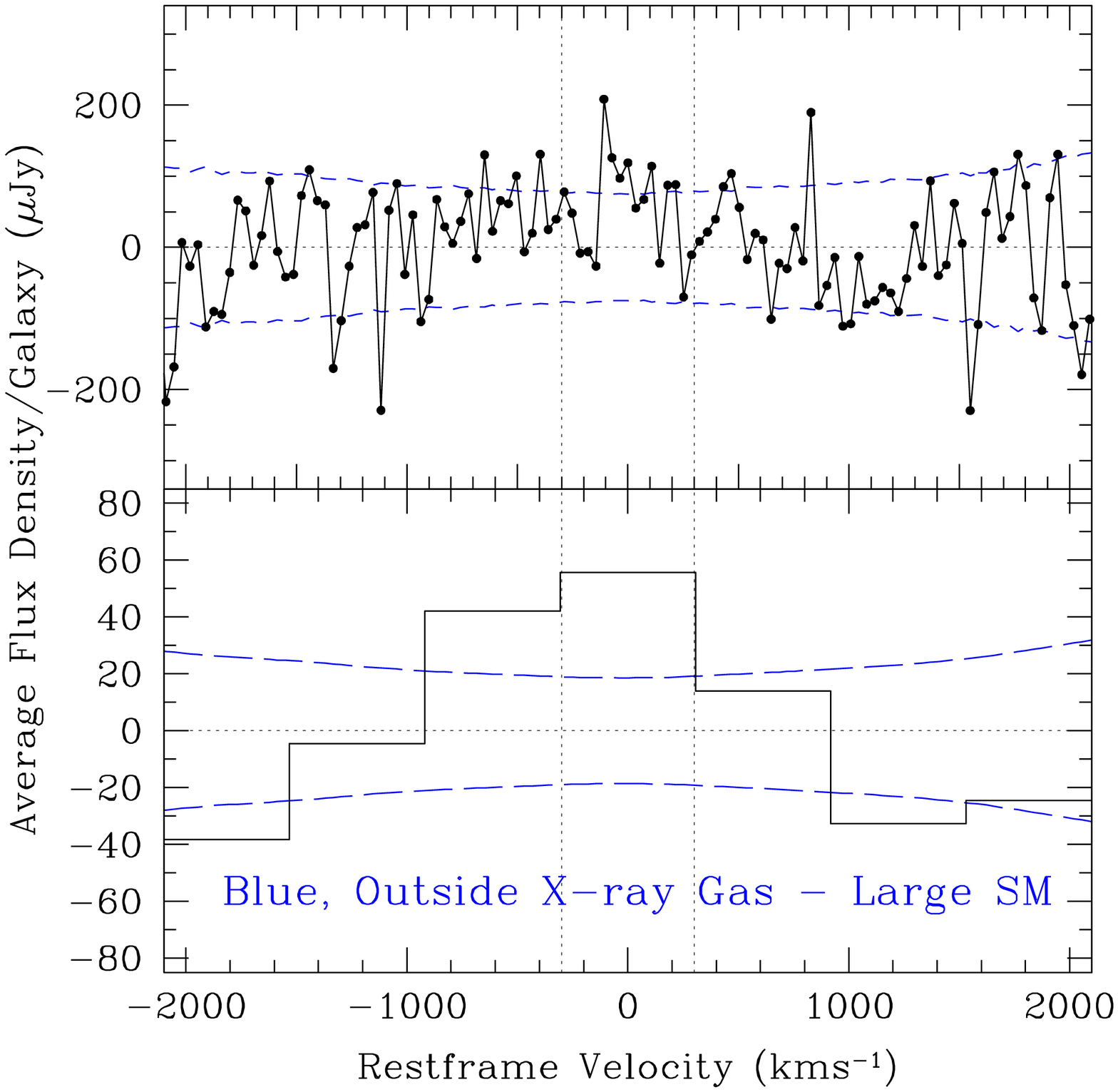}
    \includegraphics[height=7cm]{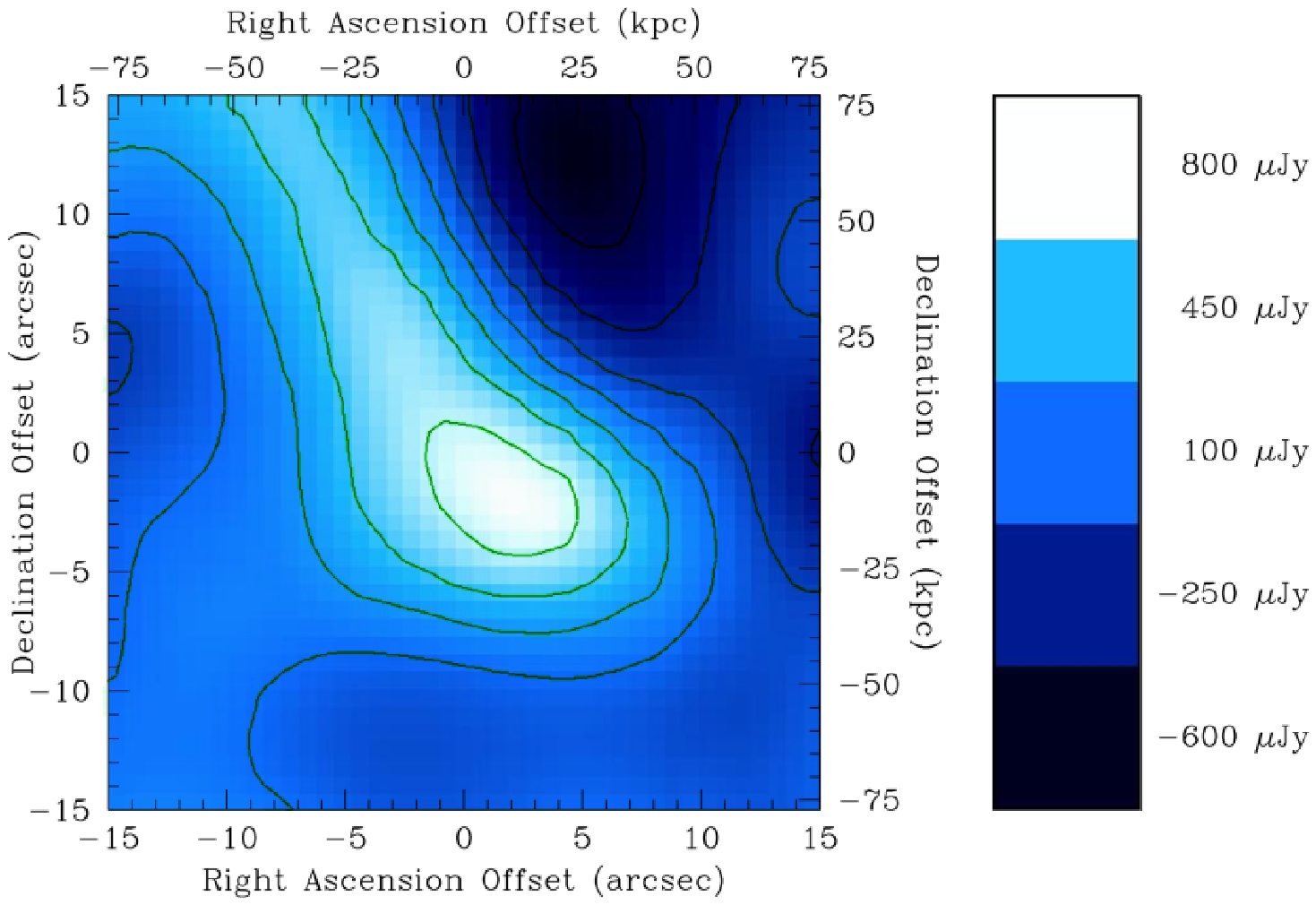}
  
   \end{center}

   \caption{The left panel shows the average \HI\ galaxy spectrum created from coadding the signal of the 94 blue galaxies outside the intracluster, X-ray gas using the large smoothing criteria.  The top spectrum has no smoothing or binning and has a velocity step size of 36.0~\kms .  The bottom spectrum has been binned to 600~\kms .  This is the velocity width that the combined \HI\ signal of the galaxies is expected to span.  For both spectra the $1\sigma$ error is shown as dashed lines above and below zero.
The right panel shows the average \HI\ image made by coadding the data cube around each of the blue galaxies outside the intracluster, X-ray gas and binning across 17~spectral channels (600~\kms).  The radio data used for this image is that smoothed to a synthesised beam size of 9.7 arcsec (50 kpc at z~=~0.373).  The image size is shown in both arcsec and in kpc on the opposite axes.  The contour levels are -500, -300, -100, 100, 300, 500 and 700~\microJy.
}


   \label{HI_out_x_ray_butchler_omeler}

\end{figure*}


In the central regions of nearby clusters, late-type galaxies are found to be \HI\ deficient compared to similar galaxies in the field \citep{haynes84}.  One would like to see if this trend is seen in the late-type galaxies inside the hot intracluster medium of the cluster core of Abell~370 at z~=~0.37.  Assuming that blue galaxies are a representative sample of the late-type galaxies, then there are 11 such galaxies within the 3$\sigma$\ extent of the X-ray gas (1.45~Mpc from the cluster centre, see Section~\ref{The_optical_properties_of_the_Abell_370_galaxies}).  Unfortunately this is an insufficient number of galaxies to coadded to enable a meaningful measurement of the gas depletion of the galaxies.  The best that one can do is consider the subsample of blue galaxies outside the X-ray significance radius of which there are 94.  For this subsample the average \HI\ mass is $(23.0 \pm 7.7) \times 10^9$~\Msun\ using the large smoothing criteria.  This is larger than that found for the blue subsample as a whole and is the highest significance detection of \HI\ 21-cm emission found in this work.  The difference between this subsample and the subsample of all blue galaxies seems to be greatest in the large smoothing \HI\ measurements; there is only a small increase in the unsmoothed and mid-smoothing values (see Table~\ref{HI_mass_measurments}).  This suggests that the blue galaxies inside the hot intracluster gas have lost \HI\ gas in their outer regions.  This is consistent with how most of the various environment mechanisms would remove the gas in galaxies (see Section~\ref{Introduction}).  This difference in average \HI\ mass for all 105 blue galaxies and the 94 blue galaxies outside the X-ray gas is not statistically significant but it does follow the expected trend of \HI\ deficiency found in late-type galaxies within nearby clusters.   

The left panel in Fig.~\ref{HI_out_x_ray_butchler_omeler} shows the weighted average \HI\ spectrum from blue galaxies outside the X-ray gas using the large smoothing criteria.  The right panel in Fig.~\ref{HI_out_x_ray_butchler_omeler} shows the coadded \HI\ image for the same galaxies after averaging the frequency channels across the velocity width of 600~\kms.  This image was made using the same method as discussed in Section~\ref{HI_all_galaxies} for the image of all galaxies (seen in Fig.~\ref{HI_spectrum_all}) except that data from the larger smoothed synthesised beam of 50~kpc (9.7~arcsec) was used.  This was done to highlight the signal from the outskirts of the coadded galaxies which are difficult to see in the smaller beam sized data.  The centre of the \HI\ image appears \around 2~arcsec different from the centre of the optical galaxies.   This is not a real astrometric difference between the radio and optical data.  Instead it is an effect of the large smoothed synthesised beam size used and the contribution of a slight positive noise spike (less than $1 \sigma$) that is located away from the centre of the image.  The good alignment between the optical and radio data can be seen in images made using the smallest synthesised beam size where only the \HI\ signal from the very central regions of the galaxies is noticeable.


\subsection{The [OII] and non-[OII] emission subsamples}
\label{The_[OII]_and_non_[OII]_emission_subsamples}

Subsamples of the Abell~370 galaxies were made based on the presence or lack of the [OII]$\lambda$3727 optical emission line in their spectra.  A measured [OII] equivalent width of 5~\AA\ was used as the cut off between the emission and non--emission subsamples (see Section~\ref{The_optical_properties_of_the_Abell_370_galaxies}).  Multiple \HI\ mass measurements for each subsample were made, as previously, and the results are shown in Fig.~\ref{HI_mass}.  Unlike the red and blue galaxies, these two subsamples seem to have similar average \HI\ gas masses in the inner regions of the galaxies.  It is in measurements including the outer regions that a difference between the two subsamples can be seen.  For the large smoothing with the [OII] emission subsample shows an increase in the average \HI\ gas content while the non-[OII] emission subsample measurement is consistent with no increase in \HI\ gas.  The 168 [OII] emission galaxies have an average \HI\ mass of $(11.4 \pm 5.2) \times 10^9$~\Msun\ using the large smoothing criteria while 156 non-[OII] emission galaxies have only $(4.3 \pm 1.8) \times 10^9$ using the unsmoothed criteria.  The \HI\ spectrum for the [OII] emission galaxies using the large smoothing criteria can be seen in panel~(b) of Fig.~\ref{HI_spectrum_multiple} and the \HI\ spectrum for the non-[OII] emission galaxies using the unsmoothed criteria in panel~(e) of this figure.

The \HI\ gas distribution in these subsamples can be seen in greater detail in the middle panel of Fig.~\ref{HI_mass_SM_comparison}.  This figure shows the change in \HI\ mass as a function of smoothed synthesised beam size for the [OII] emission and non-[OII] emission subsamples of galaxies.  The two subsamples start out with similar average \HI\ masses within the smaller synthesised beams, until they diverge at \around 50~kpc beam size.   The measurements for the non-[OII] emission subsample are consistent with no increase in \HI\ content factoring in the increasing size of the measurement errors.  The [OII] emission subsample shows an increase in the \HI\ gas content in the outskirts of the galaxies (an increase in signal in the higher beam measurements) that is similar to that seen in the blue galaxy subsample.

This similarity brings up the obvious question of how much overlap there is between the blue galaxy subsample and the [OII] emission galaxy subsample.  In the [OII] emission subsample of 168 galaxies roughly half are blue and the other half red (81 blue galaxies vs.\ 87 red galaxies).  This is the majority of the blue galaxies (there are only 24 blue galaxies in the non-[OII] emission subsample) but less than half of the total red galaxies.  This shows that the [OII] subsamples are a substantially different grouping of galaxies than the blue/red galaxy subsamples (for more on the overlap of the subsamples see Section~\ref{The_overlap_of_the_galaxy_subsamples}).

 
\begin{figure*}  

  \begin{center}  
  \leavevmode  

  \subfigure{	
    \includegraphics[width=6cm]{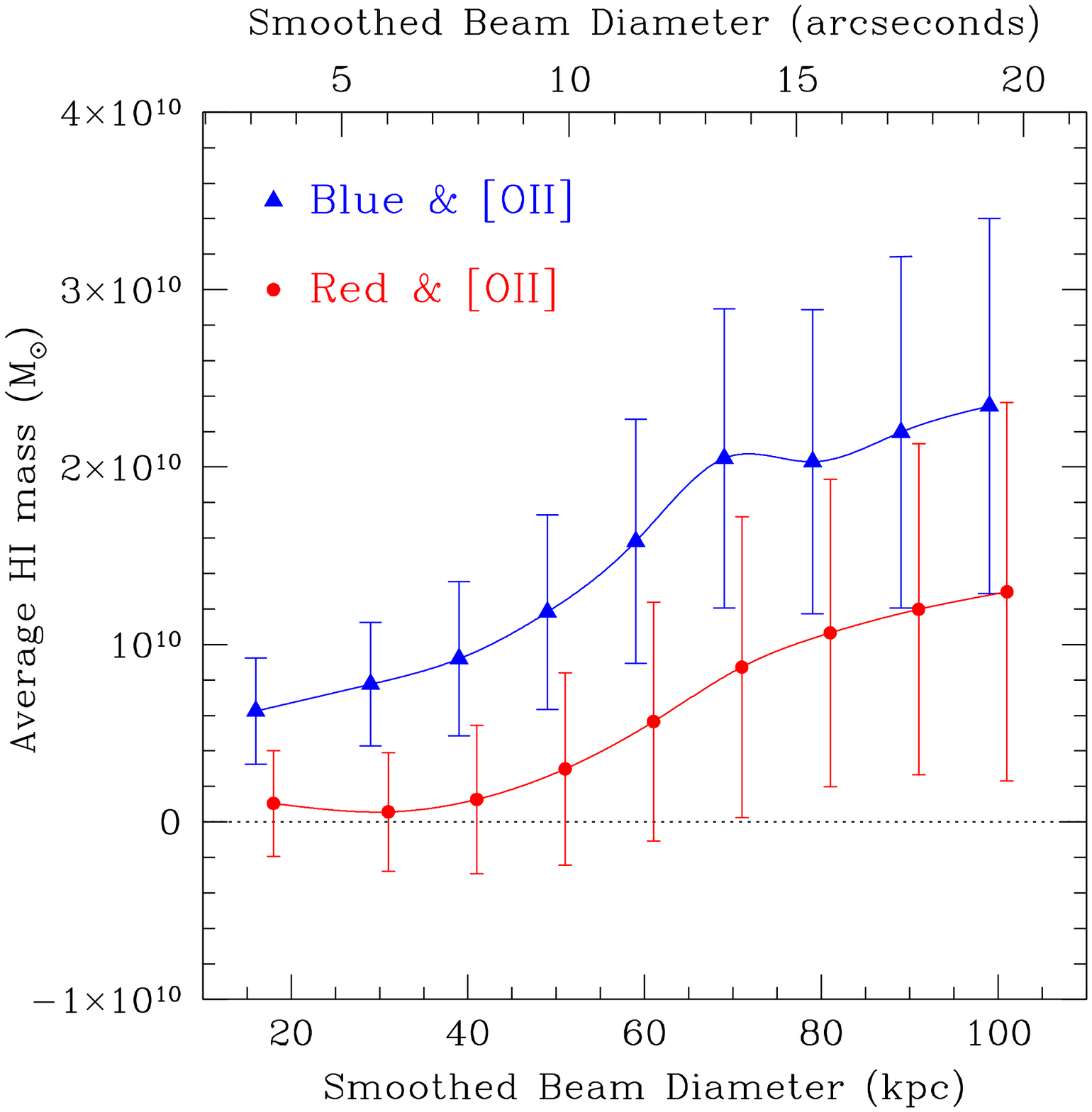}
    \includegraphics[width=6cm]{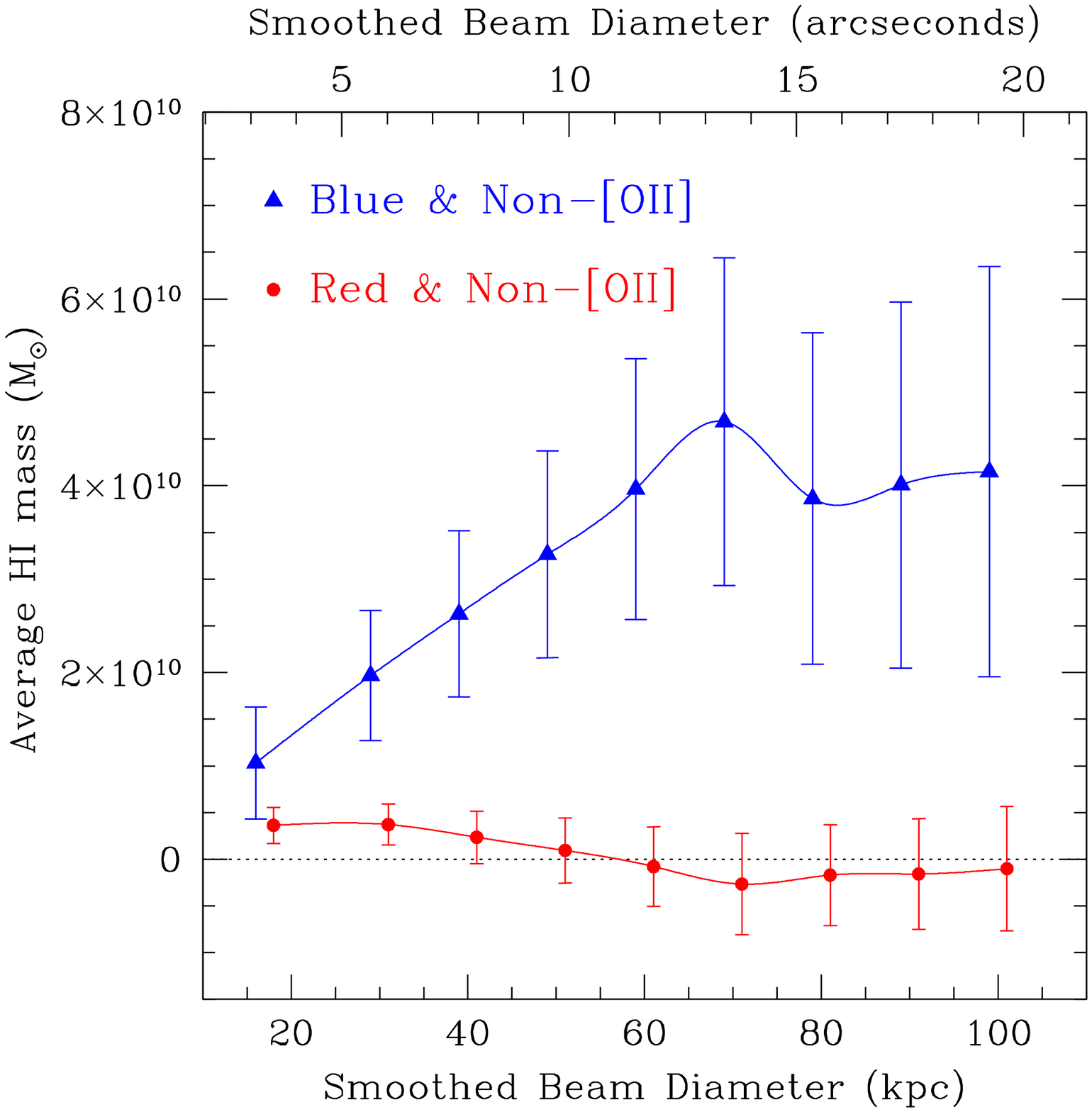}
   } 

   \end{center}

   \caption{This figure shows the average \HI\ mass measured in the different smoothed synthesised beam size data for the [OII] emission and non-[OII] emission subsamples split further into their constituent red and blue galaxies.  The left panel shows the 168 [OII] emission galaxies split up into the 87 red galaxies and 81 blue galaxies.  The right panel shows the 156 non-[OII] emission galaxies split into the 132 red galaxies and 24 blue galaxies.  The points in each panel for the two subsamples have been slightly offset in the x-direction to prevent obscuration by overlapping values.  Note that y-axis scale is substantially different between the two panels.}

   \label{HI_mass_SM_OII_non_OII}

\end{figure*}


The [OII] emission galaxies were broken up further into their blue and red galaxy subsamples and the average \HI\ mass with smoothed synthesised beam size measured as seen in the left panel of Fig.~\ref{HI_mass_SM_OII_non_OII}.  Although we are pushing the data to its limit, a clear difference can be seen.  In general the blue [OII] emission galaxies have an average higher \HI\ gas content than the red [OII] emission galaxies.  The measured \HI\ content of both subsamples increases with increasing synthesised beam sizes.  However, the red galaxies have \HI\ gas values consistent with zero \HI\ gas in their inner regions (the lower synthesised beam measurements).  It is only as one moves to the higher synthesised beam measurements that any \HI\ signal appears (at low significance).  From this data, one could suggest that the red [OII] emission subsample are galaxies that have a large, optically bright, red bulge with little or no gas surrounded by optically faint disk containing some gas, i.e. similar to Hubble type `Sa' galaxies.  The resolution of the optical imaging is not sufficient to confirm this large bulge, small disk model for these galaxies.  `Sa' galaxies in the nearby universe have median $B-V$ colour of 0.78 \citep{roberts94}, which is consistent with the colour value for the red [OII] emission galaxies (see Fig.~\ref{hist_overlap_ratio}, the middle, centre panel). 

Similarly the non-[OII] emission galaxies have been divided into their blue and red galaxies.  Of the 156 non-[OII] emission galaxies, there are 132 red galaxies and only 24 blue galaxies.   The average \HI\ mass with smoothed synthesised beam size has been measured for these subsamples of galaxies as seen in the right panel of Fig.~\ref{HI_mass_SM_OII_non_OII}.  The blue non-[OII] emission galaxies appear to contain substantial amounts of \HI\ gas with an average $\sim 40 \times 10^9$~\Msun, twice that found in the blue galaxies as a whole.  This result should be judged with caution as we are looking at subsample of only 24 galaxies with large measurement errors.  No one galaxy or handful of galaxies in this sample dominates the \HI\ measurement.  Of these blue non-emission galaxies only 3 lie within the \Rtwo\ radius of the cluster.   These galaxies span the complete spread of the $B$~band total absolute magnitudes of all blue galaxies in the sample (from -22.8 to -19.9).   The [OII] emission line in these galaxies could be present at a very weak level, i.e.~below 5~\AA.  It is unlikely though, that the line is obscured completely by dust extinction internal to the galaxies.   If this was the case the galaxies would show a redder colour.  These galaxies could be similar to E+A galaxies that have no O and B stars to create the emission lines in \HII\ regions but contain sufficient A and F stars to give the galaxies a blue colour.  Unfortunately, for these galaxies the optical spectra is not of sufficient quality to use the \Hdelta\ and \Hgamma\ absorption features to identify whether they are E+A galaxies.  The lack of [OII] emission lines suggests that the galaxies currently lack strong internal ionising radiation.  Without this radiation \HII\ gas in the galaxies could recombine to give the higher \HI\ content.  This \HII\ gas could be gas ionised in the previous star formation period (possibly during a star-burst) or could be gas currently infalling onto the galaxy.

Although the blue non-[OII] emission galaxies appear to have a strong \HI\ signal when considered alone, they do not dominate the total \HI\ signal of all non-[OII] emission galaxies.  The red non-[OII] emission galaxies dominate the average \HI\ signal from all the non-[OII] emission galaxies due to their large number and their higher weight in the coadded signal, as many are located near the cluster centre and hence close to the GMRT pointing centre.  In the right panel of Fig.~\ref{HI_mass_SM_OII_non_OII}, the red non-[OII] emission measurements are actually quite similar to the values found for both the red galaxies and the non-[OII] emission galaxies as a whole (see the left and middle panels of Fig.~\ref{HI_mass_SM_comparison}).   The red non-[OII] emission sample has a measured average \HI\ mass of $(3.6 \pm 1.9) \times 10^9$~\Msun\ in the unsmoothed data, not dissimilar to both the total red subsample and total non-[OII] emission sample values (see Table~\ref{HI_mass_measurments}).  


\subsection{The inner and outer cluster subsamples}
\label{The_inner_and_outer_cluster_subsamples}

Subsamples of the Abell~370 galaxies were made based on their location relative to the centre of the cluster.  {An inner subsample was formed from galaxies that fell within a projected distance of 2.57~Mpc (the \Rtwo\ radius) of the cluster centre and in the redshift range z~=~0.356 to 0.390 around the cluster redshift centre (see Section~\ref{The_overlap_of_the_galaxy_subsamples}).  Galaxies outside these criteria make up the outer galaxy subsample.}

For these subsamples \HI\ measurements were made as previously and the results are shown in Fig.~\ref{HI_mass}.  The inner subsample of galaxies have an average \HI\ mass similar to the red galaxy subsample; any \HI\ gas content they have is located in the central regions of the galaxies.  The 110 inner galaxies have an unsmoothed average \HI\ mass measurement of $(3.6 \pm 1.7) \times 10^9$~\Msun.  The similarity with the red galaxies is not surprising as the inner sample is  dominated by red galaxies (87 out of 110 galaxies).  The \HI\ measurements for the outer subsample are similar to that seen for the  blue galaxies.  There is a definite \HI\ signal from the central regions of the galaxies with an increased \HI\ signal when including the outskirts of the galaxies.   The 219 outer galaxies have an average \HI\ mass of $(12.1 \pm 6.1) \times 10^9$~\Msun\ using the large smoothing criteria.  The outer sample is made up of 132 red galaxies and 82 blue galaxies (this is \around 80~per~cent of all the blue galaxies around Abell~370).  Around two thirds of the outer subsample \HI\ signal is due to the blue galaxies with the final third coming from the outer red galaxies.

The galaxies within the inner regions of the cluster Abell~370 at z~=~0.37 have a lower average \HI\ mass than galaxies outside this region, following the environmental trend seen in nearby clusters.  The difference in the way the \HI\ gas is distributed in the inner and outer galaxies can be seen in greater detail in the right panel of Fig.~\ref{HI_mass_SM_comparison} which traces the change in \HI\ mass as a function of smoothed synthesised beam size for these subsamples.  The \HI\ spectrum for the outer galaxies using the large smoothing criterion can be seen in panel (c) of Fig.~\ref{HI_spectrum_multiple} and the \HI\ spectrum for the inner galaxies using the unsmoothed criterion in panel~(f) of this figure.  

Any \HI\ gas located in the inner galaxy subsample seems to be concentrated in the centre of the galaxies.  The central region of a galaxy is the place where \HI\ gas could survive for longer.   Only a small percentage of the interstellar medium in a galaxy needs to be neutral to greatly reduce the ionising flux that reaches the central regions of the galaxy.   Gas in the central regions would be packed into a small volume giving rise to higher gas densities.  This would allow for the \HI\ gas to recombine faster after being ionised, extending its lifetime.  Additionally, \HI\ gas in the centre of a galaxy would also be less affected by ram pressure stripping and galaxy harassment lying deep in the gravitational well of the galaxy.  In fact galaxy harassment may actually encourage gas in the outskirts of a galaxy to move towards the centre of the galaxy.  In the simulations of \citet{moore96}, galaxy harassment had the effect of stripping \around 10~per~cent of a galaxy's gas content while the other 90~per~cent of the gas sank to the inner few hundred parsecs.  Such an effect could be important in creating the \HI\ gas distribution seen in the Abell~370 galaxies.


\section{Comparison of the \HI\ measurements with the literature}

\label{Comparison_of_the_HI_measurements_with_the_literature}


\begin{table*} 

\centering

\begin{centering}

\begin{tabular}[b]{|c|c|c|c|c|c|}  

\hline 

\  &
\  &
Luminosity &
X-ray gas &
Velocity &
\  
\\

Galaxy &
\  &
Distance &
Temperature &
Dispersion &
R$_{200}$ 
\\

Cluster  &
Redshift  &
(Mpc) &
kT$_X$ (eV) &
(\kms) &
(Mpc) 
\\

\hline

Abell 370 &
0.373 \ &
2000 &
$7.13 \pm 1.05$ &
$1263^{+99}_{-99}$ &
$2.57^{+0.20}_{-0.20}$ 
\\

Coma cluster &   
0.023 \ &
\ \ 96.0 &
$8.38 \pm 0.34$ &  
$1010^{+51}_{-49}$ &
$2.47^{+0.12}_{-0.12}$
\\

Abell 1367 &     
0.022 \ & 
\ \ 91.3 &   
$3.50 \pm 0.18$ &    
\ $822^{+69}_{-55}$ &
$2.01^{+0.17}_{-0.13}$ 
\\

Virgo cluster &   
0.0036 & 
\ \ 17.0 &
$2.20 \pm 0.69$ &    
\ $673^{+48}_{-40}$ &
$1.66^{+0.12}_{-0.10}$ 
\\

\hline 

\end{tabular}


\caption{The cluster properties of Abell~370 and some nearby clusters.  The cluster velocity dispersion of Abell~370 is measured from the redshifts obtain in this work \citep{pracy08}.  The luminosity distance for Abell~370 and Abell~1367 has been calculated from their redshift.  For the Coma and Virgo clusters the luminosity distance is the value found in the GOLDMine database \citep{gavazzi03}.  The X-ray temperature data and other velocity dispersion measurements are from the compilation by \citet{wu99}. The \Rtwo\ values are calculated from the velocity dispersions.} 

\label{cluster_properties}  

\end{centering}
\end{table*}


Abell~370 is a large galaxy cluster with a velocity dispersion of $1263 \pm 99$~\kms.  In the nearby universe, there are few galaxy clusters of this size, which limits the available literature \HI\ 21-cm emission observations for direct comparison.  Table~\ref{cluster_properties} lists the cluster properties of Abell~370 and three nearby clusters that have extensive literature \HI\ observations.  Abell~370 and the nearby Coma cluster have similar cluster velocity dispersions and similar X-ray gas temperatures for the hot intracluster gas in their cores.  This indicates that the two clusters have similar total masses, assuming that they are both dynamically relaxed.  Coma and Abell~370 are also similar in that they both have two cD galaxies rather than the usual one seen in rich clusters.   Abell~370 is substantially more massive than the two other clusters listed in the table, Abell~1367 (also known as the Leo cluster) and the Virgo cluster.  Detailed comparisons of the \HI\ gas content in Abell~370 and these clusters are discussed below.


\subsection{\HI\ density}
\label{HI_density_section}


\begin{table*} 

\centering

\begin{centering}

\begin{tabular}[b]{ll@{}lrr}  

\hline 

\  &                
\multicolumn{2}{c}{\HI\ Gas Density} &     
Volume  &
Number of \\

\ \ \ \ Galaxy Sample  &                
\multicolumn{2}{c}{(10$^9$ \Msun\,Mpc$^{-3}$)} &     
\ \ (Mpc$^3$) &
Galaxies \ \\

\hline

\ \ \ \ {\it Abell 370 Samples} & \ & \  & \ & \ \\

All Galaxies &
0.034 & 
$\pm$ 0.018 & 
62300 &
324 \ \ \ \\

Blue Galaxies &
0.032 & 
$\pm$ 0.011 & 
62300 &
105 \ \ \ \\

[OII] Emission Galaxies &
0.031 & 
$\pm$ 0.014 & 
62300 &
168 \ \ \ \\

Red Galaxies &
0.0098 & 
$\pm$ 0.0056 & 
62300 &
219 \ \ \ \\

\ & \ & \  & \ & \ \\

All Galaxies - {\it Extrapolated}  $(\times 2.18 \pm 0.58)$ &
0.075 & 
$\pm$ 0.044 & 
62300 &
324+ \ \ \ \\

Blue galaxies - {\it Extrapolated} $(\times 2.13 \pm 0.61)$ &
0.068 & 
$\pm$ 0.030 & 
62300 &
105+ \ \ \ \\

\ & \ & \  & \ & \ \\

Inner Galaxies &
5.6 & 
$\pm$ 2.6 & 
71 &
110 \ \ \ \\

\ & \ & \ & \ & \ \\

Galaxies within 8 Mpc &
0.53 & 
$\pm$ 0.38 & 
2140 &
220 \ \ \ \\

Blue Galaxies within 8 Mpc &
0.47 & 
$\pm$ 0.20 & 
2140 &
58 \ \ \ \\

\ & \ & \  & \ & \ \\

\ \ \ \ {\it Literature Samples} & \ & \  & \ & \ \\

Cosmic Density z = 0: \HI\ 21-cm \citep{zwaan05} &
0.0510 \ &
$\pm$ 0.0083 &
- &
4315 \ \ \ \\

Cosmic Density z = 0.24: \HI\ 21-cm \citep{lah07} &
0.095 &
$\pm$ 0.044 &
- &
121 \ \ \ \\


Cosmic Density z \around\ 0.6: DLAs \citep{rao06} &
0.100 &
$\pm$ 0.037 &
- &
18 \ \ \ \\

Cosmic Density z \around\ 3.7: DLAs \citep{prochaska05} &
0.111 &
$\pm$ 0.018 &
- &
89 \ \ \ \\

\ & \ & \  & \ & \ \\

Virgo cluster within 2.5 Mpc (GOLDMine) &
2.23 &
\ \ \ \ \ - &
65 &
252 \ \ \ \\

Abell 1367 within 2.5 Mpc \citep{cortese08} &
1.066 & 
$\pm$ 0.019 &
65 &
36 \ \ \ \\

Coma cluster within 2.5 Mpc (GOLDMine) &
0.74 &
\ \ \ \ \ - &
65 &
22 \ \ \ \\

\ & \ & \  & \ & \ \\

Coma cluster within 8 Mpc (GOLDMine) &
0.066 &
\ \ \ \ \ - &
2140 &
42 \ \ \ \\

\hline 

\end{tabular}

\caption{This table lists the \HI\ gas density measured in different subsamples of galaxies around Abell~370, including extrapolations to the total \HI\ density.   Included in the table are also a number of literature values for the cosmic \HI\ gas density at a few different redshifts and the \HI\ gas density found in nearby galaxy clusters.  The volume for which the \HI\ density is measured is listed where relevant.} 

\label{table_HI_density}  

\end{centering}
\end{table*}


 
\begin{figure}  

  \begin{center}  
  \leavevmode  
	


   \includegraphics[height=8cm, bb = 18 174 592 718, angle=270]{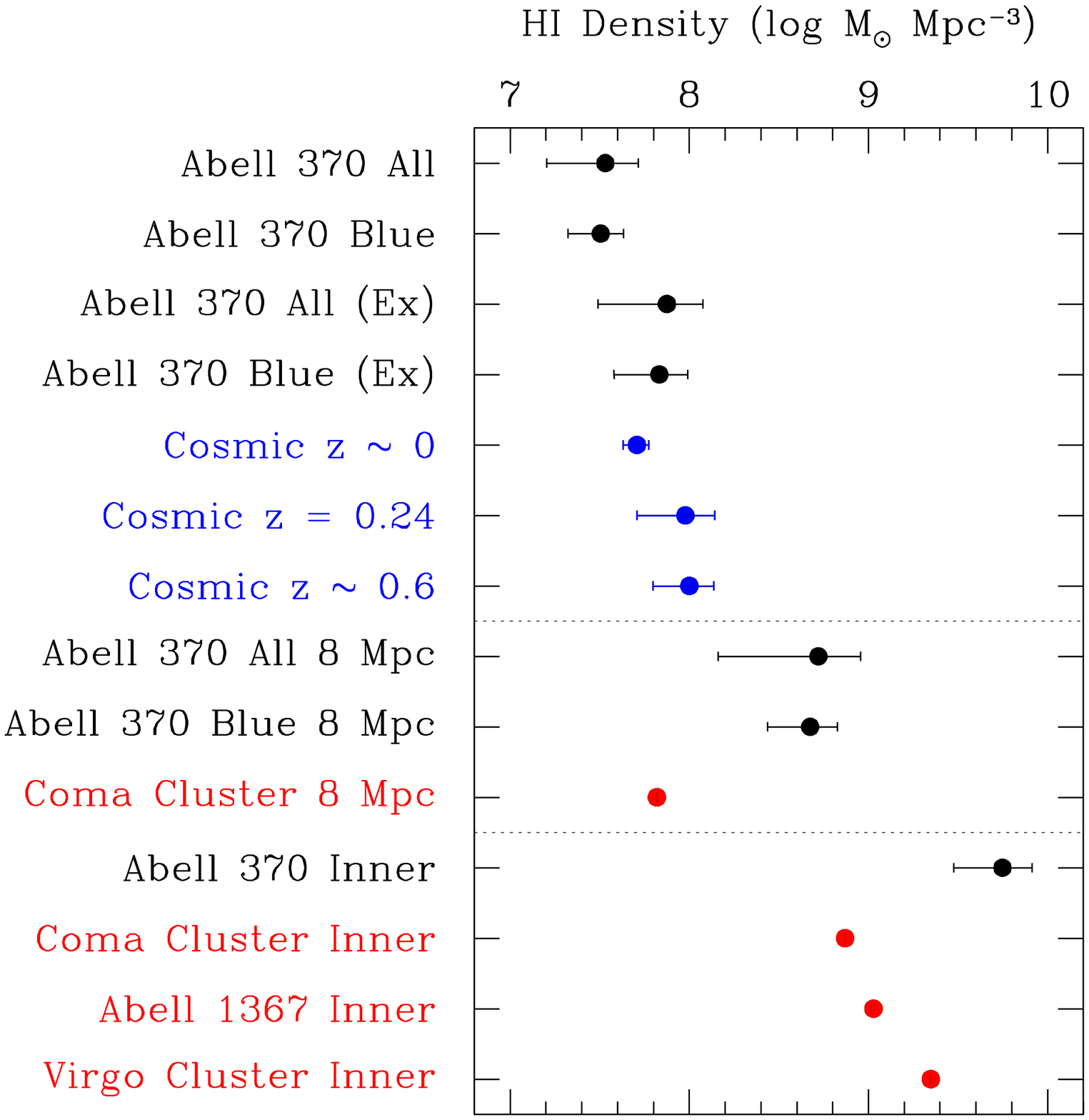}

   \end{center}

   \caption{This figure shows the measured \HI\ gas density around Abell~370 for various subsamples and that found in a variety of literature samples.  The figure is divided into three parts by faint dashed lines.  The top part shows measurements made from the total volume sampled around Abell~370 in this work.  It includes the \HI\ density measured from just the known Abell~370 galaxies as well as extrapolations to the total \HI\ density in the volume (the `Ex' values).  Also shown in the top part are cosmic \HI\ density values at variety of redshifts.  The middle part shows the \HI~density within 8~Mpc of the cluster centre of Abell~370 and a similar sized volume around the nearby Coma cluster.  The bottom part shows the \HI\ density within \around 2.5~Mpc of the cluster centre of Abell~370 and in similar volumes within nearby galaxy clusters.}

   \label{HI_density}

\end{figure}


It is necessary to compare the \HI\ measurements of the galaxies in Abell~370 with local samples in order to quantify any evolution in the \HI\ gas over the past \around 4~billion years (since z~=~0.37).  One way to quantify the gas evolution in Abell~370 is to compare the \HI\ density around the cluster with values from the literature.  The \HI\ densities calculated for a variety of subsamples and volumes around Abell~370 along with various literature values can be found in Table~\ref{table_HI_density} and can be seen plotted in Fig.~\ref{HI_density}.  The \HI\ density in a volume can be calculated using:

\begin{equation}
  \rm \rho\,_{HI} = \frac{ n_{gal} \ \overline{M}_{HI}}{ V },
\end{equation}

\noindent where $\rm \rho\,_{HI}$ is the \HI\ density,  $\rm n_{gal}$ is the number of galaxies being considered, $\rm \overline{M}_{HI}$ is the average \HI\ mass measured for these galaxies, and V is the volume in which these galaxies are contained.  

The comoving volume containing all 324 Abell~370 galaxies was calculated from the extent of the optical imaging (see Fig.~\ref{exclusion_WFI_galaxies_all}) and the redshift range spanned by the galaxies.  The total area on the sky of the optical imaging after accounting for the removal of the exclusion regions (see Fig.~\ref{exclusion_WFI_galaxies_all}) is 0.776~deg$^2$.  Not all this area has uniform optical sampling which may cause a slight underestimate of the \HI\ densities measured.  Taking into account the limits placed by the GMRT 10~per~cent beam level, the area on the sky containing the galaxies with \HI\ measurements is 0.668 deg$^2$.  The \HI\ frequency range of the GMRT observations spans from z~=~0.345 to 0.387.  Using these values, the total comoving volume of the Abell~370 \HI\ measurements is 62300~Mpc$^3$. The volume is much longer in the redshift direction than in projected distance on the sky (at most \around 15~Mpc across compared to 148~Mpc deep).  

Using this volume, a \HI\ density of $(0.034 \pm 0.018) \times 10^9$ \Msun\,Mpc$^{-3}$ is found using all 324 Abell~370 galaxies.  The \HI\ density in this volume due to just the 105 blue galaxies is $(0.032 \pm 0.011) \times 10^9$ \Msun\,Mpc$^{-3}$.  This almost equals the density calculated for all the galaxies, indicating that most of the \HI\ gas around the cluster is located in the blue galaxies.  In contrast, the \HI\ density due to just the 219 red galaxies is $(0.0098 \pm 0.0056) \times 10^9$ \Msun\,Mpc$^{-3}$.  The blue and red galaxy values do not sum exactly to give the \HI\ density from all galaxies due to the weighting schemes used when measuring the \HI\ gas.  The \HI\ density measured using only the 168 [OII] emission galaxies is $(0.031 \pm 0.014) \times 10^9$ \Msun\,Mpc$^{-3}$.  This is similar to that found for the blue galaxies, which shows that this sample also selects the majority of the \HI\ gas in galaxies within this volume.

The \HI\ density measurements listed above only include the \HI\ gas contained in the known Abell~370 galaxies; they do not take into account the many `missing', optically fainter galaxies in the volume that may contain \HI\ gas.  As such these measurement are only lower limits on the total \HI\ density in the volume.  These previous measured values have been scaled up to an estimate of the total \HI\ density in the volume.  This is done by assuming that the optical luminosity density of the galaxies is proportional to their \HI\ density.  The optical luminosity density in the $B$~band for the galaxies in each Abell~370 galaxy sample was measured.  For each of the galaxy samples a Schechter function fit using $\chi^2$ minimisation was made to their magnitude distribution.  In this function fitting, the faint end slope of the luminosity function, $\alpha$, was set at $1.35 \pm 0.10$ because the Abell~370 $B$~band magnitudes do not extend faint enough to allow an accurate determination of this parameter.  This $\alpha$ value is based on literature results for \HI\ galaxies in HIPASS \citep{zwaan05} and optical cluster galaxies in the 2dFGRS \citep{depropris03}.  The fitted Schechter functions were integrated over all magnitudes to create an estimate of the total optical luminosity density of the galaxies.  The \HI\ density for each galaxy sample was then scaled up by the ratio of this total optical luminosity density to the optical luminosity density measured from just the galaxies in the sample.  These total \HI\ density estimates are the extrapolated values listed in Table~\ref{table_HI_density} and are shown in Fig.~\ref{HI_density}.  The value of this ratio is listed in brackets in the table, next to the sample name.

Also measured was the \HI\ density for the 110 inner galaxies in Abell~370, i.e.~the \HI\ density within the cluster core.  When calculating the inner density the average \HI\ mass from the unsmoothed measurement was used because this has the highest precision and appears to contain the total \HI\ signal for these galaxies (see Section \ref{The_inner_and_outer_cluster_subsamples}).  The galaxies in this inner subsample span a projected distance on the sky of \Rtwo = 2.57~Mpc from the cluster centre and a redshift range of z~=~0.357 to 0.387, a cosmological distance of 106~Mpc.  However, these galaxies close to the cluster core have large peculiar motions.  As such the majority of these galaxies are likely to span a much smaller distance in the redshift direction than that indicated by this direct cosmological distance conversion.  A reasonable assumed distance that the majority of the inner galaxies would span in the redshift direction would be 2.57~Mpc, i.e.~similiar to the galaxies projected distance on the sky.  This is the physical volume spanned by the galaxies which should remain unchanged with time in this gravitationally bound region.  Using this value, the volume probed is 71~Mpc$^3$, which makes the \HI\ density in this region $(5.6 \pm 2.6) \times 10^9$ \Msun\,Mpc$^{-3}$.  

The \HI\ density measured for the inner subsample is substantially greater (more than 50 times higher) than even the extrapolated \HI\ densities measured for the other galaxy samples in Abell~370.  Even though the galaxies in the cluster core may have lower \HI\ gas content than similar galaxies in the field, they are packed into a very small volume, dramatically raising the \HI\ density measured there.  Galaxies in field environments may have more \HI\ gas per galaxy but they are spread over larger volumes reducing the \HI\ density found there.

 
\begin{figure}  

  \begin{center}  
  \leavevmode  
		
    \includegraphics[width=8cm]{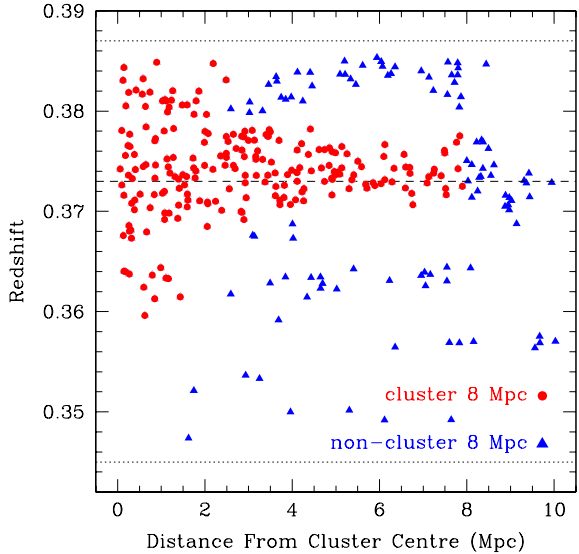}
  
   \end{center}

   \caption{This figure shows the redshift vs.\ projected distance from the cluster centre for all 324 galaxies used in the \HI\ coadding.  The circular points are those galaxies expected to lie within 8~Mpc of the cluster centre both in projected distance and in the redshift direction after taking into account the galaxies peculiar motion.  The triangular points are those galaxies expected to lie further away than 8~Mpc.  The two dotted lines are the frequency limits of the GMRT and the dashed line is the redshift of the cluster centre.}

   \label{radius_vs_redshift}

\end{figure}


In order to make a comparison with the nearby galaxy cluster of Coma, a sample of galaxies within 8~Mpc of the cluster centre of Abell~370 was selected.  Fig.~\ref{radius_vs_redshift} shows the redshifts of the Abell~370 galaxies plotted against their projected distance from the cluster centre.  {Galaxies close to the cluster core have larger peculiar velocities than those further out.  Using this fact, an envelop was drawn on the figure consisting of smoothly varying curves that is the locus boundary for those galaxies likely to lie within 8~Mpc of the cluster centre; these are the 220 circular points in Fig.~\ref{radius_vs_redshift}.}  These galaxies are assumed to all lie within the  8~Mpc radius sphere centred on the cluster core with physical volume 2140~Mpc$^3$ (again the volume spanned by these galaxies should remain relatively unchanged with time in this gravitationally bound region).  The average \HI\ mass measured for these galaxies is listed in Table~\ref{HI_mass_measurments}, as is the values for the subsample of 58 blue galaxies within this selection.  Using the large smoothing average \HI\ mass measurement of all 220 galaxies, a \HI\ density of $(0.53 \pm 0.38) \times 10^9$~\Msun\,Mpc$^{-3}$ is found (this is the value plotted in Fig.~\ref{HI_density} and listed in Table~\ref{table_HI_density}).  Using the unsmoothed average \HI\ mass measurement, a \HI\ density of $(0.37 \pm 0.14) \times 10^9$~\Msun\,Mpc$^{-3}$ is found, which has substantially higher signal to noise.   If one considers only the 58 blue galaxies in this region, using their average \HI\ mass large smoothing measurement, a \HI\ density of $(0.47 \pm 0.20) \times 10^9$~\Msun\,Mpc$^{-3}$ is found.  This is comparable to that measured for all 220 galaxies in the selection 8~Mpc radius region which suggests that again it is the blue galaxies that contain the majority of the \HI\ gas within this volume.

The first set of literature values listed in Table~\ref{table_HI_density} are \HI\ gas cosmic density values at a variety of redshifts.   These values were all converted to the \HI\ densities; some were published as neutral gas densities and included a correction for the neutral helium content.  The first value listed is the z~=~0 cosmic \HI\ density as measured in the HIPASS survey \citep{zwaan05} using \HI\ 21-cm emission from a large sample of galaxies across the entire southern sky.  The second value is the \HI\ density measured in a sample of star-forming galaxies at z~=~0.24 using coadded \HI\ 21-cm emission \citep{lah07}.  The other two values are from damped \Lya\ measurements, looking at the \HI\ absorption in quasar spectra in the rest-frame ultraviolet.   The lower redshift value is from damped \Lya\ absorbers at redshifts z~\around~0.6 (redshift range z~=~0.1 to 0.9) that have been optically selected to have Mg{\sc II} absorption, before being followed up in the ultraviolet with the {\it Hubble Space Telescope} \citep{rao06}.  The z~\around~3.7 (redshift range z~=~3.5 to 4.0) value has been measured from optical spectra of quasars primarily from the Sloan Digital Sky Survey (the \Lya\ absorption is redshifted into the optical at these redshifts) \citep{prochaska05}. 

The listed cosmic \HI\ density values beyond redshift z~=~0 are similar.  They are all around $0.10 \times 10^9$ \Msun\,Mpc$^{-3}$, which is twice the \HI\ density at z~=~0.  The z~=~0.24 and z~\around~0.6 value are the cosmic density measurements which are closest in redshift to Abell~370 at z~=~0.37.   They are plotted with the z~=~0 in the top part of Fig.~\ref{HI_density}.  The z~\around~3.7 value is the highest \HI\ cosmic density value as currently measured.  The \HI\ density measured in the larger volume samples of Abell~370 galaxies are slightly less than the \HI\ cosmic density found at z~=~0  (see the top part of Fig.~\ref{HI_density}).  However the extrapolated \HI\ density values are comparable to the \HI\ cosmic density found at z~=~0.24 and z~\around~0.6.  The errors in these measurements make it difficult to determine if there has been any substantial evolution in the \HI\ gas in galaxies from these values.  Indeed it is probably not fair to compare the \HI\ density found in this large volume around Abell~370 to the cosmic density as we are dealing with an unusual volume of the universe with a considerably higher galaxy density than the average.  Additionally the extrapolation used to scale up the \HI\ gas density is highly uncertain particularly in the inner regions of the cluster where the high galaxy density would probably effect the smaller galaxies gas content more significantly than the larger galaxies.  To do a fair test for evolution it is necessary to compare the measured \HI\ density around Abell~370 to nearby volumes with similarly high galaxy densities.

The second set of literature values listed in Table~\ref{table_HI_density} are derived from \HI\ density values for three nearby galaxy clusters.  The first value is for the Virgo cluster which has \HI\ observations from \citet{gavazzi05} which are available from the GOLDMine database \citep{gavazzi03}.  These include targeted observations of all late--type galaxies with $\rm m_{p} \le 18.0$ magnitude, which at the distance of the cluster is more than 6 magnitudes fainter than the Abell~370 observations.  There are 252 galaxies with \HI\ measurements within a radius 2.5~Mpc in the inner regions of the Virgo cluster.  For this volume, a \HI\ density of $2.23 \times 10^9$ \Msun\,Mpc$^{-3}$ was measured.  This measurement is a lower limit as it does not include a correction for any missing galaxies.  However it is likely to be close to the total \HI\ density as it contains the majority of the galaxies with significant \HI\ gas content.  

The second galaxy cluster considered is Abell~1367 which has \HI\ observations from Arecibo Galaxy Environment Survey (AGES), a blind \HI\ survey \citep{cortese08}.  Using this data, the \HI\ density within a radius 2.5~Mpc around the cluster centre was measured.  Galaxies were included in this volume if they lay with the 2.5~Mpc projected distance of the sky of the cluster centre and had redshifts between 4000~\kms\ and 9000~\kms.  There are 36 galaxies with \HI\ measurements not contaminated by RFI in this region which gives a \HI\ density of $(1.066 \pm 0.0192) \times 10^9$~\Msun\,Mpc$^{-3}$. 

The third galaxy cluster considered is the Coma cluster.  The \HI\ observations came from \citet{gavazzi06} and are compiled with optical data in the GOLDMine database \citep{gavazzi03}.  The \HI\ observations include 94~per~cent of all late--type galaxies with apparent magnitude $\rm m_{p} \le 15.7$ mag in the Coma supercluster (a much larger region surrounding the Coma cluster).  This magnitude limit is equivalent to $B$~band absolute magnitude of \around $-19.2$ at the distance of Coma cluster (the faintest Abell~370 galaxy is -19.7).  The \HI\ density was measured around the Coma cluster centre out to a radius of 2.5~Mpc and 8~Mpc.  A radius of 8~Mpc is the maximum projected distance from the Coma cluster with good observations (there were insufficient \HI\ observations to make similar 8~Mpc measurements for Abell~1367 and the Virgo cluster).  The inner 2.5~Mpc radius region of the Coma cluster has a \HI\ density of $0.74 \times 10^9$ \Msun\,Mpc$^{-3}$ from 22 \HI\ galaxies.  The larger 8~Mpc radius region of the Coma cluster has a \HI\ density of $0.066 \times 10^9$ \Msun\,Mpc$^{-3}$ from 42 \HI\ galaxies.  

The inner values for Abell~370 and the three literature clusters are shown together in the bottom part of Fig.~\ref{HI_density}.  These inner samples all come from similar sized volumes, spheres with radii \around 2.5~Mpc.  The \Rtwo\ radius for Coma and Abell~370 are close to this value but Abell~1367 has an \Rtwo\ radii of $2.01^{+0.17}_{-0.13}$~Mpc and the Virgo cluster of $1.66^{+0.12}_{-0.10}$~Mpc (see Table~\ref{cluster_properties}).  The \HI\ density within the \Rtwo\ radii for these two smaller clusters is almost twice as high in both cases.  As can be seen in bottom part of Fig.~\ref{HI_density} in the nearby cluster values, the \HI\ density increases with decreasing cluster size.  The \HI\ density in the inner Coma region is 3~times smaller than that found in the smaller irregular Virgo cluster, with the \HI\ density of Abell~1367 in between.  This clearly shows the known trend in the nearby universe, that \HI\ gas in galaxies is lower in high galaxy density environments.  The \HI\ density found for Abell~370 is \around 3 times higher than that found in Virgo and \around 8~times higher than that found in Coma, a similar sized cluster.  This is not evidence against the trend of \HI\ gas in galaxies being lower in high galaxy density environments.   This trend is seen for the Abell~370 galaxies when comparing the inner and outer subsamples (see Section~\ref{The_inner_and_outer_cluster_subsamples}).  Rather this high \HI\ density value found in the inner regions of Abell~370 compared to nearby clusters is an indication that there has been substantial evolution in the gas content of galaxies in clusters over the last \around 4~billion years since z~=~0.37.

Similarly, the region within 8~Mpc of the centre of Abell~370 has a \HI\ density that is \around 8~times higher than the similar size region around Coma.  This is a statistically significant difference as can be seen in the middle part of Fig.~\ref{HI_density} (the Coma value has no discernible random error).  As Coma and Abell~370 are galaxy clusters of similar size, this is again evidence of substantial evolution in the gas content of cluster galaxies between redshift z~=~0.37 and the present. 

In order to ensure that the striking differences between the \HI\ densities found for the Coma cluster and Abell~370 are real one must ensure that there is no significant biases on the measurements that could be distorting the result.  The Coma \HI\ gas measurements come from targeted observations rather than a blind \HI\ search of the cluster.  As such it is not impossible that an appreciable fraction of the \HI\ gas in galaxies within the cluster has been missed.  This is unlikely though, as all the optically bright late-type galaxies have been observed and these are likely to be the dominate contributors to the \HI\ density.  Even if including the missing low \HI\ mass galaxies raises the \HI\ density in Coma by a factor of two there is still considerably more gas found around Abell~370.  The unaccounted for gas due to these smaller galaxies is unlikely to be anywhere near as substantially as this particular as these galaxies with lower total galaxy masses will have a harder time holding onto their gas against the high density environmental mechanisms such as ram pressure stripping.  Additionally it is likely that appreciable amounts of the \HI~gas is missing in the Abell~370 volumes due to the relatively bright optical magnitude limit of this galaxy sample.  

The optical imaging of Abell~370 does not extend fully out to a projected radius of 8~Mpc in all directions as seen in Fig.~\ref{A370_radius_Mpc}.  This will result in missing some galaxies within this volume causing a small underestimation of the \HI\ density in this volume.  To select the Coma and Virgo galaxies the distances in the GOLDMine database were used.  These distances have had the effect of the peculiar velocities of the galaxies removed, allowing one to select the galaxies close to the cluster centres.   If these distances are incorrect for a large number of galaxies, then this could result in an underestimation of the true \HI\ densities in these clusters.  However, this is unlikely as the distances appear quite reasonable based on their redshift and spatial distribution on the sky.  The literature values for the clusters included \HI\ flux upper limits for a number of galaxies.  These galaxies were considered to have no gas when doing the density calculations.  They are unlikely to contain sufficient gas to significantly effect the results.  

Ideally one would want deep \HI\ blind observations of a number of large galaxy cluster at low redshift to compare with the Abell~370 observations.  Such published data do not exist as yet.  Due to the uncertainties on the current galaxy samples of both Abell~370 and the low redshift clusters it is difficult to say with precision the amount of difference in the \HI\ densities.  However, from the measurements it is clear that Abell~370 has considerably more gas than local clusters suggesting there has been substantial evolution in the gas content in clusters over the last 4~billion years. 

The \HI\ density of Abell~370 is markedly larger than that found in the Coma cluster in the regions considered.   The \HI\ gas density in Abell~370 is up to 8 times higher than in Coma.  The increase in the cosmic \HI\ density from z~=~0 to the largest known values at higher redshifts is at most a factor of two (see Table~\ref{table_HI_density}).  If Abell~370 were to evolve into a gas poor system like Coma in \around 4~billion years, then the rate of decrease in the gas would be considerably faster than the rate of decrease seen in the field.  {This higher rate of decrease in gas content could be caused by the combination of the higher rate of galaxy-galaxy interactions in the cluster environment and the interactions between the inter-stellar medium (ISM) of the galaxies with the inter-galactic medium (IGM) of the cluster which is denser than the IGM of the field.}


\subsection{Average \HI\ mass comparisons with the Coma cluster}

\label{Average_HI_mass_comparisons_with_the_Coma_cluster}

Comparing the \HI\ measurements of the galaxies in Abell~370 directly with local samples is difficult, primarily due to the unusual way in which the Abell~370 measurements were made, i.e.~by coadding the \HI\ 21-cm emission signal from multiple galaxies.  One way of doing this comparison is to make similar coadded average \HI\ mass measurements in nearby clusters using literature \HI\ 21-cm emission values.  To do this one needs a complete sample of the optical galaxies in a nearby cluster down to the magnitude limits of the Abell~370 galaxy sample.  Additionally one needs to know the \HI\ mass for each of these optical galaxies.  Each galaxy in the Abell~370 sample can then be randomly matched to a literature galaxy in the nearby cluster with similar absolute optical magnitude.  The average \HI\ mass of this randomly matched sample of literature galaxies can then be calculated.  This can be repeated multiple times, each time taking different randomly matched samples.  The different random samples are then combined to give a robust measure of the average \HI\ mass for a sample of galaxies with similar optical magnitude distributions to the Abell~370 galaxies.  The variation between the different random samplings provides an estimate on the error in this measurement.

This comparison was done using galaxies in the nearby Coma cluster as the GOLDMine database contains both optical and \HI\ observations for the galaxies \citep{gavazzi03} (See above for details on this galaxy sample).  Each galaxy in the Abell~370 sample was matched to a random Coma galaxy with $B$~band absolute magnitude within 0.25~mag.

The Coma data was used to make average \HI\ mass measurements to compare with the Abell~370 measurements for all 324 galaxies, for the inner subsample and for the outer subsample.  Galaxies that lay within 8.0~Mpc of the cluster centre of Coma were chosen to match against all 324 Abell~370 galaxies.   In this Coma cluster sample there were 113 galaxies within the magnitude range spanned by the 324 Abell~370 galaxies.  Of these 24~per~cent had detected \HI\ masses.  The galaxies within a radius of 2.5~Mpc of the cluster centre of Coma were matched with the inner subsample of the Abell~370 galaxies.  In this Coma inner sample there were 67 galaxies within the magnitude range of the galaxies in the Abell~370 inner sample.  Of these 19~per~cent had measured \HI\ masses.  The galaxies more than 2.5~Mpc from the cluster centre of Coma but less than 8.0~Mpc from the centre were chosen to compare with the Abell~370 outer subsample. In this Coma outer sample there are 37 galaxies within the magnitude range of the galaxies in the Abell~370 outer sample.  Of these 35~per~cent had measured \HI\ masses.  The number of Coma galaxies in each subsamples is smaller than the number of Abell~370 galaxies.  As such the matched samples will contain repeats of the Coma galaxies which may introduce a bias.  However a larger sample of nearby cluster galaxies will likely have a similar distribution so that the effect of any bias will likely be small. 


 
\begin{figure}  

  \begin{center}  
  \leavevmode  
	
   


   \includegraphics[height=8cm, bb= 18 174 392 718, angle=270]{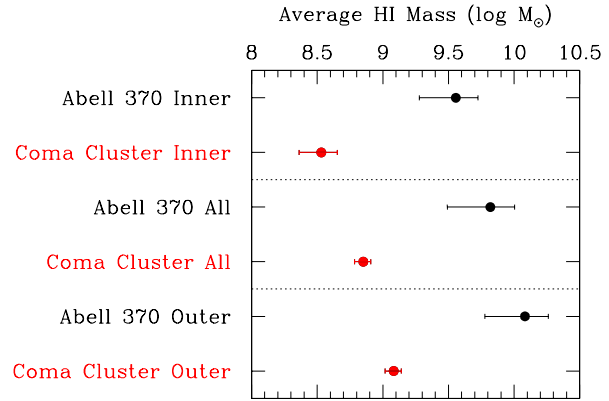}
 
   \end{center}

   \caption{This figure shows the average \HI\ mass for selected galaxy subsamples from Abell~370 and similar measurements made using literature values for the Coma cluster.  The Coma average \HI\ mass measurements are made from randomly selected samples of galaxies with similar $B$~band absolute magnitudes to the Abell~370 galaxies (see text for details).}

   \label{HI_comparison}

\end{figure}


The measured average \HI\ mass for the Abell~370 subsamples and the similar Coma measurements can be seen in Fig.~\ref{HI_comparison}.  There is a similar trend in both Coma and Abell~370, that the galaxies in the outer regions of both clusters have average \HI\ masses that are \around 3.5 times higher than that found in their inner galaxies.  This is the well known trend, that galaxies within dense cluster cores generally have less \HI\ gas content than galaxies in lower density environments \citep{haynes84}.  This is true in nearby clusters and appears to be true at z~=~0.37, \around 4~billion years in the past.  The similar ratio between the inner and outer measurements in the two clusters could suggest that the mechanism for creating this \HI\ gas reduction is of similar strength in both clusters.

Despite this trend with galaxy density seen in both clusters, the amount of \HI\ gas content of the galaxies is substantially different between the two clusters.  In all three measurements the average \HI\ mass in the Abell~370 galaxies is \around 10~times larger than that found in the Coma samples of optical galaxies with similar magnitudes.  This higher \HI\ gas content is likely due to the optically bright, blue galaxies that exist around Abell~370 that have been shown to have large quantities of \HI\ gas.  Similar galaxies do not exist in Coma.  This is seen in the difference in Butcher--Oemler blue fraction between the clusters; Abell~370 has a blue fraction of \around 0.13 while Coma has a blue fraction of \around 0.03 \citep{butcher84}, a factor of \around 4 less.  The blue galaxies in Coma have had an extra \around 4~billion years of evolution to remove their \HI\ gas through star formation or interactions in the dense galaxy environment.  The young, blue stars in such galaxies would have died out in timescales less than \around 4~billion years.  Without more \HI\ gas to supply the fuel for additional star formation, the galaxies would have dimmed and evolved to a redder colour.  The passive evolution of the galaxies in Abell~370 over \around 4~billion years would decrease their $B$~band magnitudes by up to 1~magnitude \citep{poggianti97}.  Decreasing the optical brightness of the galaxies in Abell~370 by \around 1~magnitude creates a reasonable match to the $B$~band magnitude distribution of the Coma galaxies. 

Similar average \HI\ mass comparison with either Abell~1367 or the Virgo cluster are not practical as there are insufficient galaxies optically bright enough to match to the Abell~370 galaxies. 


\subsection{\HI\ mass to light ratios}
\label{HI_mass_to_light_ratio_section}

 
\begin{figure}  

  \begin{center}  
  \leavevmode  
	
   

   \includegraphics[height=8cm, bb = 18 174 592 718, angle=270]{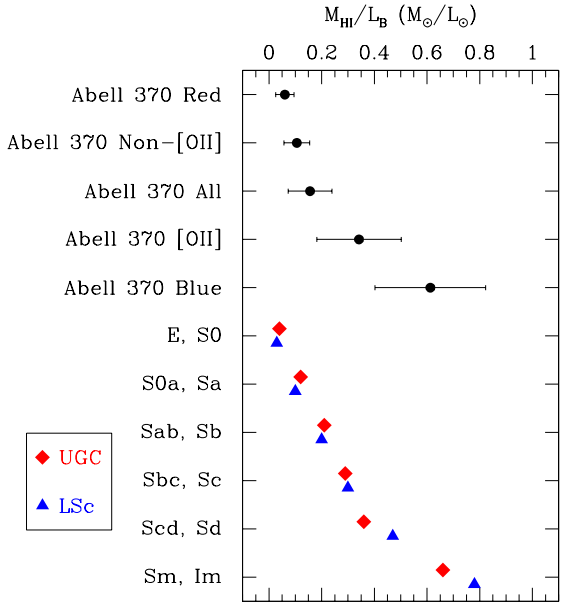}
 
   \end{center}

   \caption{This figure shows the average ratio of \HI\ mass to optical $B$~band light ratio for various subsamples of the Abell~370 galaxies.  Also plotted for comparison are the median \HI\ mass to light ratios for different morphological type galaxies from the Uppsala General Catalogue (UGC) and the Local Super Cluster (LSc) \citep{roberts94}.}

   \label{HI_mass_to_light}

\end{figure}


In the previous literature comparisons it has been shown that the Abell~370 galaxies have higher \HI\ gas content than galaxies in nearby clusters.  This raises the question whether the Abell~370 galaxies have unusual \HI\ gas properties compared to nearby galaxies.  One way of assessing this is to measure the \HI\ mass to optical light ratios seen for the Abell~370 galaxy subsamples and compare these to `normal' galaxy values.   The average ratio of \HI\ mass to the $B$~band luminosity for a variety of Abell~370 subsamples are shown in Fig.~\ref{HI_mass_to_light}.
  
For each of the Abell~370 galaxy subsamples the average rest frame $B$~band luminosity in units of the solar luminosity was calculated using an absolute $B$~band magnitude for the sun of 5.46 \citep{bessell98}.  When combining the individual galaxy $B$~band luminosities, weights were used equal to those used in the average \HI\ mass measurements to ensure that similar quantities were measured.  The weighted average and the non-weighted average $B$~band luminosities are very similar for the subsamples considered (for all 324 galaxies the difference in values was \around 7~per~cent).  This suggests that the weighting scheme used in \HI\ mass measurements does not create a bias in the results.  Using this average $B$~band luminosity and the average \HI\ mass for a subsample, the \HI\ mass to light ratio can be calculated.

The literature \HI\ mass to blue light ratios considered are for samples of galaxies with different morphological types \citep{roberts94}.  The morphologies move along the Hubble sequence starting with the early-type galaxies of ellipticals and S0s, moving across the variety of late-type galaxies in the direction of increasing spiral structure (moving from Sa to Sd galaxies), and finally reaching the irregular galaxies.  The \HI\ mass to light ratio increases fairly regularly along this sequence.  These literature measurements are the median values from two catalogues, the Uppsala General Catalogue (UGC) and the Local Super Cluster sample (LSc).  

In Fig.~\ref{HI_mass_to_light} the values for the galaxy subsamples of Abell~370 are plotted in order of increasing \HI\ mass to light ratio.   The red galaxy subsample has the lowest \HI\ mass to light ratio while the blue galaxies have the highest.   Below these are plotted the literature values from both the Uppsala General Catalogue (UGC) and the Local Super Cluster sample (LSc).  The red galaxy subsample has a \HI\ mass to light ratio similar to that for the ellipticals and S0 galaxies of the literature samples and the blue galaxy subsample has a ratio similar to those literature galaxies with the most spiral structure or which are irregular.  Unfortunately deriving morphologies for the galaxies in Abell~370 is not possible due to the poor seeing in the optical imaging combined with the small size of galaxies at a redshift of z~=~0.37.  However, it is clear from this comparison that galaxies in Abell~370 seem to follow similar trends in \HI\ mass to light ratios as nearby, `normal' galaxies and even have similar \HI\ mass to light ratios for galaxies of roughly similar types (the red and blue galaxies).   With time, the galaxies in Abell~370 will undergo passive evolution, and their optical brightness will decrease.  In order for the galaxies still to have `normal' \HI\ mass to light ratios their \HI\ gas content will need to decrease similarly during this evolution.  


\section{Star Formation Rate Results}
\label{Star_Formation_Rate_Results}

\subsection{The SFR--\HI\ mass correlation for the galaxies}
\label{The_SFR_HI_mass_galaxy_correlation}

 
\begin{figure}  

  \begin{center}  
  \leavevmode  
		
    \includegraphics[width=8cm]{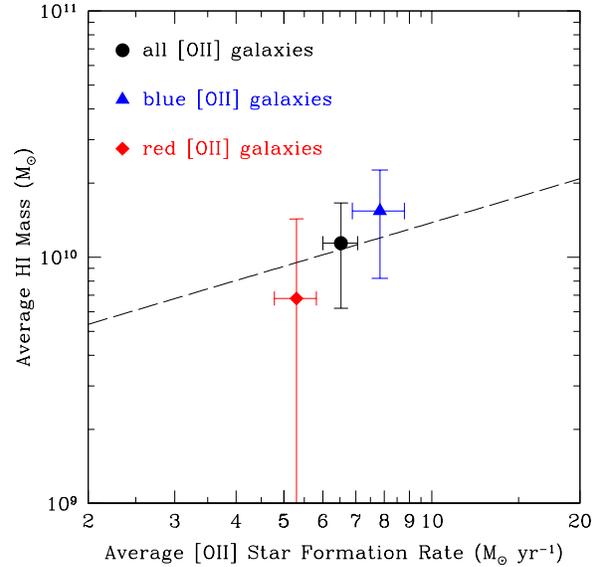}
 
   \end{center}

   \caption{
This figure shows the comparison of the average galaxy [OII] star formation rates with their average \HI\ mass.  The dashed line is the relationship seen in z~\around~0 galaxies \citep{doyle06}.  The large circular point is the average for all 168 galaxies with [OII] emission.  The triangle point is the average for the 81 blue galaxies with [OII] emission.  The diamond point is the average of the 87 red galaxies with [OII] emission. }

   \label{SFR_HI}

\end{figure}


In the local universe, there is a reasonably strong correlation between the star formation rate in a galaxy and the mass of \HI\ gas in that galaxy.  This relationship can be seen in Fig.~3 of \citet{doyle06}, where they compared the \HI\ masses of individual galaxies from HIPASS to their star formation rate derived from {\it IRAS} infrared data.   This correlation between galaxy \HI\ mass and star formation rate can be examined in the galaxies around Abell~370.  The star formation rate for the Abell~370 galaxies was derived from their [OII] luminosity as described in Section~\ref{The_optical_properties_of_the_Abell_370_galaxies}.  In Fig.~\ref{SFR_HI}, the large circular point shows the comparison of the average [OII] star formation rate against the average \HI\ for the Abell~370 galaxies with [OII] equivalent width greater than 5~\AA\ (the 168 galaxies of the [OII] emission subsample).  When calculating the average [OII] star formation rate the same weighting scheme was used as in the average \HI\ mass measurement.  The difference between this weighted average star formation rate and the standard average is small (less than 1~per~cent).

Plotted on Fig.~\ref{SFR_HI} is the linear fit to the SFR--\HI\ correlation for the local sample of galaxies \citep{doyle06}.  The average value for the Abell~370 [OII] emission galaxies lies almost on this line.  This indicates that the galaxies around Abell~370 have normal galaxy properties, in that their higher \HI\ gas contents leads naturally to higher star formation rates.  The same SFR--\HI\ correlation relationship was found to hold at z~=~0.24 in a field sample of star-forming galaxies \citep{lah07}.  These two results at z~=~0.24 and z~=~0.37 suggest that the increase in star formation rate densities seen at moderate redshifts are simply due to higher \HI\ gas content in the galaxies.  The good SFR--\HI\ correlation agreement for the Abell~370 galaxies also indicates that the assumptions and corrections made when calculating the [OII] star formation rate were reasonable.

Also shown in Fig.~\ref{SFR_HI} are the values for the average galaxy [OII] star formation rates and average galaxy \HI\ masses for the 81 galaxies that have [OII] emission and blue colours (the triangular point) as well as for the 87 galaxies that have [OII] emission but have red colours (the diamond point).  Both of these measurements agree with the \citeauthor{doyle06} line.  The expected trend that the blue [OII] galaxies have higher star formation rates and higher \HI\ masses compared to red [OII] galaxies is seen.  

From the measured star formation rate, it would take \around 1.7~billion years for the [OII] emission galaxies to turn all their \HI\ gas into stars.  This assumes: that the conversions from \HI\ gas to stars is 100~per~cent efficient, that there is no change in the star formation rate as the \HI\ gas decreases, that there is no significant amounts of molecular hydrogen gas, that galaxy harassment or \HI\ stripping have minimal effect on the gas, that there is no recycling of gas, and that there is no gas accretion onto the galaxies.  None of these assumptions are reasonable.  However, this rough time frame does show that it is possible for the \HI\ gas in the Abell~370 galaxies to be depleted in the \around 4~billion years from z~=~0.37 to the current epoch.  Abell~370 can easily evolve into a \HI\ gas poor galaxy cluster based entirely on the star formation rate seen in its galaxies.


\subsection{The SFR--radio continuum correlation}
\label{The_SFR_HI_radio_continuum_correlation}

 
\begin{figure}  

  \begin{center}  
  \leavevmode  
		
    \includegraphics[width=8cm]{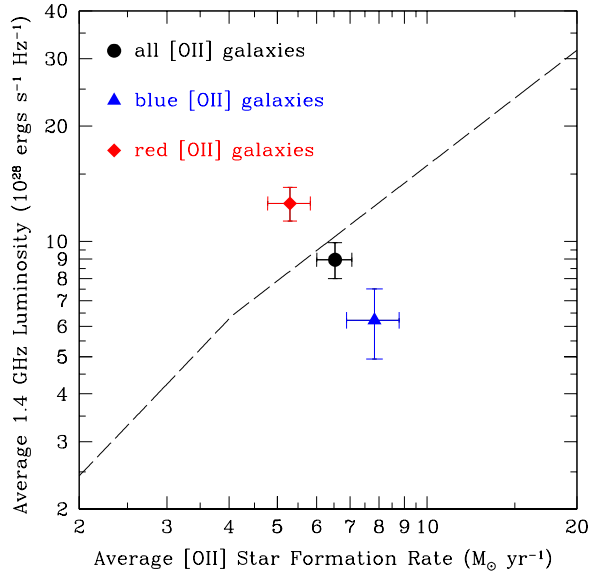}
 
   \end{center}

   \caption{
This figure shows the comparison of the average galaxy star formation rates measured from the [OII] emission line luminosity and the de-redshifted 1.4~GHz radio continuum luminosity.  The dashed line is the conversion from 1.4~GHz radio continuum to star formation rate \citet{bell03}.   The large circular point is the average for all 168 galaxies with [OII] emission.  The triangle point is the average for the 81 blue galaxies with [OII] emission.  The diamond point is the average of the 87 red galaxies with [OII] emission. }

   \label{SFR_rc}

\end{figure}


Synchrotron radio emission is generated in areas of active star formation from relativistic electrons accelerated in supernova remnants.  The luminosity of this radiation has a good correlation with other galaxy star formation rate indicators.  This radiation in nearby galaxies is often measured using 1.4~GHz observations and compared to star formation rates measured using other indicators \citep{sullivan01,bell03}.  The GMRT data also provided measurements of the radio continuum emission at 1040~MHz for objects surrounding Abell~370.  Since the Abell~370 galaxies are at a redshift of z~=~0.37, their de-redshifted radio continuum emission in the GMRT data has a frequency of 1.4~GHz.  It is therefore possible to directly measure the 1.4~GHz radio continuum emission from the Abell~370 galaxies and compare it to their measured [OII] star formation rate to see if the correlations found in the nearby universe hold at z~=~0.37.

Despite the low radio continuum RMS of only 20~\microJy, only a handful of the 168 Abell~370 galaxies with [OII] emission have radio continuum flux densities at the $5 \sigma$ level.  To increase the number of galaxies studied, the signal from all 168 galaxies was coadded using their known optical position to measure their average radio continuum flux density.  This coadding was done using a similar weighted average as used in the \HI\ measurements, to take into account the variation in noise due to the GMRT primary beam shape.  Using this method, the 168 galaxies with [OII] emission have a measured average flux density of $25.7 \pm 2.8$~\microJy.  The coadded radio continuum image appeared to be unresolved with no extended emission.  Therefore the central specific intensity value was used as a measure of the total flux density.  Converting from flux density to luminosity density using the cosmological distance and de-redshifting the radio continuum emission from z~=~0.37, gives an average 1.4~GHz radio continuum luminosity density of $\rm (8.96 \pm 0.96) \times 10^{28} \, ergs \; s^{-1} \, Hz^{-1}$ for the [OII] emission galaxies.  This measurement is the large circular point in Fig.~\ref{SFR_rc}.  

The dashed line in Fig.~\ref{SFR_rc} is the 1.4~GHz radio continuum conversion to star formation rate \citep{bell03}.  There is a change in the slope of this conversion at a star formation rate of \around 3.5 \Msun~yr$^{-1}$.  This is due to galaxies of lower mass not being able to retain all their cosmic rays accelerated in supernovae remnants and so reducing the radio continuum emission produced relative to other star formation indicators.  The value for the [OII] emission subsample lies $\sim 1 \sigma$ below this conversion line, showing reasonable agreement.  The result shown here does not depend on a handful of galaxies with high radio continuum luminosity but is a general result from all the galaxies in the subsample.

There is some intrinsic scatter (i.e.~not purely random error) around the star formation rate correlation between different indicators.  The galaxies studied in the Abell~370 analysis have been selected based on their [OII] equivalent widths.  This will create a selection bias such that we may miss those galaxies that have low [OII] luminosities but higher radio continuum that exist within the general scatter of the correlation.  If these galaxies were included in the sample the average [OII] star formation rate would slightly decrease and average radio continuum luminosity increase.  This would bring the average point in Fig.~\ref{SFR_rc} closer to the line derived by \citet{bell03}.  

Many of the individual galaxies in the [OII] emission subsample will have radio continuum luminosities below the point were the SFR correlation changes slope.  As we do not have individual measures of the radio continuum luminosity for these galaxies, it is not possible to correct for this.  If this could be corrected, the star formation rate determined from the average radio continuum of the galaxies would increase slightly, which will improve the agreement between the [OII] and radio continuum star formation rate.

It has been found that active galactic nuclei (AGN) dominate the radio continuum sources above 1.4~GHz luminosities of $\rm 10^{30}\, ergs \; s^{-1} \, Hz^{-1}$ and star formation below this value \citep{condon02}.  The highest individually measurable radio continuum luminosity for a galaxy in the Abell~370 [OII] emission subsample is $\rm (0.884 \pm 0.080) \times 10^{30} \, ergs \; s^{-1} \, Hz^{-1}$.  This brightest radio continuum source in the [OII] emission subsample falls below this transition value with the majority of galaxies having appreciably lower radio continuum luminosities.  Thus it is likely that the AGN contamination of the sample is small with a minimal effect on the measured average radio continuum luminosity.

The [OII] emission galaxies were split into the 81 blue and 87 red galaxies and the average radio continuum luminosity and average [OII] star formation rates for these subsamples were measured.  In Fig.~\ref{SFR_rc} the average values for the blue [OII] emission subsample is the triangular point and the average values for the red [OII] emission subsample is the diamond point.  An unexpected relationship is seen, with the red galaxies showing a higher radio continuum luminosity compared to the blue galaxies even though the red galaxies have a lower [OII] star formation rate.  It is not known what is causing this relationship.  It is not due to the effect of a small handful of galaxies in the red and blue subsamples; it is a general trend across both subsamples.  No simple systematic effect such as AGN contamination on the radio continuum, or metallicity and/or dust extinction effects on the [OII] star formation rate are able to explain this trend.  Since the [OII] star formation--\HI\ mass correlation seems to agree with expectations (see in Section~\ref{The_SFR_HI_mass_galaxy_correlation}), it is likely that the radio continuum is the cause of this unusual relationship.  

{A working hypothesis to explain this effect is the different timescale responsible for the production of the [OII] and radio continuum emission in galaxies.  The weak radio continuum observed in normal galaxies is emitted by an ensemble of relativistic electrons that are produced in supernova remnants.  Lifetimes of these relativistic particles can exceed the timescale associated with star bursts. The lifetime, $\rm t_{life}$, of synchrotron emitting particles capable of emitting a characteristic frequency can be expressed as:

\begin{equation}
\rm t_{life} \sim 3{\times}10^4 \, B^{-\frac{3}{2}} \, \nu_{c}^{-\frac{1}{2}} \, yrs
\end{equation}

\noindent 
where B is the magnetic field strength of the galaxy (in gauss), $\rm \nu_{c}$ is the characteristic frequency of emission and can be expressed as $\rm \nu_c \sim 4 \times 10^6 \, B \gamma^2$~Hz and $\rm \gamma$ is the electron Lorentz factor \citep{rybicki86}.  For frequencies of 1~GHz, $\rm t_{life} \sim B^{-\frac{3}{2}}$~yrs, which ranges from $3 {\times} 10^7$ to $10^9$ years for the dilute magnetic fields (1 to 10~$\rm \mu$gauss) in an ageing disk galaxy. 

Thus we suggest that red [OII] galaxies are older galaxies coming out of a burst of star formation.  They have some [OII] emission left behind as well as supernovae remnants that have produced a large reservoir of decaying, relativistic electrons in the galaxies.  The blue [OII] galaxies are younger systems, which are only currently seen to be building their stellar populations and their relativistic electron distributions.   The hypothesis is that this difference in relativistic electron distributions between the red [OII] and blue [OII] galaxies is the cause of the difference in their radio continuum measurements.}


\section{Conclusion}
\label{Conclusion}

We have measured the average \HI\ mass for a large sample of galaxies around the galaxy cluster Abell~370 at a redshift of z~=~0.37, a look-back time of \around 4~billion years.  {The average \HI\ mass measured for all 324 galaxies is $(6.6 \pm 3.5) \times 10^9$~\Msun\ while the average \HI\ mass measured for the 105 optically blue galaxies is $(19.0 \pm 6.5) \times 10^9$~\Msun.}  The \HI\ gas content of the galaxies is found to be markedly higher than that found in nearby clusters.  Abell~370 has considerably more \HI\ gas than Coma, a cluster of similar size.  The average galaxy \HI\ mass measurements in Abell~370 are \around 10 times higher than similar measurements made from galaxies in Coma.  The measured \HI\ density around the galaxy cluster Abell~370 is \around 8 higher than in Coma.  These results show there has been substantial evolution in the gas content of clusters over the past \around 4~billion years.

Despite the appreciable \HI\ gas content in the Abell~370 galaxies, there is evidence that environmental effects reduce the gas content of galaxies, similar to what is seen in nearby clusters.  The galaxies in the inner regions of the cluster (within the \Rtwo\ radius) have average \HI\ masses smaller by a factor of \around 3.5 than galaxies outside this region.  The optically blue galaxies outside the hot, intracluster medium of the cluster core have a higher average \HI\ gas mass than found from the complete sample of blue galaxies in Abell~370.  This shows that the late-type galaxies close to the cluster core in Abell~370 are \HI\ deficient like those seen in nearby clusters \citep{haynes84}.

Although the galaxies in Abell~370 have high \HI\ gas contents, they have similar galaxy properties to present day galaxies:  the Abell~370 galaxies have normal \HI\ mass to optical light ratios and have a similar correlation between their star formation rate and \HI\ mass as found in nearby galaxies.  The average star formation rate derived from [OII] emission and from de-redshifted 1.4~GHz radio continuum for the Abell~370 galaxies follows the correlation found in the local universe.   However, there is an unexpected relationship where the red [OII] emission galaxies have a higher average radio continuum luminosity than the blue [OII] emission galaxies despite having a lower average [OII] star formation rate.  This effect is the reverse of what is expected and is currently unexplained.

The data suggests that the red galaxies in Abell~370 may have discernible amounts of \HI\ gas contained within their central regions unlike nearby galaxies.  Additionally the blue galaxy population with no appreciable [OII] emission appear to contain large amounts of \HI\ gas.  Both of these results merit further investigation in future more sensitive observations.

The current rate of star formation in the Abell~370 galaxies can easily exhaust their \HI\ gas in the \around 4~billion years to the present epoch.  Abell~370 seems set to evolve into a gas poor system like nearby galaxy clusters, especially when one considers the other physical mechanisms besides star formation that may reduce the gas content of the galaxies in the dense galaxy environment.  Such a rapid rate of decrease in \HI\ gas in the volume around Abell~370 would be significantly faster than the rate of decrease seen in field environments. 

The final evolved state of the galaxy cluster Abell~370 has not been considered.  This is a complex problem involving looking at the effect of the current measured star formation rates in the galaxies, the effect of passive evolution on the stellar population of the galaxies, the possible growth in the cluster core mass, the motion of the galaxies in the cluster potential and the precise effects of the environment on each galaxy which will change with time as the galaxies move within the cluster. 


One question that has not been answered is why Abell~370 has such a large \HI\ gas content compared to nearby galaxy clusters.  In order to produce the mass in stars seen in galaxies it is necessary for there to be replenishment of the \HI\ gas to fuel sufficient star formation \citep{hopkins08}.  This replenishment must come from ionised hydrogen (\HII\ gas) around the galaxy becoming sufficiently dense and cool to condense into \HI\ gas which can then fuel star formation in a galaxy.  At some point this process stops in clusters so that they evolve into the \HI\ poor systems that are seen today.  There are two possible models to explain this decrease in \HII\ condensation: the `reservoir' model and the `accretion' model.  

In the `reservoir' model there is an envelope of ionised gas surrounding a galaxy that can fall onto the galaxy over time and condense into \HI\ gas.  A galaxy would slowly use up the ionised gas component of this envelope that could condense to \HI\ gas which would eventually halt star formation in that galaxy.  Interactions in higher galaxy density regions would cause galaxies to use up this envelop faster by either driving this gas into the galaxies to condense faster or remove it from the halo of the galaxies entirely (i.e.~strangulation). 

In the `accretion' model ionised gas moves from the inter-galactic medium onto a galaxy where it reaches sufficient density to condense into \HI\ gas.  This accreted ionised gas would not have initially been gravitationally bound to the galaxy unlike the ionised gas in the `reservoir' model.  The expansion of the universe with time would slow the rate at which this inter-galactic ionised gas can fall onto galaxies, reducing the production of \HI\ gas and hence the rate of star formation in the galaxies.   The rate of gas replenishment in this mechanism will be proportional to the density of the inter-galactic material which will be inversely proportional to the physical volume size of an element of comoving volume (the change in the size of universe).  At z~=~0.37, the redshift of Abell~370, a comoving volume element of the universe would have a physical volume that is \around 40~per~cent its current size, a significant difference.  Once the expansion of the universe has caused this gas accretion to drop to a low level, the rate of interactions in high galaxy density environments would quickly use up \HI\ gas in the galaxies creating the \HI\ gas poor clusters we see today.  Galaxies in lower density regions with slow star formation rates would be able to retain their existing \HI\ gas for longer.  

Distinguishing between these two models is not easy as both involve ionised \HII\ gas which is notoriously hard to measure.  The best hope for understanding what is occurring will come from comparing how the star formation rate and \HI\ gas content of galaxies in clusters and in lower density `field' environments change with time. 


\section*{Acknowledgments}

We thank the staff of the GMRT who have made these observations possible. The GMRT is run by the National Centre for Radio Astrophysics of the Tata Institute of Fundamental Research. We are grateful to the staff of the Anglo Australian Observatory and the staff of the Siding Spring Observatory for their assistance.  
This research was supported under Australian Research Council's Discovery Projects funding scheme (project number 0559688).
This research has made use of the GOLDMine Database.
We are indebted to
Andrew Hopkins,
Agris Kalnajs, 
Stefan Keller,
Wolfgang Kerzendorf,
Bruce Peterson
and Eduard Westra
for their valuable help.



\label{lastpage}


\begin{thebibliography}{99}

\bibitem[\protect\citeauthoryear{{Baldwin}, {Phillips}, \&
  {Terlevich}}{{Baldwin} et~al.}{1981}]{baldwin81}
{Baldwin} J.~A., {Phillips} M.~M.,  {Terlevich} R., 1981, PASP, 93, 5

\bibitem[\protect\citeauthoryear{{Balogh} et~al.}{{Balogh}
  et~al.}{2004}]{balogh04}
{Balogh} M., et~al., 2004, MNRAS, 348, 1355

\bibitem[\protect\citeauthoryear{{Balogh} et~al.}{{Balogh}
  et~al.}{1998}]{balogh98}
{Balogh} M.~L., et~al., 1998, ApJ, 504, L75

\bibitem[\protect\citeauthoryear{{Balogh} et~al.}{{Balogh}
  et~al.}{1999}]{balogh99}
{Balogh} M.~L., {Morris} S.~L., {Yee} H.~K.~C., {Carlberg} R.~G.,  {Ellingson}
  E., 1999, ApJ, 527, 54

\bibitem[\protect\citeauthoryear{{Basilakos} et~al.}{{Basilakos}
  et~al.}{2007}]{basilakos07}
{Basilakos} S., {Plionis} M., {Kova{\v c}} K.,  {Voglis} N., 2007, MNRAS, 378,
  301

\bibitem[\protect\citeauthoryear{{Becker}, {White}, \& {Helfand}}{{Becker}
  et~al.}{1995}]{becker95}
{Becker} R.~H., {White} R.~L.,  {Helfand} D.~J., 1995, ApJ, 450, 559

\bibitem[\protect\citeauthoryear{{Bell}}{{Bell}}{2003}]{bell03}
{Bell} E.~F., 2003, ApJ, 586, 794

\bibitem[\protect\citeauthoryear{{Bessell}, {Castelli}, \& {Plez}}{{Bessell}
  et~al.}{1998}]{bessell98}
{Bessell} M.~S., {Castelli} F.,  {Plez} B., 1998, A\&A, 333, 231

\bibitem[\protect\citeauthoryear{{Bravo-Alfaro} et~al.}{{Bravo-Alfaro}
  et~al.}{2000}]{bravo-alfaro00}
{Bravo-Alfaro} H., {Cayatte} V., {van Gorkom} J.~H.,  {Balkowski} C., 2000, AJ,
  119, 580

\bibitem[\protect\citeauthoryear{{Broeils} \& {Rhee}}{{Broeils} \&
  {Rhee}}{1997}]{broeils97}
{Broeils} A.~H.,  {Rhee} M.-H., 1997, A\&A, 324, 877

\bibitem[\protect\citeauthoryear{{Butcher} \& {Oemler}}{{Butcher} \&
  {Oemler}}{1984}]{butcher84}
{Butcher} H.,  {Oemler} A., 1984, ApJ, 285, 426

\bibitem[\protect\citeauthoryear{{Calzetti}}{{Calzetti}}{1997}]{calzetti97}
{Calzetti} D., 1997, AJ, 113, 162

\bibitem[\protect\citeauthoryear{{Carlberg} et~al.}{{Carlberg}
  et~al.}{1997}]{carlberg97}
{Carlberg} R.~G., et~al., 1997, ApJ, 485, L13

\bibitem[\protect\citeauthoryear{{Catinella} et~al.}{{Catinella}
  et~al.}{2008}]{catinella08} Catinella, B., Haynes, M.~P., Giovanelli, R., Gardner, J.~P., \& Connolly, A.~J.\ 2008, \apjl, 685, L13 

\bibitem[\protect\citeauthoryear{{Cayatte} et~al.}{{Cayatte}
  et~al.}{1990}]{cayatte90}
{Cayatte} V., {van Gorkom} J.~H., {Balkowski} C.,  {Kotanyi} C., 1990, AJ, 100,
  604

\bibitem[\protect\citeauthoryear{{Chengalur}, {Braun}, \&
  {Wieringa}}{{Chengalur} et~al.}{2001}]{chengalur01}
{Chengalur} J.~N., {Braun} R.,  {Wieringa} M., 2001, A\&A, 372, 768

\bibitem[\protect\citeauthoryear{{Chung} et~al.}{{Chung}
  et~al.}{2007}]{chung07}
{Chung} A., {van Gorkom} J.~H., {Kenney} J.~D.~P.,  {Vollmer} B., 2007, ApJ,
  659, L115

\bibitem[\protect\citeauthoryear{{Condon}, {Cotton}, \& {Broderick}}{{Condon}
  et~al.}{2002}]{condon02}
{Condon} J.~J., {Cotton} W.~D.,  {Broderick} J.~J., 2002, AJ, 124, 675

\bibitem[\protect\citeauthoryear{{Cortese} et~al.}{{Cortese}
  et~al.}{2008}]{cortese08}
{Cortese} L., et~al., 2008, MNRAS, 383, 1519

\bibitem[\protect\citeauthoryear{{Couch} \& {Sharples}}{{Couch} \&
  {Sharples}}{1987}]{couch87}
{Couch} W.~J.,  {Sharples} R.~M., 1987, MNRAS, 229, 423

\bibitem[\protect\citeauthoryear{{De Propris} et~al.}{{De Propris}
  et~al.}{2003}]{depropris03}
{De Propris} R., et~al., 2003, MNRAS, 342, 725

\bibitem[\protect\citeauthoryear{{De Propris} et~al.}{{De Propris}
  et~al.}{2004}]{depropris04}
{De Propris} R., et~al., 2004, MNRAS, 351, 125

\bibitem[\protect\citeauthoryear{{Diaferio} et~al.}{{Diaferio}
  et~al.}{2001}]{diaferio01}
{Diaferio} A., et~al., 2001, MNRAS, 323, 999

\bibitem[\protect\citeauthoryear{{Doyle} \& {Drinkwater}}{{Doyle} \&
  {Drinkwater}}{2006}]{doyle06}
{Doyle} M.~T.,  {Drinkwater} M.~J., 2006, MNRAS, 372, 977

\bibitem[\protect\citeauthoryear{{Dressler}}{{Dressler}}{1980}]{dressler80}
{Dressler} A., 1980, ApJ, 236, 351

\bibitem[\protect\citeauthoryear{{Dressler} et~al.}{{Dressler}
  et~al.}{1997}]{dressler97}
{Dressler} A.,  et~al., 1997, ApJ, 490, 577

\bibitem[\protect\citeauthoryear{{Dressler} et~al.}{{Dressler}
  et~al.}{2004}]{dressler04}
{Dressler} A.,  et~al., 2004, ApJ, 617, 867

\bibitem[\protect\citeauthoryear{{Farouki} \& {Shapiro}}{{Farouki} \&
  {Shapiro}}{1981}]{farouki81}
{Farouki} R.,  {Shapiro} S.~L., 1981, ApJ, 243, 32

\bibitem[\protect\citeauthoryear{{Fasano} et~al.}{{Fasano}
  et~al.}{2000}]{fasano00}
{Fasano} G.,  et~al., 2000, ApJ, 542, 673

\bibitem[\protect\citeauthoryear{{G{\' o}mez} et~al.}{{G{\' o}mez}
  et~al.}{2003}]{gomez03}
{G{\' o}mez} P.~L.,  et~al., 2003, ApJ, 584, 210

\bibitem[\protect\citeauthoryear{{Gavazzi} et~al.}{{Gavazzi}
  et~al.}{2003}]{gavazzi03}
{Gavazzi} G., {Boselli} A., {Donati} A., {Franzetti} P.,  {Scodeggio} M., 2003,
  \aap, 400, 451

\bibitem[\protect\citeauthoryear{{Gavazzi} et~al.}{{Gavazzi}
  et~al.}{2005}]{gavazzi05}
{Gavazzi} G., {Boselli} A., {van Driel} W.,  {O'Neil} K., 2005, A\&A, 429, 439

\bibitem[\protect\citeauthoryear{{Gavazzi} et~al.}{{Gavazzi}
  et~al.}{2006}]{gavazzi06}
{Gavazzi} G., {O'Neil} K., {Boselli} A.,  {van Driel} W., 2006, A\&A, 449, 929

\bibitem[\protect\citeauthoryear{{Gunn} \& {Gott}}{{Gunn} \&
  {Gott}}{1972}]{gunn72}
{Gunn} J.~E.,  {Gott} J.~R.~I., 1972, ApJ, 176, 1

\bibitem[\protect\citeauthoryear{{Hashimoto} et~al.}{{Hashimoto}
  et~al.}{1998}]{hashimoto98}
{Hashimoto} Y., {Oemler} A.~J., {Lin} H.,  {Tucker} D.~L., 1998, ApJ, 499, 589

\bibitem[\protect\citeauthoryear{{Haynes}, {Giovanelli}, \&
  {Chincarini}}{{Haynes} et~al.}{1984}]{haynes84}
{Haynes} M.~P., {Giovanelli} R.,  {Chincarini} G.~L., 1984, ARA\&A, 22, 445

\bibitem[\protect\citeauthoryear{{Hopkins}}{{Hopkins}}{2004}]{hopkins04}
{Hopkins} A.~M., 2004, ApJ, 615, 209

\bibitem[Hopkins et al.(2008)]{hopkins08} Hopkins, A.~M., 
McClure-Griffiths, N.~M., \& Gaensler, B.~M.\ 2008, \apjl, 682, L13 

\bibitem[\protect\citeauthoryear{{Kauffmann} et~al.}{{Kauffmann}
  et~al.}{2004}]{kauffmann04}
{Kauffmann} G., et~al., 2004, MNRAS, 353,
  713

\bibitem[\protect\citeauthoryear{{Kennicutt}}{{Kennicutt}}{1983}]{kennicutt83}
{Kennicutt} R.~C., Jr., 1983, ApJ, 272, 54

\bibitem[\protect\citeauthoryear{{Kewley}, {Geller}, \& {Jansen}}{{Kewley}
  et~al.}{2004}]{kewley04}
{Kewley} L.~J., {Geller} M.~J.,  {Jansen} R.~A., 2004, AJ, 127, 2002

\bibitem[\protect\citeauthoryear{{Kim}, {Goobar}, \& {Perlmutter}}{{Kim}
  et~al.}{1996}]{kim96}
{Kim} A., {Goobar} A.,  {Perlmutter} S., 1996, PASP, 108, 190

\bibitem[\protect\citeauthoryear{{Lah} et~al.}{{Lah} et~al.}{2007}]{lah07}
{Lah} P., 2007, MNRAS, 376, 1357

\bibitem[\protect\citeauthoryear{{Larson}, {Tinsley}, \& {Caldwell}}{{Larson}
  et~al.}{1980}]{larson80}
{Larson} R.~B., {Tinsley} B.~M.,  {Caldwell} C.~N., 1980, ApJ, 237, 692

\bibitem[\protect\citeauthoryear{{Lewis} et~al.}{{Lewis}
  et~al.}{2002}]{lewis02}
{Lewis} I.,  et~al., 2002, MNRAS, 334, 673

\bibitem[\protect\citeauthoryear{{Lilly} et~al.}{{Lilly}
  et~al.}{1996}]{lilly96}
{Lilly} S.~J., {Le Fevre} O., {Hammer} F.,  {Crampton} D., 1996, ApJ, 460, L1

\bibitem[\protect\citeauthoryear{{Madau} et~al.}{{Madau}
  et~al.}{1996}]{madau96}
{Madau} P., et~al., 1996, MNRAS, 283, 1388

\bibitem[\protect\citeauthoryear{{McGaugh} et~al.}{{McGaugh}
  et~al.}{2000}]{mcgaugh00}
{McGaugh} S.~S., {Schombert} J.~M., {Bothun} G.~D.,  {de Blok} W.~J.~G., 2000,
  ApJ, 533, L99

\bibitem[\protect\citeauthoryear{{Meyer} et~al.}{{Meyer}
  et~al.}{2004}]{meyer04}
{Meyer} M.~J.,  et~al., 2004, MNRAS, 350, 1195

\bibitem[\protect\citeauthoryear{{Moore} et~al.}{{Moore}
  et~al.}{1996}]{moore96}
{Moore} B., {Katz} N., {Lake} G., {Dressler} A.,  {Oemler} A., 1996, Nature,
  379, 613

\bibitem[\protect\citeauthoryear{{Moore}, {Lake}, \& {Katz}}{{Moore}
  et~al.}{1998}]{moore98}
{Moore} B., {Lake} G.,  {Katz} N., 1998, ApJ, 495, 139

\bibitem[\protect\citeauthoryear{{Niklas}, {Klein}, \& {Wielebinski}}{{Niklas}
  et~al.}{1997}]{niklas97}
{Niklas} S., {Klein} U.,  {Wielebinski} R., 1997, A\&A, 322, 19

\bibitem[\protect\citeauthoryear{{Ota} \& {Mitsuda}}{{Ota} \&
  {Mitsuda}}{2004}]{ota04}
{Ota} N.,  {Mitsuda} K., 2004, A\&A, 428, 757

\bibitem[\protect\citeauthoryear{{Pimbblet} et~al.}{{Pimbblet}
  et~al.}{2002}]{pimbblet02}
{Pimbblet} K.~A., et~al., 2002, MNRAS, 331, 333

\bibitem[\protect\citeauthoryear{{Poggianti}}{{Poggianti}}{1997}]{poggianti97}
{Poggianti} B.~M., 1997, \aaps, 122, 399

\bibitem[\protect\citeauthoryear{{Poggianti} et~al.}{{Poggianti}
  et~al.}{1999}]{poggianti99}
{Poggianti} B.~M.,  et~al., 1999, ApJ, 518, 576

\bibitem[\protect\citeauthoryear{{Poggianti}}{{Poggianti}}{2004}]{poggianti04}
{Poggianti} B., 2004, {Evolution of galaxies in clusters}, in {Dettmar} R.,
  {Klein} U.,  {Salucci} P. (eds.), Baryons in Dark Matter Halos

\bibitem[Poggianti et al.(2008)]{poggianti08} Poggianti, B.~M., et 
al.\ 2008, \apj, 684, 888 

\bibitem[\protect\citeauthoryear{{Pracy} et~al.}{{Pracy}
  et~al.}{2009}]{pracy08}
{Pracy} M.~B., {Lah} P., {Colless} M., {de Propris} R.,  {Owers} M., 2009,
  {in preparation}

\bibitem[\protect\citeauthoryear{{Prochaska}, {Herbert-Fort}, \&
  {Wolfe}}{{Prochaska} et~al.}{2005}]{prochaska05}
{Prochaska} J.~X., {Herbert-Fort} S.,  {Wolfe} A.~M., 2005, ApJ, 635, 123

\bibitem[\protect\citeauthoryear{{Quilis}, {Moore}, \& {Bower}}{{Quilis}
  et~al.}{2000}]{quilis00}
{Quilis} V., {Moore} B.,  {Bower} R., 2000, Science, 288, 1617

\bibitem[\protect\citeauthoryear{{Rao}, {Turnshek}, \& {Nestor}}{{Rao}
  et~al.}{2006}]{rao06}
{Rao} S.~M., {Turnshek} D.~A.,  {Nestor} D.~B., 2006, ApJ, 636, 610

\bibitem[\protect\citeauthoryear{{Richstone}}{{Richstone}}{1976}]{richstone76}
{Richstone} D.~O., 1976, ApJ, 204, 642

\bibitem[\protect\citeauthoryear{{Roberts} \& {Haynes}}{{Roberts} \&
  {Haynes}}{1994}]{roberts94}
{Roberts} M.~S.,  {Haynes} M.~P., 1994, ARA\&A, 32, 115

\bibitem[Rybicki \& Lightman(1986)]{rybicki86} Rybicki, G.~B., \& Lightman, A.~P.\ 1986, Radiative Processes in Astrophysics, by George B.~Rybicki, Alan P.~Lightman, pp.~400.~ISBN 0-471-82759-2.~Wiley-VCH , June 1986.,  

\bibitem[\protect\citeauthoryear{{Schlegel}, {Finkbeiner}, \&
  {Davis}}{{Schlegel} et~al.}{1998}]{schlegel98}
{Schlegel} D.~J., {Finkbeiner} D.~P.,  {Davis} M., 1998, ApJ, 500, 525

\bibitem[\protect\citeauthoryear{{Schmidt} et~al.}{{Schmidt}
  et~al.}{1998}]{schmidt98}
{Schmidt} B.~P., 1998, ApJ, 507, 46

\bibitem[\protect\citeauthoryear{{Solanes} et~al.}{{Solanes}
  et~al.}{2001}]{solanes01}
{Solanes} J.~M.,  et~al., 2001,
  ApJ, 548, 97

\bibitem[\protect\citeauthoryear{{Sullivan} et~al.}{{Sullivan}
  et~al.}{2001}]{sullivan01}
{Sullivan} M., et~al., 2001, ApJ, 558, 72

\bibitem[\protect\citeauthoryear{{Toomre} \& {Toomre}}{{Toomre} \&
  {Toomre}}{1972}]{toomre72}
{Toomre} A.,  {Toomre} J., 1972, ApJ, 178, 623

\bibitem[\protect\citeauthoryear{{Verheijen}}{{Verheijen}}{2004}]{verheijen04}
{Verheijen} M.~A.~W., 2004, {Galaxy evolution in dense environments: a concise
  HI perspective}, in IAU Colloq. 195: Outskirts of Galaxy Clusters: Intense
  Life in the Suburbs, p. 394

\bibitem[\protect\citeauthoryear{{Verheijen} et~al.}{{Verheijen}
  et~al.}{2007}]{verheijen07}
{Verheijen} M., et~al., 2007, ApJ, 668, L9

\bibitem[\protect\citeauthoryear{{Wieringa}, {de Bruyn}, \&
  {Katgert}}{{Wieringa} et~al.}{1992}]{wieringa92}
{Wieringa} M.~H., {de Bruyn} A.~G.,  {Katgert} P., 1992, A\&A, 256, 331

\bibitem[\protect\citeauthoryear{{Wu}, {Xue}, \& {Fang}}{{Wu}
  et~al.}{1999}]{wu99}
{Wu} X.-P., {Xue} Y.-J.,  {Fang} L.-Z., 1999, ApJ, 524, 22

\bibitem[\protect\citeauthoryear{{Zwaan}}{{Zwaan}}{2000}]{zwaan00}
{Zwaan} M.~A., 2000, Ph.D. thesis, PhD Thesis, Groningen: Rijksuniversiteit,
  2000 152 p.~Proefschrift, Rijksuniversiteit Groningen, 2000

\bibitem[\protect\citeauthoryear{{Zwaan} et~al.}{{Zwaan}
  et~al.}{1997}]{zwaan97}
{Zwaan} M.~A., {Briggs} F.~H., {Sprayberry} D.,  {Sorar} E., 1997, ApJ, 490,
  173

\bibitem[\protect\citeauthoryear{{Zwaan}, {van Dokkum}, \& {Verheijen}}{{Zwaan}
  et~al.}{2001}]{zwaan01}
{Zwaan} M.~A., {van Dokkum} P.~G.,  {Verheijen} M.~A.~W., 2001, Science, 293,
  1800

\bibitem[\protect\citeauthoryear{{Zwaan} et~al.}{{Zwaan}
  et~al.}{2005}]{zwaan05}
{Zwaan} M.~A., {Meyer} M.~J., {Staveley-Smith} L.,  {Webster} R.~L., 2005,
  MNRAS, 359, L30

\end{thebibliography}
\end{document}